%% file: main.tex
\documentclass[final,abstractlogo]{ise}

\usepackage{nicefrac}
\usepackage{colortbl}
\usepackage{pifont}
\usepackage{fontawesome5}
\usepackage{twemojis}
\usepackage{tikz}
\usetikzlibrary{arrows.meta}
\usepackage{pgf-pie}
\usepackage{forest}
\useforestlibrary{edges}
\usepackage{float}
\usepackage{longtable}
\usepackage{tabularx}
\usepackage[framemethod=tikz]{mdframed}
\setcitestyle{authoryear,round}
\makeatletter
\renewcommand{\hyper@natlinkbreak}[2]{#1}
\makeatother

\definecolor{targetframework}{HTML}{D9EEF7}
\definecolor{targetmemory}{HTML}{E2F0D9}
\definecolor{targetskill}{HTML}{FCE4D6}
\definecolor{targetmodel}{HTML}{E4DFEC}
\definecolor{targetworkflow}{HTML}{FFF2CC}
\definecolor{targetenvironment}{HTML}{F4CCCC}
\definecolor{tableheader}{HTML}{1F4E78}
\definecolor{definitionblue}{HTML}{1F4E78}
\definecolor{definitionbackground}{HTML}{F0F7FD}
\definecolor{tablerow}{HTML}{F7F9FC}
\definecolor{conceptRequired}{RGB}{0,176,0}
\definecolor{conceptOptional}{RGB}{255,0,0}
\definecolor{signalE}{HTML}{D95B47}
\definecolor{signalT}{HTML}{3D938D}
\definecolor{signalD}{HTML}{D9A441}
\definecolor{signalR}{HTML}{557EAA}
\definecolor{signalH}{HTML}{6B925F}
\definecolor{signalQ}{HTML}{8870A2}
\hypersetup{
  colorlinks=true,
  linkcolor=tableheader,
  citecolor=blue!55!black,
  urlcolor=blue!55!black,
  pdftitle={Self-Evolving Coding Agents},
  pdfauthor={Hao Zhou, Haichuan Hu, Tianyu Luo, Ye Shang, Quanjun Zhang},
  pdfsubject={ISE-formatted adaptation of arXiv:2608.03392v2}
}

\newcommand{\conceptrequired}{%
  \textcolor{conceptRequired}{{\usefont{U}{msa}{m}{n}\symbol{"58}}}}
\newcommand{\conceptoptional}{\textcolor{conceptOptional}{\ding{55}}}

\newcommand{\githublink}[1]{%
  \href{#1}{\textcolor{black}{\faGithub}}}
\newcommand{\hflogo}{%
  \raisebox{-0.13em}{\includegraphics[height=1em]{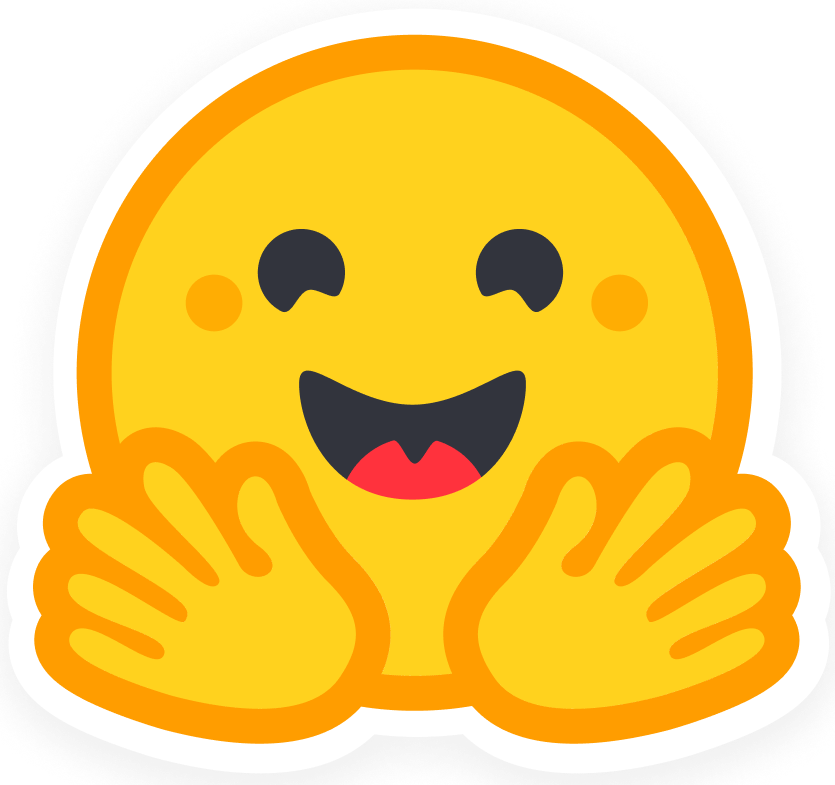}}}
\newcommand{\hflink}[1]{%
  \href{#1}{\hflogo}}
\newcommand{\ghhflinks}[2]{%
  \githublink{#1}\hspace{0.18em}\hflink{#2}}
\newcommand{\externallink}[1]{%
  \href{#1}{\textcolor{blue!55!black}{\raisebox{-0.12ex}{\normalsize\faIcon{external-link-alt}}}}}

\newmdenv[
  linecolor=tableheader!48,
  linewidth=0.75pt,
  roundcorner=2pt,
  backgroundcolor=black!1,
  innerleftmargin=7pt,
  innerrightmargin=7pt,
  innertopmargin=7pt,
  innerbottommargin=2pt,
  skipabove=8pt,
  skipbelow=10pt,
  splittopskip=6pt,
  splitbottomskip=4pt
]{productcatalog}
\newmdenv[
  leftline=true,
  rightline=false,
  topline=false,
  bottomline=false,
  linecolor=targetenvironment!68!black,
  linewidth=2pt,
  backgroundcolor=targetenvironment!24,
  innerleftmargin=9pt,
  innerrightmargin=9pt,
  innertopmargin=7pt,
  innerbottommargin=7pt,
  skipabove=8pt,
  skipbelow=9pt
]{futureharnessquote}

\title{Self-Evolving Coding Agents}
\shorttitle{Self-Evolving Coding Agents}

\author{%
  {\sffamily\small\bfseries\color{ISEInk}
    Hao Zhou$^{1}$ \quad Haichuan Hu$^{1}$ \quad
    Tianyu Luo$^{2}$ \quad Ye Shang$^{2}$}\\
  {\sffamily\small\bfseries\color{ISEInk}
    Chunrong Fang$^{2}$ \quad Zhenyu Chen$^{2}$ \quad
    Liang Xiao$^{1}$ \quad Quanjun Zhang$^{1}$}\\[3pt]
  $^{1}$Nanjing University of Science and Technology\\
  $^{2}$Nanjing University\\[3pt]
  \texttt{125106010779@njust.edu.cn},
  \texttt{huhaichuan2024@gmail.com},
  \texttt{Ty\_L191025@outlook.com}\\
  \texttt{yeshang@smail.nju.edu.cn},
  \texttt{fangchunrong@nju.edu.cn},
  \texttt{zychen@nju.edu.cn}\\
  \texttt{xiaoliang@mail.njust.edu.cn},
  \texttt{quanjunzhang@njust.edu.cn}
}

\date{August 20, 2026}
\project{\href{https://github.com/iSEngLab/Awesome-Self-Evolving-Coding-Agents}{iSEngLab/Awesome-Self-Evolving-Coding-Agents}}

\newcommand{\surveyheading}[1]{%
  \phantomsection
  \addcontentsline{toc}{subsubsection}{#1}%
  \textbf{#1.}}
\newcommand{\surveyparagraph}[3]{%
  \phantomsection
  \addcontentsline{toc}{#1}{#2}%
  \paragraph{#3}}

\begin{document}
\raggedbottom

\maketitle

\begin{abstract}
Large language models (LLMs) have moved software automation beyond code
generation toward coding agents that inspect repositories, invoke tools, run
tests, and repair failures. Yet many agents remain episodic, using feedback
to solve the current task without changing how they approach the next one.
Software engineering offers a particularly suitable setting for overcoming
this limitation. Execution provides concrete outcomes and diagnostics, while
coding trajectories and versioned repositories make experience traceable and
often reproducible. Self-evolving coding agents use such experience to update
the reusable components that shape their later work. This survey distinguishes
these agent updates from changes to task code and identifies assets,
architecture, and model parameters as progressively deeper intervention
targets. As intervention deepens, a lesson can influence more decisions and
become harder to attribute or reverse. We examine coding signals and adoption
timing alongside this taxonomy to show how similar feedback can produce
different updates and when those changes begin to shape later execution.
Early adoption can use fresh feedback but may rely on incomplete evidence,
while waiting permits broader validation but delays reuse. Finally, we
examine benchmarks and emerging products, finding that practical systems
favor inspectable assets and editable harnesses while evaluation still
centers largely on individual task outcomes. Passing tests support the
current software artifact under checked conditions; showing that an agent
has improved requires evidence that retained changes benefit later decisions.

\end{abstract}

\begin{figure}[H]
  \centering
  \includegraphics[width=\linewidth]{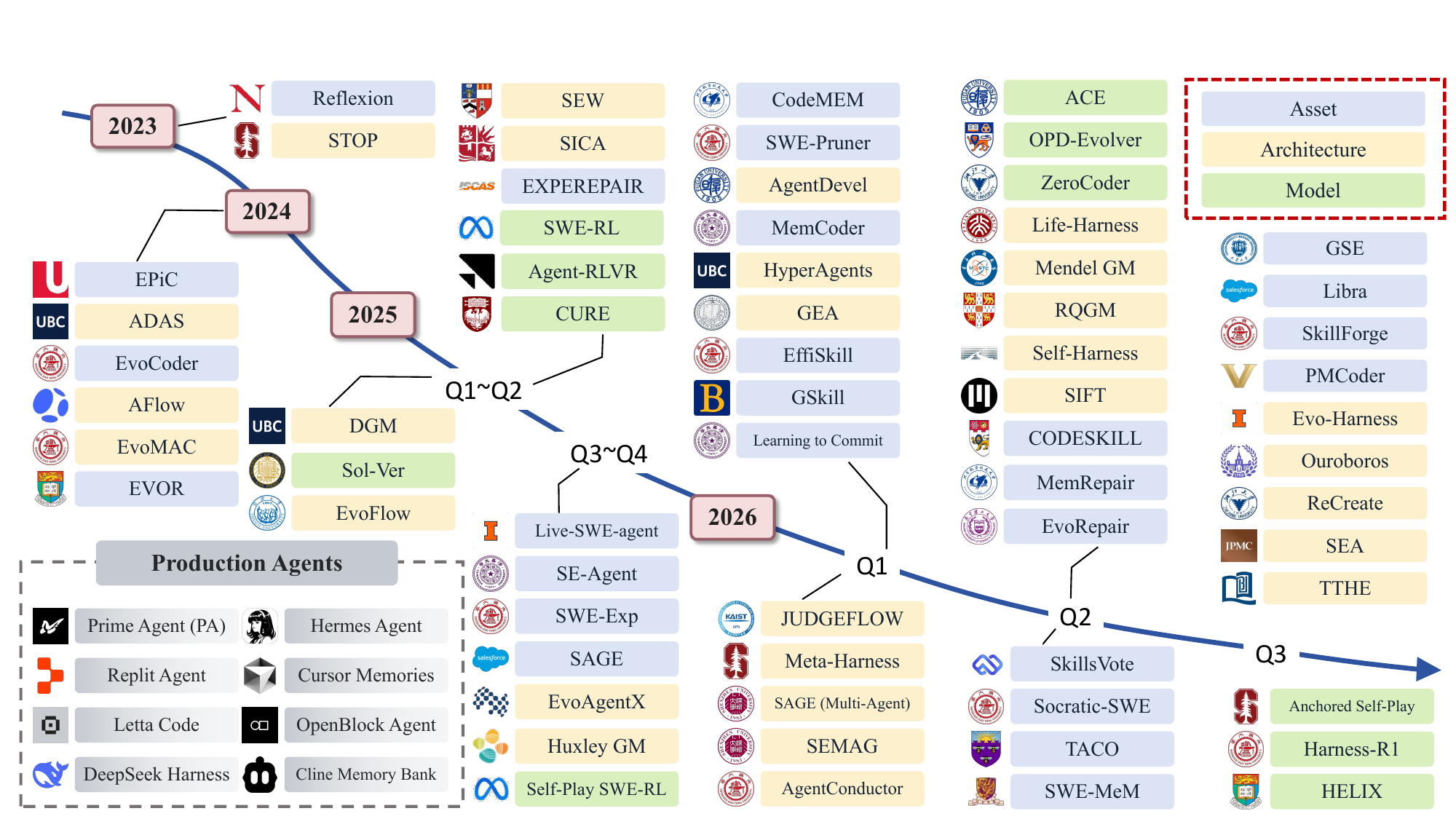}
  \caption{Evolutionary landscape of representative agent methods and coding products from 2023 to 2026. Colors indicate the primary update target in assets, architecture, or model parameters.}
  \label{fig:evolutionary-landscape}
\end{figure}

\clearpage
\setcounter{tocdepth}{2}
\begingroup
  \fontsize{12}{16}\selectfont
  \setlength{\parskip}{2pt}
  \hypersetup{linktoc=all}
  \tableofcontents
\endgroup
\clearpage

\section{Introduction}

\begin{figure}[H]
  \centering
  \includegraphics[width=\linewidth]{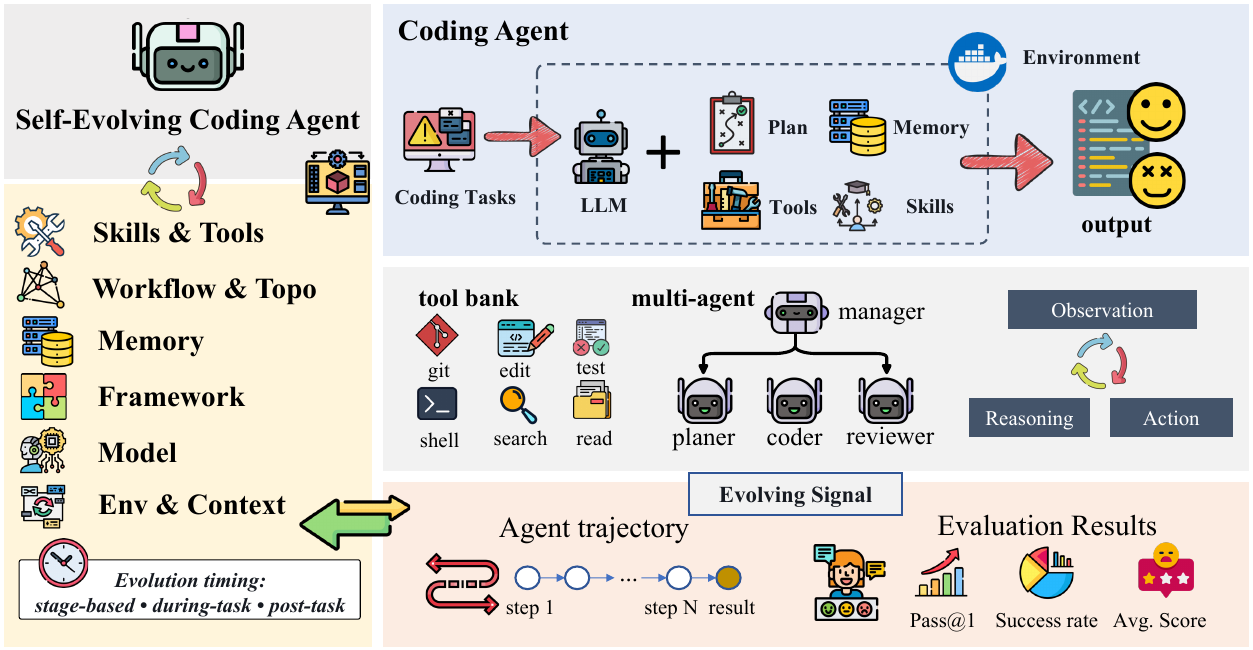}
  \caption{Overview of self-evolving coding agents.}
  \label{fig:overview}
\end{figure}

Large language models have moved software automation beyond code generation
\citep{chen2021evaluating,austin2021program} toward agents that inspect
repositories, use command-line tools, revise files, and test candidate
patches. SWE-agent, OpenHands, AutoCodeRover, and related repair systems
exemplify this shift toward repository-level execution
\citep{yang2024sweagent,wang2024openhands,zhang2024autocoderover,repairagent2024,masai2024}. Benchmarks
based on real issues make the resulting repairs measurable, while tying each
outcome to a particular repository and environment
\citep{jimenez2023swebench,swebenchpro2025,zhang2025swebenchlive}.
Figure~\ref{fig:evolutionary-landscape} places this development alongside
the broader rise of self-evolving agents. The unresolved step is what happens
beyond the immediate repair. An agent may use a failed test to fix one patch
yet carry no reusable change into later tasks, a limitation that motivates
experience-driven issue resolution~\citep{chen2026sweexp}.
Figure~\ref{fig:overview} sketches the feedback and update loop at issue.

Coding offers unusually concrete evidence for studying this step. Tests
provide executable outcomes; compiler messages and stack traces narrow
failures; and trajectories record the searches, edits, and retries that
preceded a patch
\citep{swegym2025,jain2025r2egym,debug2fix2026,executionagent2024}.
Issues, diffs, and versioned repositories connect those observations to the
software conditions under which they arose~\citep{jimenez2023swebench}.
Recorded dependencies and tool versions can make many observations
re-executable, although changing environments limit that reproducibility
\citep{hu2025repo2run,eliseeva2025envbench}. Specific, traceable feedback is
still incomplete feedback. A passing test checks only selected behavior, and
functional success need not establish patch quality or security
\citep{chen2024patchquality,wei2025patcheval}.
Interpreting this evidence requires distinguishing changes to the program
being built from changes to the agent that builds it. The first change is
visible in a patch and its validation results.
The second requires an experience-derived update to a reusable agent
component, followed by its use in later execution. Earlier agent research
showed how reflection and stored experience can inform later attempts
\citep{shinn2023reflexion,zhao2024expel,wang2025awm}; coding-specific
systems now retain repair knowledge across issues~\citep{chen2026sweexp}.
We use
\emph{self-evolving coding agent} for a system whose retained updates improve
subsequent coding behavior. A successful patch may yield an overgeneralized
memory, while an unsuccessful attempt may contain a useful diagnosis. The
current artifact's result therefore cannot by itself establish the value of
an agent update.

The same debugging experience can enter an agent at different depths.
SWE-Exp and Repository Memory retain information for later retrieval, while
CODESKILL and Socratic-SWE turn trajectories into reusable procedures
\citep{chen2026sweexp,wang2026repositorymemory,li2026codeskill,xiao2026socraticswe}.
Live-SWE-agent creates callable utilities during issue resolution
\citep{xia2025livesweagent}. Other methods revise the execution structure,
as in source-level agent evolution, workflow search, harness repair, or team
coordination
\citep{zhang2025darwin,zhang2024aflow,chen2026harnessfix,hu2024evomac}; Agent-RLVR and
SWE-RL instead update model parameters using coding feedback
\citep{da2025agentrlvr,selfplaysoftwareagents2026}. We organize these targets
as assets, architecture, and model parameters. The ordering reflects typical
intervention depth: a resource can often be edited locally, whereas changes
to control flow or learned policy can affect more decisions. Depth does not
guarantee benefit.
The update target alone does not explain how evolution occurs. Coding signals
determine what evidence a method consumes, while adoption timing determines
how much evidence is available before a change enters execution. A utility
can be created within a task, team instructions can change between development
rounds, experience can be retained after an issue, and a model update can
aggregate trajectories across tasks
\citep{xia2025livesweagent,hu2024evomac,chen2026sweexp,da2025agentrlvr}.
Earlier adoption
allows a rapid response to local feedback; later adoption permits broader
validation but delays reuse.

Evaluation must determine whether a retained change helped after reuse,
rather than crediting it for the task that produced it. Issue-resolution
benchmarks measure patches, while sequential agent benchmarks make persistence
and later-task behavior part of the protocol
\citep{jimenez2023swebench,swebenchpro2025,jiang2026seaeval,xue2026pastbench,zhang2026skillflow}. Coding tasks
and their tests may also be used to construct updates, select candidates, and
report final performance. Reusing related cases across these roles can make
adaptation to an evaluator look like general progress. Studies of harness
evolution further show why search budgets and harness choice matter when
attributing a gain
\citep{wang2026rethinkingharnesseval,vats2026scaffoldeffect,bensghaier2026harnessevolution}.
An evolution study should therefore identify what persists, compare updated
and frozen agents under matched conditions on held-out tasks, and report
regressions and costs alongside gains. Those controls separate improvement
from extra attempts, computation, or familiarity with evaluation cases.

\surveyparagraph{subsection}{Relation to Existing Surveys}{Relation to Existing Surveys}
Existing surveys examine either agents that evolve across domains or agents
that perform software-engineering tasks. General self-evolving-agent surveys
discuss programming alongside many other applications, without centering
the distinctive evidence produced by tests, diagnostics, and versioned
repositories~\citep{gao2026surveyselfevolvingagents,fang2025selfevolvingaisurvey}.
Software-engineering agent surveys instead organize development tasks,
architectures, and collaboration; retaining an experience-derived agent
update is not their inclusion criterion
\citep{liu2024agent4sesurvey,wang2025agentsse}. This distinction matters
because coding work can change both the project under repair and the agent
that repairs it. \emph{Code as Agent Harness} recognizes code as agent
infrastructure, but studies that substrate across multiple domains
\citep{ning2026codeharness}. Our unit of analysis is the persistent change
to a coding agent produced by software-engineering experience. We ask which
component retains it, what software evidence motivates it, when it is
adopted, and whether its reuse improves later coding behavior. A correct
patch establishes a result for one task; it does not establish that the
agent learned a useful capability. Table~\ref{tab:related-surveys} compares
these boundaries.

\begin{table}[H]
  \caption{Comparison of related surveys by focus, persistent updates, and the role of coding.}
  \label{tab:related-surveys}
  \centering
  \scriptsize
  \setlength{\tabcolsep}{3pt}
  \renewcommand{\arraystretch}{1.20}
  \begin{tabularx}{\linewidth}{@{}>{\raggedright\arraybackslash}p{0.28\linewidth}
                      >{\raggedright\arraybackslash}p{0.27\linewidth}
                      >{\raggedright\arraybackslash}p{0.17\linewidth}
                      >{\raggedright\arraybackslash}X@{}}
    \toprule
    \bfseries Survey & \bfseries Main focus & \bfseries Persistent agent update &
    \bfseries Role of coding \\
    \midrule
    \emph{General self-evolving agents}~\citep{gao2026surveyselfevolvingagents} &
    Evolution mechanisms across domains & Central &
    One application domain \\
    \emph{Comprehensive self-evolving AI agents}~\citep{fang2025selfevolvingaisurvey} &
    Feedback-driven agent-system optimization & Central &
    One specialized domain \\
    \emph{LLM agents for SE}~\citep{liu2024agent4sesurvey} &
    LLM agent designs and SE tasks & Not required &
    Core task setting \\
    \emph{Agents in SE}~\citep{wang2025agentsse} &
    Agent applications and roles in SE & Not required &
    Core task setting \\
    \emph{Code as Agent Harness}~\citep{ning2026codeharness} &
    Code-based agent infrastructure &
    One mechanism & Cross-domain execution substrate \\
    \midrule
    \rowcolor[rgb]{0.88,0.95,1}
    \bfseries Our Survey & Coding-agent updates and later benefit &
    Required; tested after reuse &
    Update and evaluation evidence \\
    \bottomrule
  \end{tabularx}
\end{table}

\enlargethispage{\baselineskip}
This survey makes three contributions that connect the target of evolution,
the coding evidence behind an update, and the evaluation of its later effects:
\begin{itemize}[leftmargin=*]
  \item \textbf{Definition and target taxonomy.} We distinguish changes to
  the coding agent from changes to the code it produces. We then classify
  persistent agent updates as changes to assets, architecture, or model
  parameters, ordered by intervention depth.
  \item \textbf{Coding signals and update mechanisms.} We classify the
  evidence available during software development and examine how methods
  turn it into persistent updates. We also compare when updates are adopted,
  showing how timing affects the evidence available before reuse.
  \item \textbf{Evaluation of agent improvement.} We distinguish success on
  the current coding task from improvement caused by a retained update. Our
  analysis of benchmarks identifies the matched, held-out, and longitudinal
  comparisons needed to measure later gains, regressions, and costs.
\end{itemize}

\newpage
The remainder of this survey is organized as follows.
Section~\ref{sec:background} defines coding agents and self-evolving agents,
then specifies the retention, reuse, and improvement requirements that
distinguish self-evolving coding agents. Section~\ref{sec:taxonomy}
introduces the target-centered taxonomy and explains the different depths
of asset, architecture, and model updates, then profiles the surveyed methods
by year and publication outlet. Section~\ref{sec:assets-self-evolution}
reviews how memory, skills, tools, and context evolve.
Section~\ref{sec:architecture-self-evolution} examines changes to harnesses,
workflows, and multi-agent coordination, while
Section~\ref{sec:model-self-evolution} addresses parameter adaptation.
Section~\ref{sec:coding-evolution-signals} analyzes coding signals, how
methods turn them into persistent changes, and when those changes are adopted.
Section~\ref{sec:benchmarks} reviews benchmarks and the controls needed to
measure later benefit. Section~\ref{sec:products} examines how products
expose and govern updates. Section~\ref{sec:challenges} discusses open
problems in feedback validity, reversibility, and generalization, and
Section~\ref{sec:conclusion} concludes.

\section{Background and Definitions}
\label{sec:background}

Coding agents perform software development tasks by interacting with
repositories, tools, and execution environments, while the agents themselves
are built from code, configurations, and model parameters. Code thus plays a
dual role: it is both a product of task solving and part of the system that
solves the task. This dual role allows coding capabilities to act on both task
artifacts and the agent itself, but modifying code does not inherently constitute
self-evolution. The key distinction is whether a change extends beyond improving
the current solution to persistently changing how subsequent tasks are solved.
Self-evolving agents embody this experience- and feedback-driven adaptation of
the agent itself. Self-evolving coding agents ground that process in coding
practice: task solving generates experience, experience drives agent updates,
and retained updates guide subsequent coding activity. The object of evolution
is the agent, not merely the programs it generates or modifies.

\subsection{Coding Agents}
\label{sec:coding-agents}

A coding agent is an LLM-based system that pursues software development
objectives through actions in a development environment. It decides what
information to obtain, what changes to make, and how to assess their effects.
Repository-level tasks require reasoning across existing files and dependencies
\citep{jimenez2023swebench}, often with context gathered during task execution.
SWE-agent and OpenHands support this process through interfaces for repository
navigation, editing, and execution \citep{yang2024sweagent,wang2024openhands}.
Interaction serves both to modify software and to investigate its behavior:
even an unsuccessful repair can provide evidence that guides the next attempt.

The surrounding system shapes what the model can observe and do, and how
its decisions are organized. CodeAct expresses actions through executable
code \citep{codeact2024}, AutoCodeRover retrieves context through structured
code search \citep{zhang2024autocoderover}, and Debug2Fix exposes runtime
behavior through interactive debugging \citep{debug2fix2026}. MetaGPT
coordinates specialized development roles \citep{hong2023metagpt}, while
MASAI assigns distinct objectives and strategies to modules \citep{masai2024}.
SGAgent bridges fault localization and patch generation with explicit,
context-informed repair suggestions \citep{zhang2026sgagent}; Trae Agent
organizes candidate generation, pruning, and selection \citep{traeagent2025}.
The value of these arrangements lies in the decisions they enable, alongside
their computational cost. Agentless offers a useful comparison: a fixed
localization, repair, and validation pipeline can also achieve competitive
repair results \citep{xia2024agentless}.

\subsection{Self-Evolving Agents}
\label{sec:self-evolving-agents}

A self-evolving agent uses experience and feedback to update its own reusable
state or components, allowing past interactions to shape subsequent behavior.
Updates can affect memory, skills, tools, architecture, or model parameters
\citep{fang2025selfevolvingaisurvey,gao2026surveyselfevolvingagents}.
Experience both informs current actions and provides a basis for revising
how the agent will act again. These revisions persist as part of its memory,
capabilities, or decision process beyond the interaction that produced them.

Experience can support adaptation without changing model weights: ExpeL
extracts reusable insights \citep{zhao2024expel}, Voyager accumulates
composable executable skills \citep{wang2023voyager}, and Agent Workflow
Memory induces reusable routines from task trajectories \citep{wang2025awm}.
A-MEM further revises existing memories and their links as new experiences
arrive \citep{xu2025amem}. Adaptation also reaches control logic through
G\"odel Agent's modification of its own implementation \citep{yin2025godelagent}.
Policy learning offers another route: WebRL builds an online training
curriculum from unsuccessful web interactions \citep{qi2025webrl}, while
Tool-R0 co-evolves a task generator and a tool-using solver through self-play
reinforcement learning \citep{acikgoz2026toolr0}. These mechanisms turn
experience into changes that subsequent execution can use. A retained record
contributes to self-evolution when it participates in this update-and-reuse
process; storage alone is insufficient. Reuse may occur within a continuing
task or across tasks, and human feedback or approval can accompany
system-generated updates. Retention and reuse establish an adaptation
mechanism, while reliable gains require evidence from evaluating the
updated agent.

\subsection{Self-Evolving Coding Agents}
\label{sec:self-evolving-coding-agents}

Coding agents acquire experience as they inspect repositories, execute
programs, and revise solutions. This experience can help complete the current
task, but it can also inform changes to the system performing the task.
Self-evolving coding agents connect software task execution with improvement
of the agent itself, turning coding experience into retained updates whose
reuse improves subsequent coding performance. Task solving supplies both
software artifacts and experience from which the agent can develop further
capabilities.

\begin{tcolorbox}[
  enhanced,
  title={Definition},
  fonttitle=\bfseries\large,
  coltitle=white,
  colframe=definitionblue,
  colbacktitle=definitionblue,
  colback=definitionbackground,
  boxrule=1.1pt,
  titlerule=0pt,
  arc=7pt,
  outer arc=7pt,
  boxsep=0pt,
  left=10pt,right=10pt,
  top=8pt,bottom=8pt,
  toptitle=4pt,bottomtitle=4pt,
  before skip=9pt,after skip=9pt]
A coding agent that uses experience and feedback from coding interactions
to update its own assets, architecture, or model parameters, and improves
its problem-solving capability, reliability, or efficiency in subsequent
coding activity by retaining and reusing those updates.
\end{tcolorbox}

To distinguish changes to the task solution from changes to the agent,
we represent the reusable agent configuration at adaptation round $t$ as
\begin{equation}
  \mathcal{A}_t=(\mathcal{S}_t,\mathcal{H}_t,\theta_t),
  \label{eq:agent-configuration}
\end{equation}
where $\mathcal{S}_t$ denotes assets (memory, skills, tools, and context),
$\mathcal{H}_t$ denotes architecture (harness, workflow, and multi-agent
organization), and $\theta_t$ denotes model parameters. These components
correspond to the three categories of evolution targets in our taxonomy and
jointly shape the agent's capabilities and decision process. For a software
task $q$ in environment $\mathcal{E}$, execution is represented by
\begin{equation}
  (z_t,\tau_t,f_t)=
  \operatorname{Execute}(\mathcal{A}_t,q,\mathcal{E}).
  \label{eq:coding-execution}
\end{equation}
Here, $z_t$ is the software artifact produced or revised, $\tau_t$ records
the agent's actions and resulting observations, and $f_t$ denotes feedback
on execution and its outcomes. The artifact is the immediate result of task
solving; the trajectory and feedback also provide experience for revising
the system that produced it.

An update mechanism $\mathcal{U}$ uses this experience to propose changes
to the reusable agent configuration:
\begin{equation}
  \widetilde{\mathcal{A}}_{t+1}
  =\mathcal{U}(\mathcal{A}_t,\tau_t,f_t).
  \label{eq:agent-evolution}
\end{equation}
The candidate may revise one or several components of $\mathcal{A}_t$ while
leaving the others unchanged. The tilde distinguishes a proposed
configuration from one adopted for subsequent execution. A proposal may be
accepted, revised, or discarded; for a single candidate, adoption is
expressed as
\begin{equation}
  \mathcal{A}_{t+1}=
  \begin{cases}
    \widetilde{\mathcal{A}}_{t+1}, & \text{if the candidate is accepted},\\
    \mathcal{A}_t, & \text{otherwise}.
  \end{cases}
  \label{eq:agent-adoption}
\end{equation}
Acceptance retains the candidate's changes in $\mathcal{A}_{t+1}$, making
them available to guide later execution. Rejection preserves the existing
configuration, while revision produces another candidate for consideration.

A single coding task can involve several rounds of adaptation, and the
effects of one update may extend to multiple later tasks; $t$ therefore
indexes adaptation rounds. An agent may use experience gained during an
ongoing task to adjust its remaining work \citep{xia2025livesweagent}, or
retain those adjustments for future tasks. Persistence allows experience
to have an effect beyond the operation that produced the update, giving
a coding attempt the potential to change how the agent solves problems
later. Yet retained changes may encode useful strategies or mistaken
generalizations from local experience. Only when their subsequent use
improves problem-solving capability, reliability, or efficiency can the
agent be said to have progressed through experience, rather than merely
accumulated changes.

\surveyparagraph{subsubsection}{The operational boundary}{The operational boundary}
We identify self-evolving coding agents using four criteria that examine
how code-related experience changes the agent itself and whether reusing
those changes improves subsequent coding performance:
\begin{enumerate}[label=(\arabic*),leftmargin=2.2em,itemsep=0.45em]
  \item \textbf{Update target.}
  The change must affect an identifiable component of $\mathcal{A}_t$.
  Repairing a failing patch or making a target program faster changes the
  task artifact $z_t$. Self-Refine's code-refinement loop revises generated
  outputs \citep{madaan2023selfrefine}; AlphaEvolve's search over candidate
  algorithms likewise optimizes task programs \citep{novikov2025alphaevolve}.
  These operations alone constitute task solving.

  \item \textbf{Update source.}
  The system must generate updates from code-related experience and feedback,
  such as repository interactions, execution traces, or test results.

  \item \textbf{Update reuse.}
  Updates must be retained as agent components and actually used in later
  coding execution, within the same task or across tasks. Storage alone is
  insufficient.

  \item \textbf{Capability improvement.}
  Reusing the updates must improve the agent's problem-solving capability,
  reliability, or efficiency in subsequent coding activity, supported by
  comparisons with the pre-update agent under comparable conditions.
\end{enumerate}

\surveyparagraph{subsubsection}{From retained experience to behavioral influence}{From retained experience to behavioral influence}
We analyze improvement through five links: \emph{experience collection,
component update, subsequent use, behavioral change, and effect evaluation}.
These links help locate where an evolution mechanism succeeds or breaks down.
Collection supplies observations, while updating converts them into reusable
components. This transition requires deciding what to retain and which
conditions govern its applicability. Subsequent use brings the update into
execution, but use alone does not establish contribution. A retrieved memory
may repeat available information, and an invoked skill may recommend an action
the agent would have taken anyway. Let $\pi_{\mathcal A}(\cdot\mid x)$ denote
the agent's action-selection distribution at a matched decision state $x$.
The behavioral difference induced by an update can be expressed as
\begin{equation}
  \delta_t(x)=d_{\mathrm{TV}}\!\left(
    \pi_{\mathcal A_{t+1}}(\cdot\mid x),
    \pi_{\mathcal A_t}(\cdot\mid x)
  \right),
  \label{eq:behavioral-change}
\end{equation}
where $d_{\mathrm{TV}}$ is the total variation distance between probability
distributions. Here, actions may include tool calls, code edits, or other
execution choices. A value of $\delta_t(x)=0$ indicates unchanged action
probabilities at that state, while $\delta_t(x)>0$ indicates a behavioral
change without establishing its benefit. To assess improvement, let
$\mathcal D$ denote an evaluation distribution over tasks or matched
continuation states within the intended scope of reuse, $B$ a common
execution budget, and $s(\mathcal A,q;B)$ a performance score on case $q$,
with higher values indicating better performance. The expected gain is
\begin{equation}
  \Delta J_t=\mathbb E_{q\sim\mathcal D}\!\left[
    s(\mathcal A_{t+1},q;B)-s(\mathcal A_t,q;B)
  \right].
  \label{eq:evolution-gain}
\end{equation}
The expectation also averages over execution randomness. Evidence for
$\Delta J_t>0$ supports an average improvement under the specified scope
and budget. A larger behavioral difference need not yield a larger gain.
Coding introduces a further gap between the scope of feedback and the scope
of reuse. Tests evaluate a concrete software artifact, whereas an agent update
may influence many future decisions. A successful patch does not by itself
validate a reusable debugging rule derived from it. Effective evolution
therefore requires preserving relevant conditions when abstracting experience
and aligning $\mathcal D$ with the claimed scope of reuse. The five links
expose where useful evidence may be lost, overgeneralized, or fail to influence
subsequent behavior.

\surveyparagraph{subsubsection}{How coding shapes evolution}{How coding shapes evolution}
Coding interactions connect the agent's problem-solving decisions with actual
execution outcomes, providing a verifiable basis for experience formation.
Test failures expose limitations of the current solution while also providing
evidence for examining the decisions made during task solving and their
consequences. Analysis of historical repairs further supports the identification
of effective strategies and their applicable conditions, turning experience
from individual tasks into reusable repair knowledge
(e.g., SWE-Exp~\citep{chen2026sweexp}, EvoRepair~\citep{hu2026evorepair},
and Repository Memory~\citep{wang2026repositorymemory}).
Translating experience into agent updates involves using this knowledge to
shape subsequent learning and execution. Repair patterns can be abstracted
into reusable guidance, while recurring execution problems can inform the
proposal and validation of changes to how the agent operates
(e.g., CODESKILL~\citep{li2026codeskill}, Socratic-SWE~\citep{xiao2026socraticswe},
Self-Harness~\citep{zhang2026selfharness}, and AutoSaddler~\citep{taxonomy260823041}).
Experience is context-dependent, and its reuse beyond the original conditions
requires further execution evidence to assess and refine the applicability
of the retained knowledge.
Table~\ref{tab:conceptual-comparison} summarizes the defining requirements
of the three agent concepts.

\begin{table}[H]
  \caption{Defining requirements across the three agent concepts.}
  \label{tab:conceptual-comparison}
  \centering
  \footnotesize
  \setlength{\tabcolsep}{3pt}
  \renewcommand{\arraystretch}{1.28}
  \arrayrulecolor{black}
  \begin{tabularx}{\linewidth}{lcccc>{\centering\arraybackslash}X}
    \toprule
    \bfseries Agent Type &
    \bfseries Coding Tasks &
    \bfseries Agent Updates &
    \bfseries Updates from Coding &
    \bfseries Retention and Reuse &
    \bfseries Verified Coding Gains \\
    \midrule
    Coding Agents &
    \conceptrequired & \conceptoptional & \conceptoptional &
    \conceptoptional & \conceptoptional \\
    Self-Evolving Agents &
    \conceptoptional & \conceptrequired & \conceptoptional &
    \conceptrequired & \conceptoptional \\
    \midrule
    \rowcolor{black!8}
    \bfseries Self-Evolving Coding Agents &
    \conceptrequired & \conceptrequired & \conceptrequired &
    \conceptrequired & \conceptrequired \\
    \bottomrule
  \end{tabularx}
\end{table}

\section{Taxonomy of Self-Evolving Coding Agents}
\label{sec:taxonomy}

\begin{figure}[!htbp]
  \centering
  \resizebox{\linewidth}{!}{\input{fig/taxonomy-coding-agents}}
  \caption{Hierarchical taxonomy of self-evolving coding agents by the primary
  persistent target of evolution: Assets Self-Evolution (blue), Architecture
  Self-Evolution (orange), and Model Self-Evolution (purple). The leaf nodes
  cover all 108 papers in Section 1 of the companion repository's taxonomy,
  using abbreviated titles and citations to identify each work.}
  \label{fig:taxonomy-coding-agents}
\end{figure}
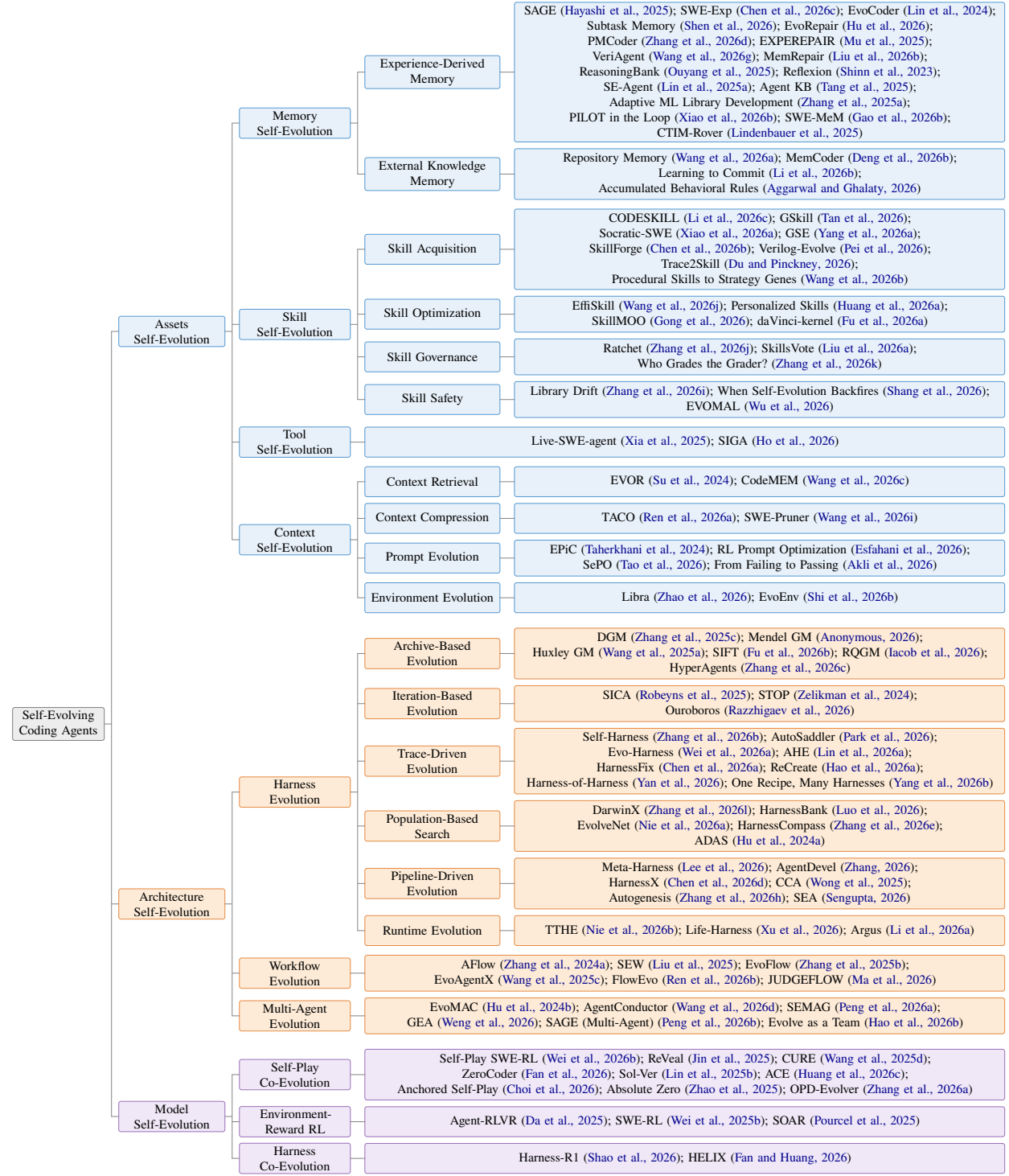

Coding experience can improve an agent by changing the resources it can use,
the organization of its execution, or the parameters underlying its decisions.
Following $\mathcal{A}_t=(\mathcal{S}_t,\mathcal{H}_t,\theta_t)$, we organize
these targets into \textbf{Assets Self-Evolution}, \textbf{Architecture
Self-Evolution}, and \textbf{Model Self-Evolution}.
Figure~\ref{fig:taxonomy-coding-agents} presents the three categories,
their subcategories, and representative works.
Assets evolution changes
\emph{what is available}, for example by adding a tool, revising a skill, or
refining memory and context. Architecture evolution changes \emph{how execution
is organized}, including the harness, workflow, and collaboration among agents.
Model evolution changes \emph{the parameterized policy} that shapes decisions
within this system. This ordering reflects progressively deeper intervention
in the agent's implementation. Assets are typically modular resources that
can be added, revised, or replaced within an existing execution structure.
Architecture updates reach into the code or configuration governing execution,
changing control flow, stage dependencies, or coordination among agents.
Model updates reach further into the learned parameters underlying action
selection. The progression thus moves from modifying available resources,
through restructuring the system that uses them, to updating its learned
decision mechanism. Tool updates may also involve code, but they revise
available operations, while architecture updates change how the system
organizes and controls execution.

\subsection{Assets Self-Evolution}
\label{sec:assets-overview}

Assets comprise the reusable information and capabilities that support coding
activity. We represent the asset state as
\begin{equation}
  \mathcal{S}_t =
  \left(\mathcal{M}_t,\mathcal{K}_t,\mathcal{T}_t,\mathcal{C}_t\right),
  \qquad
  \widetilde{\mathcal{S}}_{t+1} =
  \mathcal{U}_{\mathcal{S}}(\mathcal{A}_t,\tau_t,f_t),
  \label{eq:asset-evolution}
\end{equation}
where $\mathcal{M}_t$, $\mathcal{K}_t$, $\mathcal{T}_t$, and $\mathcal{C}_t$
denote memory, skills, tools, and context, respectively. Memory retains
knowledge acquired from prior interactions or external sources; skills organize
knowledge into reusable procedures; tools provide executable operations; and
context determines the information, instructions, and environmental settings
available during execution. Assets self-evolution changes what the agent can
recall, apply, invoke, or condition its reasoning on, with representative
examples including SWE-Exp~\citep{chen2026sweexp},
CODESKILL~\citep{li2026codeskill}, and
Live-SWE-agent~\citep{xia2025livesweagent}. These components may share content
while serving different functions: knowledge retained in memory can enter the
current context, and a learned procedure can be realized through an executable
tool.

\subsection{Architecture Self-Evolution}
\label{sec:architecture-overview}

Architecture determines how the agent's capabilities are organized into task
execution. We describe it through three complementary aspects,
\begin{equation}
  \mathcal{H}_t = \left(h_t,w_t,g_t\right),
  \qquad
  \widetilde{\mathcal{H}}_{t+1} =
  \mathcal{U}_{\mathcal{H}}(\mathcal{A}_t,\tau_t,f_t),
  \label{eq:architecture-evolution}
\end{equation}
where $h_t$ denotes the harness, $w_t$ the workflow, and $g_t$ the multi-agent
organization. The harness governs runtime interaction and execution control;
the workflow specifies the arrangement and dependencies of task-solving stages;
and the multi-agent organization determines how responsibilities and
information are distributed among participating agents. Architecture
self-evolution changes how available capabilities are coordinated, allowing
experience to revise the procedure through which subsequent tasks are solved,
as exemplified by DGM~\citep{zhang2025darwin}, AFlow~\citep{zhang2024aflow}, and
EvoMAC~\citep{hu2024evomac}. These architectural aspects are not independent:
a change in collaboration may alter the workflow, while its implementation may
require modifications to the harness.
For example, moving validation ahead of a costly rewrite changes which
assumptions reach implementation without adding a new capability. Requiring
a reviewer to inspect an intermediate result similarly changes a decision
rule while leaving the communication graph fixed. Architecture therefore
includes both structural connections and the rules governing their use.
Even a small edit to these rules can affect many later decisions, making
the scope of the change more informative than its code size.

\subsection{Model Self-Evolution}
\label{sec:model-overview}

Model self-evolution incorporates coding experience into the parameters
underlying action selection. Let $x$ denote the information available at a
decision point and $a$ an agent action. Parameter adaptation changes the policy
exercised within the surrounding agent system,
\begin{equation}
  \begin{gathered}
    \widetilde{\theta}_{t+1} =
    \mathcal{U}_{\theta}(\mathcal{A}_t,\tau_t,f_t), \\
    \pi_{\theta_t}(a\mid x;\mathcal{S}_t,\mathcal{H}_t)
    \;\longrightarrow\;
    \pi_{\widetilde{\theta}_{t+1}}
    (a\mid x;\mathcal{S}_t,\mathcal{H}_t).
  \end{gathered}
  \label{eq:model-evolution}
\end{equation}
Holding assets and architecture fixed in this expression isolates the effect
of parameter changes without requiring them to remain fixed throughout
evolution. Our taxonomy includes self-play co-evolution, environment-reward
reinforcement learning, and harness co-evolution. These directions organize
learning through coupled task generation and solving, feedback from
environmental interaction, or adaptation between the model and its runtime
harness. Representative approaches include Self-Play
SWE-RL~\citep{selfplaysoftwareagents2026} and Agent-RLVR~\citep{da2025agentrlvr}.
Their shared target is the learned policy, through which experience can
influence subsequent decisions without retrieving each contributing
experience explicitly.

\subsection{Research Landscape}
\label{sec:research-landscape}

We describe the publication landscape of the 108 methods in
Figure~\ref{fig:taxonomy-coding-agents}, counting each once by its primary
update target. Benchmarks, products, and separate empirical studies are
excluded. First availability follows arXiv dates for 103 papers and catalogue
years for five. The 2026 quarterly view uses arXiv months or documented
release quarters for four papers without arXiv links. Venue labels follow
the catalogue and are kept separate from first availability: 13 papers
appeared on arXiv before their venue year.

\begin{figure}[H]
  \centering
  \includegraphics[width=\linewidth]{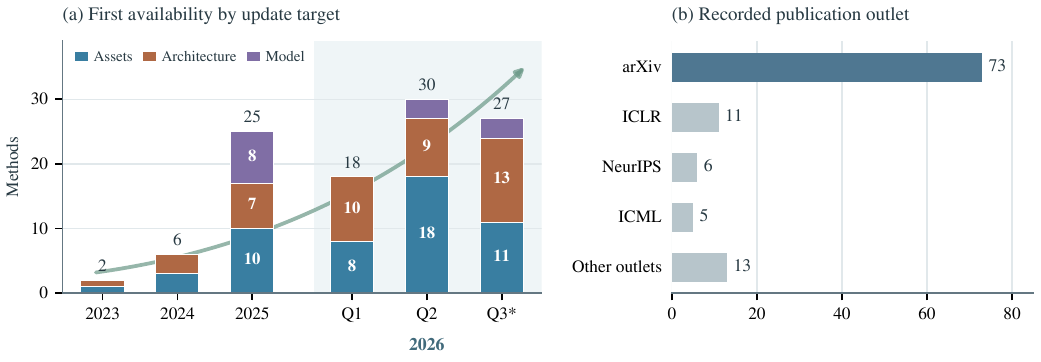}
  \caption{Publication landscape of 108 methods. (a) First availability by
  update target; Q3* is incomplete as of September 2026. (b) Recorded outlets;
  ``arXiv'' indicates no other listed venue.}
  \label{fig:publication-landscape}
\end{figure}

Figure~\ref{fig:publication-landscape} shows two papers in 2023, six in 2024,
25 in 2025, and 75 in 2026, with 18, 30, and 27 in the first three quarters
of 2026. The first half of 2026 alone contains 48 methods, nearly twice the
2025 total, showing marked acceleration within this corpus. The 2026 entries
comprise 37 asset, 32 architecture, and six model updates; across all years,
these categories contain 51, 43, and 14 papers, respectively. Assets and
architecture thus account for 69 of 75 entries in 2026, compared with 17 of
25 in 2025, suggesting growing attention to reusable resources and editable
execution structures. These counts describe research activity rather than
effectiveness. The curated collection and incomplete Q3 also preclude
conclusions about field-wide growth or a slowdown after Q2.

The outlet view records 73 papers as arXiv, 11 as ICLR, six as NeurIPS, five as
ICML, and 13 under other labels. Here, arXiv means no other venue is recorded;
catalogue labels may lag publication. This distribution describes dissemination
at the snapshot date, while Section~\ref{sec:benchmarks} examines evidence
of actual improvement.

The following sections examine the mechanisms and representative work in
assets self-evolution (Section~\ref{sec:assets-self-evolution}),
architecture self-evolution (Section~\ref{sec:architecture-self-evolution}),
and model self-evolution (Section~\ref{sec:model-self-evolution}).

\begingroup
\clubpenalty=10000
\widowpenalty=10000
\displaywidowpenalty=10000
\section{Assets Self-Evolution}
\label{sec:assets-self-evolution}

Asset self-evolution turns coding experience and external knowledge into
resources that remain available for subsequent execution. The evolving state is
$\mathcal S_t=(\mathcal M_t,\mathcal K_t,\mathcal T_t,\mathcal C_t)$.
The main design choices concern the unit of experience retained, its
representation, and the conditions under which it is revised or reused.
Memory can preserve a diagnosis and its supporting evidence; a skill can
organize that knowledge into a procedure; a tool can implement an operation;
and context can expose the information needed at the next decision point.
These roles connect the four asset types while giving each a distinct update
target. Their evolution can involve consolidation, correction, or removal as
well as the addition of new resources. Table~\ref{tab:asset-methods} compares
representative methods by their asset targets, update triggers, and the
operations used to construct and maintain those assets.

\input{tables/assets-methods}

\subsection{Memory Self-Evolution}
\label{sec:memory-self-evolution}

Memory evolution updates $\mathcal M_t$ within the asset state $\mathcal S_t$.
Its central design choice is the unit of
knowledge that survives an interaction, together with the conditions under
which it should be recalled. We distinguish experience-derived memory from
external knowledge memory by the provenance of that knowledge. Both concern
explicit storage outside model parameters, but the former abstracts agent executions, whereas
the latter incorporates repository history or human guidance. For comparison,
we represent a memory item as $m_i=(v_i,e_i,\Omega_i)$, where $v_i$ is the retained
knowledge, $e_i$ its supporting evidence, and $\Omega_i$ its applicability
conditions. These fields may be explicit metadata or implicit in the memory
text. An agent-derived diagnosis and a developer-provided convention can share
this representation while drawing on different sources of $e_i$.

\surveyheading{Experience-Derived Memory}
The shortest reuse horizon is a later attempt at the same problem.
Reflexion retains verbal interpretations of feedback in an episodic buffer to
guide subsequent trials~\citep{shinn2023reflexion}. SAGE instead abstracts an
initial rollout into a plan that preserves dependencies and constraints before
re-executing the same task~\citep{hayashi2025sage}. SE-Agent revises, recombines,
and refines pilot trajectories, using relationships among previous attempts to
explore alternative solutions~\citep{taxonomy250802085}. These methods
demonstrate different ways of making earlier computation useful to later
execution. Their same-task improvements should be distinguished from evidence
that a retained memory benefits unseen tasks, and from improvements obtained
simply by increasing the number of attempts.

Across tasks, abstraction must balance reusable guidance against the details
that make a past solution applicable. SWE-Exp distills successful and failed
repair trajectories into a bank spanning issue comprehension and concrete code
changes~\citep{chen2026sweexp}. EvoCoder addresses issue reproduction through a
hierarchical pool separating common experience from repository-specific
experience~\citep{lin2024evocoder}. EXPEREPAIR separates concrete repair
demonstrations from reflective insights in episodic and semantic memories,
combining both when constructing an inference
prompt~\citep{mu2025experepair}. These designs expose complementary dimensions
of memory organization. Repository scope determines where a lesson applies,
while abstraction level determines how much of the original repair must be
reconstructed before it can guide a decision.

Retrieval also needs to match the agent's current activity. Subtask Memory
aligns storage, retrieval, and updating with functional subtasks, addressing
the mismatch between similar issue descriptions and different stage-specific
reasoning needs~\citep{shen2026subtaskmemory}. PMCoder couples a phase planner
with episodic memory, so the phase conditions retrieval and memory statistics
inform stuck detection and replanning~\citep{zhang2026pmcoder}. Its matched
component ablations and trajectory analyses connect better state management
with fewer repeated failures. Here the evidence concerns retaining useful
observations during a long repair episode; transferring the mechanism across
backbones does not itself demonstrate transfer of accumulated memories across
issues. Together, these methods suggest that relevance depends on the decision
being made, rather than issue similarity alone. A relevance-based retrieval
rule can make this dependence explicit,
\begin{equation}
  \mathcal R_t^{\mathcal M}(x)
  =\operatorname{Top}_{k_M}\!\left(
    \{m_i\in\mathcal M_t:\Omega_i(x)=1\};\,r(m_i,x)
  \right),
  \label{eq:memory-retrieval}
\end{equation}
where $x$ is the current decision state, $r(m_i,x)$ a relevance score, and
$\operatorname{Top}_{k_M}$ returns up to $k_M$ highest-scoring eligible items.
The condition $\Omega_i(x)=1$ represents an applicability check, such as a
matching repository, dependency version, or task phase. This is an analytical
retrieval rule; individual methods may implement eligibility and relevance
jointly or use soft scores. It separates the existence of an item in
$\mathcal M_t$ from its inclusion in the recalled set $\mathcal R_t^{\mathcal M}(x)$.

Domain-specific feedback further changes what should be remembered. EvoRepair
combines accumulation within a vulnerability with reuse across vulnerabilities,
updating its experience bank through quality-aware
scoring~\citep{hu2026evorepair}. MemRepair distinguishes historical fixes,
security patterns, and refinement trajectories, preserving repository
conventions alongside reusable defenses and lessons from failed
validation~\citep{taxonomy260517444}. In RTL generation, VeriAgent converts
simulation and synthesis feedback into memory nodes with triggering conditions,
guidance, and metadata; its manager inserts, refines, or discards
nodes~\citep{taxonomy260317613}. The adaptive system for ML library development
retains verified generated programs as demonstrations and selects them by
difficulty, using successful examples to support remaining tasks in an
unfamiliar programming language~\citep{taxonomy250202534}. These mechanisms
retain different evidence, from repair failures to physical design signals and
executable examples, rather than treating every trajectory as an interchangeable
text record.

Memory can also mediate experience sharing and control its own information
cost. ReasoningBank extracts strategies from self-judged successes and failures,
then incorporates new experience after retrieval-guided execution; its MaTTS
extension uses additional computation to diversify that
experience~\citep{taxonomy250925140}. Agent KB shares structured trajectories
across heterogeneous frameworks, combining planning-time retrieval with
feedback-time diagnostic guidance and a disagreement
gate~\citep{tang2025agentkb}. PILOT distills procedures and failure modes into
skills and memory during live supervisor--worker
execution~\citep{taxonomy260826530}. SWE-MeM learns when, what, and how to
compress histories through synthesized training trajectories and memory-aware
GRPO~\citep{gao2026swemem}. PILOT couples retained assets with supervision,
while SWE-MeM combines memory management with parameter learning. Their gains
therefore reflect changes beyond memory contents.
More retained experience is not necessarily better. CTIM-Rover adds general
and repository-specific cross-task memories to AutoCodeRover but does not
outperform its baseline in the evaluated configurations; the authors identify
distracting memories and exemplars as a likely cause~\citep{taxonomy2025realm130}.
This negative result makes selective recall part of the improvement mechanism.
Memory may change action selection, giving $\delta_t(x)>0$ at some states,
without producing a positive expected gain $\Delta J_t$. Evaluation should
therefore separate memory quality and
retrieval effects from added search, supervision, or training.

\surveyheading{External Knowledge Memory}
Repository history records decisions that the agent did not necessarily make
itself. Repository Memory uses historical commits, linked issues, and summaries
of actively changing code regions to improve
localization~\citep{wang2026repositorymemory}. This supplements a code snapshot
with associations between past problems and likely modification sites.
MemCoder structures human commit history into intent-to-code mappings, refines
execution using verification feedback, and incorporates human-validated
solutions into long-term knowledge~\citep{deng2026memcoder}. The distinction is
between recalling where and why developers changed code and retaining the
agent's own trial-and-error record, although a single memory system can
eventually contain both.
External supervision can also reveal project conventions that tests do not
fully express. Learning to Commit attempts historical issues, compares its
changes with actual commits, and distills the differences into reusable
guidance about style, internal APIs, and architectural
constraints~\citep{taxonomy260326664}. Its chronological split evaluates future
pull requests, directly aligning the reuse claim with later repository work,
although the reported study centers on one repository. Accumulated Behavioral
Rules converts accepted review comments into persistent instructions and a
pre-submission checklist~\citep{aggarwal2026behavioralrules}. Its production
observations support cross-session retention of corrections, but the small
session sample and absence of a controlled baseline limit causal claims about
improvement. Such artifacts overlap with skills and context in their eventual
use; their distinguishing contribution here is converting developer knowledge
into a maintained source of guidance.

Across both sources, effective memory requires preserving the relationship
between $v_i$, $e_i$, and $\Omega_i$. A compact description of maintenance is
\begin{equation}
  \widetilde{\mathcal M}_{t+1}
  =\bigl(\mathcal M_t\setminus\mathcal D_t^{\mathcal M}\bigr)
    \cup\mathcal W_t^{\mathcal M},
  \label{eq:memory-maintenance}
\end{equation}
where $\mathcal D_t^{\mathcal M}\subseteq\mathcal M_t$ contains entries selected
for removal or replacement and $\mathcal W_t^{\mathcal M}$ contains proposed
new or revised entries. Merging several records removes their originals and
writes a consolidated item; retirement removes an item without replacement.
The tilde denotes a candidate state subject to adoption. This representation
allows memory to improve through correction and consolidation even when
$|\widetilde{\mathcal M}_{t+1}|\leq|\mathcal M_t|$. A repository convention
may cease to hold after refactoring, while a general repair strategy may require
different local actions. Consequently, the evaluation distribution
$\mathcal D$ used to estimate $\Delta J_t$ should reflect whether reuse
is claimed within an episode, across issues in one repository, or across
repositories. That alignment determines whether memory has supported local
continuity, accumulated project expertise, or broader transfer.

\subsection{Skill Self-Evolution}
\label{sec:skill-evolution}

Skill self-evolution turns coding knowledge into reusable guidance for action,
updating the skill collection $\mathcal K_t$. A skill specifies a
procedure and the conditions under which it should be applied, potentially
combining instructions with executable resources. Its contribution therefore
depends on more than whether its content accurately summarizes a past task.
The procedure must remain applicable, be selected when needed, and improve
the decisions made by the receiving agent. Following Figure~\ref{fig:taxonomy-coding-agents},
we examine acquisition, optimization, governance, and safety as complementary
concerns across the skill lifecycle. We write $k=(\Omega_k,p_k,\mathcal Q_k)$
for its applicability conditions, procedure, and resource requirements,
respectively. The requirements $\mathcal Q_k$ can identify tools, dependencies,
or other skills needed to execute $p_k$.

\surveyheading{Skill Acquisition}
Acquisition begins with choosing the experience from which a procedure can be
abstracted. A candidate $\widetilde k=\operatorname{Distill}(\tau_t,f_t;\mathcal K_t)$
captures a proposed procedure together with its conditions and requirements;
conditioning on $\mathcal K_t$ allows the distiller to account for existing
guidance. CODESKILL~\citep{li2026codeskill} learns a management policy that
extracts task-level procedures and event-driven responses from coding
trajectories. It combines rubric feedback with the incremental execution
benefit of generated skills for a frozen downstream agent, and checks whether
the trajectory follows the supplied guidance. This connects acquisition to
the earlier distinction between subsequent use and behavioral contribution.
Socratic-SWE~\citep{xiao2026socraticswe} gives extracted skills a different
role. Its registry summarizes recurring failures and effective repairs to
guide the synthesis of verifiable repository tasks, which then train the
Solver. Skills thus influence future behavior through curriculum construction
as well as through direct instructions, illustrating interaction between asset
and model evolution.

Repository-specific methods can actively generate the experience they need.
GSkill~\citep{tan2026gskill} combines SWE-smith tasks with reflective
evolutionary optimization of repository skill documents, evaluating candidate
guidance through agent rollouts. SkillForge~\citep{taxonomy260818933} instead
creates practice issues by reimplementing test-covered repository
functionality, then distills the resulting solutions into skills associated
with repository entities. Both use a repository as a source of practice, but
their retained guidance differs in how it is organized for later use. Broader
reuse also requires compatibility with skills already acquired.
GSE~\citep{yang2026gse} maintains a Skill Relation Graph, consolidates
related local updates, and uses replay verification to limit overfitting and
regression. Acquiring a useful procedure therefore includes deciding how its
scope and relationships fit the existing collection.

Hardware coding exposes the importance of the available verifier.
Verilog-Evolve~\citep{taxonomy260526498} derives modular guidance from logged
refinement histories, with creation, improvement, or skipping informed by
verifier reports. Simulation, synthesis, and timing proxies provide distinct
evidence about what a useful procedure must preserve.
Trace2Skill~\citep{taxonomy260521810} uses an oracle, mutator, and selector loop
to turn rollout failures into task-specific skills for later search, editing,
validation, and recovery. Its bounded dense verifier feedback provides
diagnostic observations while withholding hidden harnesses and reference
solutions. The two approaches emphasize cross-session accumulation and
test-time adaptation, respectively, so their reuse claims require different
evaluation scopes. Representation also affects whether the acquired knowledge
becomes actionable. \emph{From Procedural Skills to Strategy
Genes}~\citep{taxonomy260415097} finds in scientific code-solving experiments
that compact control-oriented representations and distilled failure warnings
can outperform expanded documentation packages. More detailed recollection
need not yield more effective procedural guidance.

\surveyheading{Skill Optimization}
Optimization concerns both the quality of procedures and the way they are
selected and applied. To isolate a skill collection's contribution, let
$\mathcal A_t[\mathcal K]$ denote the agent with its skill collection replaced
by $\mathcal K$ and all other components held fixed. Define
$G_t(\mathcal K)=\mathbb E_{q\sim\mathcal D}[s(\mathcal A_t[\mathcal K],q;B)]$
using the same evaluation distribution and execution budget as before, with
the expectation also averaging over execution randomness. A candidate's
marginal gain is
\begin{equation}
  \Delta_t^{\mathcal K}(k\mid\mathcal K)
  =G_t(\mathcal K\cup\{k\})-G_t(\mathcal K).
  \label{eq:skill-marginal-gain}
\end{equation}
For a revision, $\mathcal K$ excludes the version being replaced, so the old
and new versions can be assessed against the same base collection. A skill's
utility depends on this base; overlap, conflict, or complementarity can make
its contribution differ from its standalone performance.
EffiSkill~\citep{wang2026effiskill} mines concrete
Operator Skills and higher-level Meta Skills from slow/fast program pairs,
then combines diagnosis, retrieval, and planning to optimize unseen programs
without runtime feedback. Its offline library provides a comparison point
for systems that continually revise skills during execution.
daVinci-kernel~\citep{taxonomy260616497} closes that loop for GPU kernels,
jointly training skill selection, kernel generation, and skill summarization
on a shared model backbone. New skills enter its library only after execution
confirms reproducible speedups. Its improvements therefore involve both an
evolving skill collection and parameter adaptation, requiring care when
attributing gains to $\mathcal K_t$ alone.

The optimization objective also matters.
SkillMOO~\citep{taxonomy260409297} evolves skill bundles using LLM-proposed
edits and Pareto selection over task success and inference cost. Such a
tradeoff can be expressed through the nondominated candidate collections
\begin{equation}
  \mathfrak P_t=
  \operatorname{ParetoMax}_{\mathcal K\in\mathfrak C_t}
  \bigl(G_t(\mathcal K),-\operatorname{Cost}_t(\mathcal K)\bigr),
  \label{eq:skill-pareto-selection}
\end{equation}
where $\mathfrak C_t$ is the set of candidate collections and
$\operatorname{Cost}_t$ measures their average execution cost under the common
budget cap $B$. A collection is nondominated if no candidate is at least as
good on both objectives and strictly better on one. This formulation allows
several useful tradeoffs instead of imposing a single performance--cost
weight. SkillMOO's analysis
finds pruning and substitution prominent among successful edits, showing that
improvement can come from reducing guidance. Specialization likewise needs
evidence of future relevance. \emph{Do Personalized Skills Help Coding
Agents?}~\citep{huang2026personalizedskills} extracts developer preferences
from interaction histories, but reports small, inconsistent improvements from
personalization in simulator-based replay, while pooled generic skills yield
more consistent gains. Together, these studies suggest that optimization
should match procedural specificity and cost to the intended reuse
distribution $\mathcal D$, rather than assume
that longer or more personalized skills are better.

\surveyheading{Skill Governance}
Governance determines which skills remain available as a collection grows.
For an existing skill $k\in\mathcal K_t$, its removal contribution is
\begin{equation}
  \gamma_t(k)=G_t(\mathcal K_t)-G_t(\mathcal K_t\setminus\{k\}).
  \label{eq:skill-removal-contribution}
\end{equation}
A small or negative estimate of $\gamma_t(k)$ motivates review for redundancy or
retirement. The score remains conditional on the other skills and the
evaluation distribution; it does not establish that $k$ is useless in every
setting. Both $\gamma_t(k)$ and $\Delta_t^{\mathcal K}$ require estimates from
comparable executions, making evaluation uncertainty part of governance.
SkillsVote~\citep{taxonomy260518401} links collection and recommendation to
post-execution attribution. It profiles environment requirements and
verifiability, searches structured skill folders, and separates skill-guided
subtasks from agent exploration and environmental effects before admitting
reusable discoveries. Ratchet~\citep{zhang2026ratchet} focuses on retirement,
combining measured contribution with a bounded active library and constrained
synthesis. Its analysis distinguishes judge errors that increase sampling
requirements from false-positive judgments that can disable retirement under
the analyzed rule. A governance policy thus depends on the reliability of the
evidence used to retain or remove skills.

\emph{Who Grades the Grader?}~\citep{taxonomy260712790} makes this dependence
explicit by evolving inspectable evaluation functions alongside the skill
library. Anchored reference judgments constrain the evaluator, while its
agreement with hidden ground truth is checked separately. Removing these
anchors can produce an always-pass evaluator even when downstream scores
appear satisfactory. This extends the adoption decision for
$\widetilde{\mathcal A}_{t+1}$ to the criterion governing adoption itself.
Improving skills and validating their evaluator are related but distinct
requirements, particularly when execution does not supply a complete oracle.

\surveyheading{Skill Safety}
Persistence can amplify errors as well as useful experience.
\emph{Library Drift}~\citep{zhang2026librarydrift} diagnoses how unrestricted
accumulation degrades retrieval and introduces irrelevant guidance. Its
contribution logs, outcome-based retirement, and active-library cap make
degradation observable and governable. \emph{When Self-Evolution
Backfires}~\citep{taxonomy260805810} identifies a further propagation mechanism.
Defective skills can influence the distillation of later skills, so removing
the original entry leaves contaminated descendants. Its Verifier-as-Gatekeeper
screens structural validity, behavioral harmlessness, and semantic
consistency, then selects skill subsets by marginal gain before admission.
Control must therefore cover individual skills and their interactions within
the collection. An illustrative admission rule for a candidate $k$ combines
estimated benefit with specified checks,
\begin{equation}
  \operatorname{Admit}_t(k)
  =\mathbf 1\!\left[
    \widehat{\Delta}_t^{\mathcal K}(k\mid\mathcal K_t)>\epsilon
  \right]
  \mathbf 1\!\left[
    \operatorname{Check}_t(k,\mathcal K_t)=1
  \right],
  \label{eq:skill-admission}
\end{equation}
where $\mathbf 1[\cdot]$ is an indicator, $\epsilon\geq0$ a gain threshold,
and $\operatorname{Check}_t$ represents the chosen validity, compatibility,
and safety checks. The hat marks an empirical gain estimate. For a replacement,
the gain compares the new and old versions against the same base collection.
This rule
expresses a conservative admission policy; passing finite checks remains
evidence under those checks, rather than a guarantee of safety in all future
uses.

EVOMAL~\citep{taxonomy260825776} demonstrates an adversarial counterpart in
coding agents. A retrieved malicious skill can become an authoring template,
causing harmful content to be copied into newly stored and executed skills.
Agent-authored descendants can persist after the planted entries are removed;
the proposed counter-prompt reduces copying-based self-poisoning in the
evaluated setting. These findings make provenance and admission consequential
throughout evolution. In terms of the five links introduced earlier, a
contaminated update can survive retention, influence subsequent behavior, and
generate the evidence for further updates. Safe skill evolution must therefore
assess both immediate behavior and the content that a skill helps propagate
into future versions of $\mathcal K_t$.

\subsection{Tool Self-Evolution}
\label{sec:tool-self-evolution}

Tool self-evolution updates the executable utilities in $\mathcal T_t$.
Where a skill describes a procedure, a tool provides a callable implementation.
The retained update is a utility or adapter that supports subsequent execution,
rather than the project code delivered for the current task.

Live-SWE-agent interleaves issue resolution with tool creation and revision
within a minimal bash-based agent~\citep{xia2025livesweagent}.
Reflection on environment feedback prompts the agent to build utilities for
editing, code search, or specialized analysis, then invoke and refine them as
the task proceeds. Its main reuse scope is later steps within the same issue,
illustrating how execution experience can expand available operations without
an offline evolution phase.

SIGA evolves a simulator-facing adapter around a fixed coding model and
execution loop~\citep{ho2026siga}. An offline proposer uses prior trajectories
and rewards to revise the task primer, procedural memory, and auxiliary skills,
retaining candidates selected by validation performance. These updates span
several assets. Its main evolved comparison disables skill invocation to keep
the runtime tool configuration fixed, so the gains support improved adapter
contents rather than newly added tools. Together, the two methods distinguish
expanding executable operations from improving their domain interface.
In either case, a working utility alone does not establish a positive
downstream gain $\Delta J_t$.

\subsection{Context Self-Evolution}
\label{sec:context-self-evolution}

Context self-evolution targets $\mathcal C_t$, the information, instructions,
and environmental settings through which an agent encounters a coding task.
The reusable change may be selected repository evidence, a compression rule,
an optimized prompt, or a repository-side information structure. These changes
shape what reaches the decision process and how it is presented. The central
distinction is between constructing context for one invocation and retaining
an experience-informed change that influences subsequent execution. The methods
below occupy different points along this boundary and support different scopes
of reuse.

\surveyheading{Context Retrieval}
Retrieval evolves when execution changes the evidence available to later
queries or the way relevant evidence is assembled. EVOR synchronously updates
queries and a heterogeneous knowledge base containing documentation, web
content, code snippets, and execution feedback~\citep{su2024evor}. Generated
programs or their error messages enter the retrieval pool, so later attempts
can condition on evidence produced by earlier execution. Its loop retains more
than a revised answer, although successful execution of a retrieved snippet
does not establish its semantic correctness. CodeMEM addresses continuity
across iterative repository-level generation~\citep{wang2026codemem}. It uses
AST structure to preserve relevant code context, incorporates newly retrieved
blocks when needed, and maintains code-centric session memory to detect
forgotten corrections. Its context-retrieval role overlaps with memory because
retention supplies the evidence from which later context is assembled. These
methods illustrate reuse within an evolving interaction; their reported gains
should not automatically be interpreted as transfer to unrelated repositories.
To distinguish retained assets from their use at a decision point, let
$\mathcal B_t$ be the available evidence pool, $p_t(x)$ the instructions for
state $x$, and $\phi_t$ the selection or compression mechanism. These are
components of the retained context assets $\mathcal C_t$. The assembled
context can be written as
\begin{equation}
  \begin{gathered}
    c_t(x)=\operatorname{Assemble}\!\left(
      p_t(x),\phi_t(\operatorname{Retrieve}(x,\mathcal B_t);x)
    \right),\\
    |c_t(x)|_{\mathrm{tok}}\leq L.
  \end{gathered}
  \label{eq:context-construction}
\end{equation}
The lowercase $c_t(x)$ denotes one constructed context and $L$ its token
budget. Updating $\mathcal B_t$, $p_t$, or $\phi_t$ changes what later
constructions can draw on. A different $c_t(x)$ produced from unchanged
assets can instead reflect a different query or task state. This distinction
locates the retained update underlying a context-adaptation claim.

\surveyheading{Context Compression}
Compression determines which observations remain visible when terminal output
or repository content exceeds the useful context budget. TACO discovers and
refines structured filtering rules from terminal trajectories, pairing
applicability conditions with retention and removal patterns
\citep{ren2026taco}. Effective rules are written back to a global pool and
selected for later tasks, making the compression behavior itself reusable.
SWE-Pruner instead conditions a trained neural skimmer on an explicit goal
supplied by the coding agent and selects relevant source lines
\citep{wang2026swepruner}. It demonstrates task-adaptive context selection,
but adaptation of the selected lines does not by itself establish continual
updating of the skimmer from deployment experience. This distinction separates
TACO's retained rule evolution from the application of a learned compression
component. For an original context $c_t(x)$ with nonzero token length and its
compressed version $c'_t(x)$, the fractional token reduction is
\begin{equation}
  \rho_t(x)=1-
    \frac{|c'_t(x)|_{\mathrm{tok}}}{|c_t(x)|_{\mathrm{tok}}}.
  \label{eq:context-compression-ratio}
\end{equation}
A positive $\rho_t(x)$ indicates token savings, but can coexist with
$\Delta J_t<0$ if compression removes information needed for correct action.
Comparisons should therefore consider both token reduction and downstream
performance under a common budget. The relevant tradeoff depends on which
observations are removed, as well as their volume.

\surveyheading{Prompt Evolution}
Prompt updates change the instructions conditioning a frozen code generator.
For a problem prompt $p_q$, writing $p'_q=r_t(p_q)$ separates the revised
prompt instance $p'_q$ from the transformation rule $r_t$. The instance may
be reused in another attempt at $q$, whereas $r_t$ can be applied to prompts
for other problems if its conditions hold.
EPiC evolves variants of a problem prompt, using the test performance of
generated code to guide selection~\citep{taherkhani2024epic}. It illustrates
feedback-driven prompt search within a problem, without requiring a durable
instruction shared across problems. Reinforcement-learning-based prompt
optimization learns a PPO policy that chooses among direct generation,
lexical mutation, and semantic rewriting using unit-test rewards
\citep{taxonomy260519102}. Here, the generator remains frozen, while the
optimizer learns how to produce the context updates. SePO moves reuse to the
system-prompt level, evolving both task-agent prompts and the prompt agent's
own instructions through an archive of candidates~\citep{tao2026sepo}.
The reusable artifact thus includes guidance for performing further prompt
optimization. From Failing to Passing evolves natural-language transformation
rules and combines them with execution-feedback repair in DUALFIX
\citep{taxonomy260705121}. Its cross-problem reuse and cross-model transfer
tests examine whether an effective transformation generalizes beyond the
failures used to discover it. These approaches distinguish an improved prompt
instance from a reusable prompt-transformation mechanism, an important
difference when defining the evaluation distribution $\mathcal D$.

\surveyheading{Environment Evolution}
The agent's information surface can also change on the repository side.
Libra introduces hierarchical Markdown catalogs and rewrites them after a
frozen solver fails to localize information in response to synthetic queries
\citep{zhao2026libra}. The catalogs persist as navigable indices for later
queries and solvers. Its transfer experiments across models and problem sets
test whether the improved environment remains useful beyond the optimization
interactions. EvoEnv explores a different environmental target, synthesizing
reusable Python environments that generate reasoning instances, compute
references, and score answers~\citep{taxonomy260514392}. Candidates enter an
active training pool after validity, semantic consistency, difficulty, and
novelty checks, and the shared generator--solver policy is updated through
reinforcement learning. This is environment--model co-evolution for reasoning
training, rather than repository configuration or direct evidence of improved
software issue resolution. The comparison identifies two distinct reuse
channels. Libra changes what future solvers can navigate, whereas EvoEnv
changes the executable sources of future training experience. Their common
asset-level feature is a retained environment artifact, while the downstream
learning mechanisms and the claims supported by evaluation differ. For a pool
$\mathfrak E_t$ of retained task-generating environments, their effect on
training can be represented by
\begin{equation}
  \mathcal D_t^{\mathrm{train}}
  =\sum_{e\in\mathfrak E_t}\alpha_t(e)\mathcal D_e,
  \label{eq:environment-task-distribution}
\end{equation}
where $\mathcal D_e$ is the task distribution generated by environment $e$,
$\alpha_t(e)\geq0$, and $\sum_e\alpha_t(e)=1$. Revising environments or their
sampling weights changes the experience available for subsequent learning.
This abstraction applies to generative training environments such as EvoEnv;
repository catalogs such as Libra instead change the evidence available to
the solver on an existing task. Environment adaptation can thus influence
either the information used during execution or the distribution from which
future learning experience is obtained.

Across these four asset types, the common challenge is to preserve useful
experience in a form that can influence the right later decision.
Memory emphasizes the evidence retained, skills the procedures made reusable, tools
the operations made executable, and context the information exposed during
execution. An asset may be present while $\delta_t(x)=0$ at a given state.
Even when $\delta_t(x)>0$, improvement requires separate evidence that
$\Delta J_t>0$ over the intended scope of reuse. Comparisons should therefore
isolate the contribution of retained assets from additional search, training,
or feedback, and account for their maintenance costs and failure modes.

\par
\endgroup

\begingroup
\clubpenalty=10000
\widowpenalty=10000
\displaywidowpenalty=10000
\section{Architecture Self-Evolution}
\label{sec:architecture-self-evolution}

Architecture self-evolution incorporates coding experience into the mechanisms that
organize execution. Within $\mathcal H_t=(h_t,w_t,g_t)$, an update may change runtime
control in the harness, dependencies among workflow stages, or the allocation of
responsibilities and information among agents. These targets offer different ways to
address the same failure. A missed validation step may motivate a submission check in
the harness, an earlier testing stage in the workflow, or a revised handoff to a
reviewing agent. The resulting improvement can come from arranging existing
capabilities more effectively, even when the available tools and model parameters
remain unchanged. The distinction concerns the functional role of the update, since a
workflow or collaboration change may also require harness code. Following
Figure~\ref{fig:taxonomy-coding-agents}, we examine how experience becomes a retained
execution rule, task-solving structure, or coordination mechanism.

\subsection{Harness Evolution}
\label{sec:harness-evolution}

Harness evolution updates the executable layer $h_t$ that mediates model
interaction, tool execution, feedback handling, and termination. In practice,
harness optimization may also revise prompts, memory, skills, and tools in
$\mathcal S_t$. We classify changes to reusable content or available operations
as asset updates, and changes to execution organization and control as
architecture updates. A method can combine both, so gains from an evolved
harness should be interpreted in light of the components actually modified.

Updating execution control gives experience a direct route into agent behavior.
A debugging lesson stored
as a skill still depends on retrieval and the agent's decision to follow it;
a harness intervention can check its activating condition whenever execution
reaches the relevant point. Its influence therefore depends on where it is
installed, which states activate it, and what it changes when activated. Its key
opportunity is to convert a recurring failure into an intervention at the point
where that failure arises. For example, a repeated invalid tool call may motivate
argument validation before execution, while an unproductive retry loop may
motivate a recovery rule after feedback. To make this intervention explicit,
let $z$ be the runtime state, including execution history and remaining budget,
and let $b$ be a control decision such as dispatching a call, retrying, or
terminating. A harness induces a transition kernel
\begin{equation}
  (b,z')\sim P_{h_t}(\cdot\mid z;\mathcal S_t,\theta_t),
  \qquad
  \widetilde h_{t+1}=\operatorname{Apply}(h_t,e_t),
  \label{eq:harness-control-edit}
\end{equation}
where $e_t$ is a proposed code or configuration edit and $z'$ is the resulting
runtime state. This description locates the update in the rules governing
execution, without prescribing how a method must search for edits. The six
mechanisms below emphasize different aspects of that search and can coexist
within one system.

\noindent\surveyheading{Archive-Based Evolution}
An archive makes earlier executable agents available as starting points for
further improvement. This matters because a modification can have value beyond
its immediate task score. A branch that currently performs poorly may contain
a useful implementation choice, expose a recurring failure, or support a later
change that would be difficult to discover from the current best agent.
DGM operationalizes this idea by sampling agents from an expanding archive,
modifying their source code, and evaluating the resulting variants on coding
tasks~\citep{zhang2025darwin}. Retaining alternative branches allows search to
continue from earlier designs instead of requiring every useful path to pass
through the latest agent. The archive retains executable implementations
alongside their traces, evaluation records, and lineage information. Search
uses this history both to select a parent and to construct a candidate
modification. Retention for further search and adoption for task execution
are separate decisions. An archive can preserve an informative candidate
without making it the deployed agent.

Choosing a parent raises a different question from choosing an agent to solve
the next task. Huxley GM identifies a mismatch between current coding
performance and the ability to produce useful descendants, and uses estimated
clade-level metaproductivity to guide tree
search~\citep{huxley2025godel}. To express this distinction independently of
a particular estimator, let
$u(\mathcal A)=\mathbb E_{q\sim\mathcal D}[s(\mathcal A,q;B)]$ measure
current performance. Under a specified search protocol and current archive
$\mathcal Z_t$, let
$\operatorname{Desc}_{C}(\mathcal A)$ be the random set of descendants
generated from $\mathcal A$ with evolution budget $C$. A conceptual measure
of its value as a starting point is
\begin{equation}
  \Psi_t(\mathcal A;C)=
  \mathbb E\!\left[
    \max_{\mathcal A'\in\{\mathcal A\}\cup
      \operatorname{Desc}_{C}(\mathcal A)} u(\mathcal A')
    \;\middle|\;\mathcal Z_t
  \right].
  \label{eq:harness-expansion-value}
\end{equation}
The evolution budget $C$ pays for candidate construction and assessment,
whereas $B$ controls each agent's task execution. A high $u(\mathcal A)$
need not imply a high $\Psi_t(\mathcal A;C)$. The latter depends on which
descendants the search can reach and afford to evaluate. It is an idealized
quantity; observed descendant results provide evidence for estimating such
potential. This distinction explains why selecting only the current
highest-scoring parent can discard useful routes to improvement.

Mendel GM focuses on what evidence should inform the next modification. Its
reaction-norm mutation compares one agent's trajectories across tasks, while
cross-lineage hybridization uses another lineage's trajectory on the same
task~\citep{mendel2026godel}. These comparisons reveal different information.
Repeated failures across tasks can suggest a recurring weakness in the harness;
different outcomes on a shared task can suggest a strategy worth transferring.
Writing $\mathcal E_{i,q}$ for an agent--task execution record and
$\Omega_t$ for the observed agent--task pairs, the two evidence views are
\begin{equation}
  \mathcal E_t^{\mathrm{tasks}}(i)
    =\{\mathcal E_{i,q}:(i,q)\in\Omega_t\},
  \qquad
  \mathcal E_t^{\mathrm{agents}}(q)
    =\{\mathcal E_{j,q}:(j,q)\in\Omega_t\}.
  \label{eq:harness-comparative-evidence}
\end{equation}
The first holds the agent fixed, and the second holds the task fixed. Neither
requires every agent to run on every task. These views turn the archive into
a source of comparative diagnoses as well as candidate implementations.
Executing the modified agent under matched conditions tests these proposed
explanations.

The archive also creates an allocation problem because evaluating every
possible modification can consume the evolution budget before useful branches
are explored. SIFT uses an LLM judge to rank candidate patches and concentrates
expensive benchmark execution on promising candidates~\citep{fu2026sift}.
By reducing full evaluations, screening can make a wider search affordable,
but its usefulness depends on which candidates it excludes. A judge that repeatedly favors familiar edits can suppress the
unusual branches that archive-based exploration is intended to retain.
The relevant comparison therefore concerns the quality of agents discovered
under a common evolution budget, alongside the cost and reliability of
candidate screening.

The proposal and evaluation mechanisms can themselves evolve. HyperAgents
places the task agent and its modifying meta agent in one editable program,
allowing experience to revise both task execution and the procedure producing
future updates~\citep{taxonomy260319461}. By modifying the proposal
mechanism itself, improvement can change which modifications become reachable. RQGM also permits evaluator changes between epochs while
keeping the utility fixed within each epoch~\citep{taxonomy260626294}.
If the internal objective changes from $u_t$ to $u_{t+1}$, a higher reported
score can reflect both a different agent and a different criterion. Comparison
under the common $\mathcal D$, $s$, and $B$ remains necessary to interpret
the result as $\Delta J_t$. Across these methods, the archive's contribution
depends on how retained variants change parent selection, comparative diagnosis,
proposal generation, or allocation of evaluation effort. Its size alone does
not capture that contribution.

\noindent\surveyheading{Iteration-Based Evolution} Iterative methods make the agent's
implementation a recurring coding task. SICA uses reflection and editing tools to
revise the program orchestrating its model calls and actions, retaining code changes
for later executions~\citep{robeyns2025selfimproving}. STOP makes this relationship
explicit through a seed program improver that applies its own language-model-assisted
search to itself, changing how subsequent programs are generated and selected while
leaving model weights fixed~\citep{zelikman2024stop}. Ouroboros extends implementation
revision into continued operation. Explicit improvement tasks and problems encountered
during ordinary work produce reviewed commits that become the runtime for later
work~\citep{razzhigaev2026ouroboros}. Its frozen benchmark snapshots and live
deployment lineage provide different evidence, respectively about a particular system
version and the operation of a continuing update process. The defining transition is
adoption into later execution; repeatedly editing candidate code does not establish
that this transition occurred.

Iterative revision also couples the quality of the current agent to the
quality of its future edits. A change that improves repository inspection
may help both ordinary issue resolution and inspection of the agent's own
implementation. However, these two uses expose different failure cases.
Evaluating later self-modification attempts can reveal whether an update
improves the revision process as well as the task solver. Retaining versioned
implementations makes this distinction inspectable and supports returning
to an earlier revision when a later change degrades execution.

\noindent\surveyheading{Trace-Driven Evolution} Execution traces can reveal which runtime
mechanism prevented an agent from using its available capabilities effectively.
Self-Harness clusters recurring failures, proposes small changes tied to those
failures, and retains modifications through regression
testing~\citep{zhang2026selfharness}. AHE makes editable components observable at file
level, condenses trajectories into inspectable evidence, and pairs edits with
predictions that can be checked against subsequent outcomes~\citep{lin2026ahe}.
HarnessFix provides more explicit localization by aligning runtime data and control
flow with harness artifacts, then converting recurring diagnoses into scoped repair
specifications~\citep{chen2026harnessfix}. AutoSaddler combines mini-batch diagnosis
and structured patching with development-set selection and separate final testing,
distinguishing repair of observed failures from evidence that an update
generalizes~\citep{taxonomy260823041}. ReCreate uses experience retrieval and a
reasoning--creation process to generate scaffold edits, abstracting individual
interactions into reusable domain patterns~\citep{taxonomy2026acllong1432}. Across
these methods, the important conversion is from an observed failure to an actionable
hypothesis about the execution mechanism responsible for it.

A diagnosis becomes operational only when it specifies where and under what
conditions execution should change. AHE illustrates this with an intervention
that records protected output paths after successful validation and intercepts
later commands that would delete or reset those outputs~\citep{lin2026ahe}.
HarnessFix retains both applicability and non-applicability conditions with
repair outcomes, helping later edits respect the scope of earlier
experience~\citep{chen2026harnessfix}. To describe this class of local change,
let $z$ include the runtime state and pending decision at a chosen intervention
point. An intended guarded transformation can be written as
\begin{equation}
  I_e(z)=
  \begin{cases}
    R_e(z), & g_e(z)=1,\\
    z, & g_e(z)=0,
  \end{cases}
  \label{eq:harness-scoped-intervention}
\end{equation}
where $g_e$ is the activating condition and $R_e$ applies the intended
correction before execution continues. This abstraction makes applicability
part of the update rather than an assumption that every failure-derived rule
should apply globally. It also exposes two failure modes. A correct repair
can be ineffective because its condition rarely activates, or harmful because
that condition captures states where the original behavior was appropriate.
Self-Harness addresses a related decision at proposal time by selecting
evidence-supported mechanisms that its editable surfaces can address, then
testing candidates for regressions~\citep{zhang2026selfharness}. Proposal
eligibility, runtime activation, and candidate admission thus govern different
parts of the improvement process.

The resulting updates need not have identical scope. One Recipe, Many Harnesses finds
that a shared evolution procedure yields both transferable strategies and adaptations
specific to language ecosystems and model weaknesses~\citep{yang2026onerecipe}.
Evo-Harness compiles single-shot executions into general and topic-specific skill
guidance that is retrieved for later tasks~\citep{wei2026evoharness}. Although
described as harness learning, its retained Markdown skills primarily update
procedural assets under our functional distinction. Harness-of-Harness illustrates a
further boundary by organizing existing harnesses into repeated planning,
implementation, and independent testing, retaining project histories to inform
subsequent development~\citep{taxonomy260901481}. Its demonstrated progression
concerns the software project and accumulated project state; this does not by itself
establish persistent revision of the underlying execution architecture. These cases
make the update target essential to interpretation, even when methods share the
language of harness improvement.

\surveyheading{Population-Based Search} Maintaining several candidates allows discoveries
from different workloads or search paths to inform subsequent harnesses. DarwinX uses
a frozen model and selects harness variants under a preserve-and-extend criterion,
retaining alternative lineages for recombination~\citep{zhang2026darwinx}. HarnessBank
organizes successful harnesses by semantic characteristics in a gene bank, then
recombines and screens offspring through staged
verification~\citep{luo2026harnessbank}. EvolveNet distributes a shared harness to
deployments with local workloads and combines the resulting program adaptations using
their scopes and supporting evidence, without centralizing the raw
workloads~\citep{nie2026evolvenet}. HarnessCompass instead emphasizes task-agnostic
edit constraints, feedback from the executing agent, and separate component
optimization before consolidation~\citep{zhang2026harnesscompass}. ADAS provides an
earlier code-based search formulation in which a meta-agent programs new agent designs
using an expanding archive of prior discoveries~\citep{taxonomy240808435}. These methods overlap with archive-based evolution, emphasizing the
selection and combination of useful changes across candidates.

Combining useful changes introduces a further problem of compatibility.
Two edits that each help their original parent may conflict when placed in
one harness, for example if they change the same retry rule or assume different
representations of execution state. Population diversity is valuable when it
preserves complementary solutions and the evidence needed to reconcile them.
The unit of inheritance therefore matters, whether a method transfers a complete
harness, a component implementation, or a scoped program change. Evaluating
these units only in isolation can miss failures created by their composition.

\surveyheading{Pipeline-Driven Evolution} An explicit development pipeline separates
proposing a modification from releasing the resulting agent. Meta-Harness gives its
proposer filesystem access to earlier candidate code, scores, and execution traces,
preserving detailed evidence for subsequent harness edits~\citep{lee2026metaharness}.
AgentDevel maintains a canonical release line, translating observed failure symptoms
into executable diagnoses and assessing candidates through newly solved cases and
regressions~\citep{taxonomy260104620}. CCA uses a meta-agent to construct, test, and
refine agents on a modular SDK, complementing its runtime orchestration and persistent
notes~\citep{wong2025cca}. HarnessX composes typed harness primitives and evolves them
from traces; its broader framework also turns trajectories into model-training
signals, connecting architecture and model updates~\citep{chen2026harnessx}.
Autogenesis separates versioned resources from the protocol for proposing, assessing,
and committing changes, with lineage and rollback spanning several component
types~\citep{taxonomy260415034}. SEA places admission gates with anytime-valid
certificates around a versioned harness and steering adapter, with guarantees tied to
the gate's statistical assumptions and error budget~\citep{sengupta2026sea}. Across
these systems, the pipeline determines which evidence is available when a candidate
becomes persistent.

A release pipeline also preserves the link between a modification and the
evidence used to accept it. Retaining a candidate's source version, motivating
traces, predicted effects, and evaluation conditions makes later regressions
easier to diagnose. This becomes especially useful when improvements accumulate
across rounds, because a failure may result from an interaction between accepted
edits. Tests of whether the modified harness runs successfully and measurements
of whether it improves downstream decisions answer different questions. A
release decision benefits from both, together with evidence about which earlier
successes were lost.

Regression-sensitive admission can be expressed using a fixed, nonempty
validation set $\mathcal V$. Let $y(h,q)\in\{0,1\}$ record whether the agent
with harness $h$ solves case $q$ under matched evaluation conditions. For a
candidate $\widetilde h$, the fractions of newly solved and newly failed cases are
\begin{equation}
  F^+(\widetilde h,h)=\frac{1}{|\mathcal V|}
    \sum_{q\in\mathcal V}\mathbf 1[y(h,q)=0,\ y(\widetilde h,q)=1],
  \qquad
  F^-(\widetilde h,h)=\frac{1}{|\mathcal V|}
    \sum_{q\in\mathcal V}\mathbf 1[y(h,q)=1,\ y(\widetilde h,q)=0].
  \label{eq:harness-regression-flips}
\end{equation}
The empirical success-rate change is $F^+-F^-$. Reporting both terms reveals
whether an aggregate gain hides losses on previously solved cases. These are
finite-set diagnostics; stochastic outcomes require repeated evaluation, and
passing an admission gate does not establish benefit beyond its validation scope.

\surveyheading{Runtime Evolution} TTHE updates harnesses during evaluation using unlabeled
execution traces, maintaining candidates and committing a program through
execution-derived proxy signals so that it governs subsequent
inputs~\citep{nie2026tthe}. Argus retains verified runtime state, including reusable
procedures and routing decisions, across bounded missions with a fixed model; its
longitudinal evidence is observational rather than a controlled learning
ablation~\citep{li2026argus}. Life-Harness offers a complementary runtime interface
design. It converts training-trajectory failures into interventions for interface
contracts, procedures, action validation, and recovery, then keeps the resulting
harness fixed on unseen tasks~\citep{xu2026lifeharness}. Its evaluation concerns
deterministic interactive environments beyond coding. These cases separate when an
intervention operates from when it is learned. A runtime check can be the product of
offline evolution, while online evolution requires an update that persists beyond the
execution that generated it.

Online updating further couples learning to the changing distribution of
experience. An adopted retry or termination rule changes which states later
executions visit, and therefore which failures the next update can observe.
The absence of an earlier error may indicate a successful repair, or simply
that the revised harness no longer reaches the relevant situation. Runtime
evaluation should therefore track both task outcomes and the circumstances
in which interventions activate. In terms of the earlier five links, harness
evolution must connect a diagnosed failure to an executable change, its actual
activation, and evidence of benefit over the intended reuse scope.

\subsection{Workflow Evolution}
\label{sec:workflow-evolution}

Workflow evolution updates $w_t$, the arrangement of task-solving stages and
their dependencies. Its value lies in changing when evidence becomes available
and which decisions it can influence. Moving validation before an expensive
rewrite can expose a mistaken assumption before it propagates into code;
routing a failing test back to localization can reopen a decision that a fixed
repair loop would leave unchanged. Such changes can improve the use of existing
capabilities. We describe a workflow as $w_t=(V_t^{w},E_t^{w},\ell_t)$, where
$V_t^{w}$ contains stages, $E_t^{w}$ describes control and information
dependencies, and $\ell_t$ assigns instructions and operations. A proposed
structural or configuration edit $\eta_t$ produces
\begin{equation}
  \widetilde w_{t+1}=\operatorname{Edit}(w_t;\eta_t).
  \label{eq:workflow-structure-edit}
\end{equation}
Edits can add or remove stages, redirect dependencies, or change branch and
loop conditions. A workflow can therefore evolve by removing redundant work
or improving feedback routing as well as by adding new stages.

\surveyheading{Structure Search and Diagnosis}
AFlow searches code-represented workflows using Monte Carlo Tree Search,
reusable operators, execution feedback, and experience from previous
modifications~\citep{zhang2024aflow}. Its search explores alternative
arrangements of operations, making the task-solving procedure itself a candidate
for evaluation. SEW separately evolves workflow structure and agent prompts,
including mutation prompts, and studies how textual representations affect
workflow validity~\citep{liu2025sew}. These methods highlight two coupled
choices, the structure to execute and the representation through which that
structure can be modified. A search space must make useful changes expressible
while preserving executable dependencies. JUDGEFLOW addresses where to edit.
Its judge inspects traces, ranks the responsibility of logic blocks, and directs
the optimizer toward problematic blocks~\citep{taxonomy260107477}. This is
useful because the stage reporting a failure may have inherited an earlier
mistake. For example, a debugging stage may repeatedly revise a patch when the
underlying problem is incorrect file localization. Trace-based diagnosis can
suggest changing the upstream stage or the feedback path. The diagnosis
guides a proposal; executing the revised workflow tests whether that proposal
improves the overall procedure.

\surveyheading{Diversity and Reusable Workflow Components}
EvoFlow maintains heterogeneous workflows through tag-based parent retrieval,
crossover, mutation, and niching selection~\citep{taxonomy250207373}.
Retaining different structures allows workflow choice to reflect task complexity
and inference cost. A short synthesis task may benefit little from a review
loop that is useful for a difficult repair, so one aggregate score can conceal
which tasks benefit from which structures. EvoAgentX provides a modular platform
for generating, executing, evaluating, and optimizing workflows, integrating
TextGrad, AFlow, and MIPRO to support changes to prompts, tool configurations,
and topology~\citep{wang2025evoagentx}. Its scope shows why improvements to a
workflow system may involve several component types. FlowEvo makes the interaction
with assets explicit by compiling successful workflows into executable skills,
retaining them in a persistent bank, and retrieving them for direct execution or
later workflow construction~\citep{taxonomy260721596}. It also tracks downstream
utility to suppress skills associated with negative transfer. A discovered
procedure can thus move between $w_t$ and the skill state $\mathcal K_t$.
Retained skills expand the building blocks available to later workflow search,
while new workflows provide settings in which those skills are exercised and
reassessed. This creates a feedback loop between learning reusable procedures
and discovering how to organize them.

Selecting an existing branch exercises a workflow; retaining an
experience-derived revision constitutes an update. To examine a stage's
contribution, let $\mathcal A^{w}$ use workflow $w$ and $w\setminus v$ be a
valid bypass of stage $v$, with the other agent components fixed. Under matched
evaluation conditions,
\begin{equation}
  \Gamma(v\mid w)=\mathbb E_{q\sim\mathcal D}\!\left[
    s(\mathcal A^{w},q;B)-s(\mathcal A^{w\setminus v},q;B)
  \right].
  \label{eq:workflow-stage-contribution}
\end{equation}
This contribution depends on the surrounding workflow and bypass design.
A stage may help because it supplies evidence that another stage needs;
its effect cannot be inferred from its own output quality alone. A small
average contribution may also hide benefits on rare failures or redundancy
with another stage. The common budget $B$ constrains extra computation, while
recording actual cost and branch activation helps explain how an evolved
workflow achieves its result. The central evaluation question is whether
reorganizing execution makes useful evidence influence decisions at the right
time and at an acceptable cost.

\subsection{Multi-Agent Evolution}
\label{sec:multi-agent-evolution}

Multi-agent evolution concerns $g_t$, including the participating agents, their responsibilities, and the mechanisms through which they exchange and act on information. While a workflow describes dependencies among stages, a team description also identifies who performs them and how participants coordinate. A useful abstraction is
\begin{equation}
g_t=(V_t^{g},E_t^{g},\rho_t^{g},p_t^{g}),
\label{eq:team-organization}
\end{equation}
where $V_t^{g}$ is the participant set, $E_t^{g}$ contains communication relations,
$\rho_t^{g}$ assigns responsibilities, and $p_t^{g}$ specifies interaction and
decision rules. The latter matters because the same communication graph can
support different handoff requirements, review responsibilities, or
disagreement-resolution procedures. For example, a reviewer may merely suggest
changes or require unresolved test failures to be addressed before a patch is
accepted. These arrangements give the same message different consequences.
Team evolution therefore concerns how distributed evidence becomes a shared
decision, including who can challenge an assumption and who must act on the result.

\surveyheading{Feedback-Driven Team Revision} EvoMAC obtains feedback from a testing team
and uses textual backpropagation to revise the coding team's agent instructions and
connections~\citep{hu2024evomac}. These updates are exercised in subsequent rounds of
the same software-development task. Meta-Team, introduced in \emph{Evolve as a Team},
preserves individual execution contexts and enables post-task discussion to combine
evidence distributed across participants~\citep{taxonomy260529790}. It updates
individual scaffolds, teammate profiles, and team-level composition and rules,
committing validated changes to a reusable team scaffold. Their comparison highlights
two distinct reuse scopes, adaptation within a task and retained changes for later
tasks. They also expose a difficulty in assigning responsibility for team failures.
The last agent to produce an incorrect answer may have acted on an incomplete
handoff, while the relevant warning remained in another participant's context.
For example, a tester may observe an environment mismatch that a developer
interprets as a code defect. Improving the developer's instructions alone may
leave this failure intact. Useful revision requires relating what each agent
observed, what it communicated, and how the recipient used that information.
This can reveal whether to improve an individual role, change a handoff, or
require explicit resolution of conflicting evidence.

\surveyheading{Adaptive Topology and Mixed Update Targets} AgentConductor trains an
orchestrator to generate layered interaction graphs that reflect task difficulty and
communication cost~\citep{wang2026agentconductor}. At inference, the orchestrator is
frozen but can regenerate a topology after execution feedback. Its persistent learning
resides in the orchestrator parameters, while its runtime effect is architectural
adaptation. SEMAG uses hierarchical planning, coding, debugging, and discussion,
together with an optional mechanism that selects updated backbone models using
retrieved information and sampled performance~\citep{peng2026semag}. Its
execution-dependent escalation demonstrates adaptive coordination, although traversing
predefined stages alone does not establish a learned revision of coordination rules.
SAGE provides another boundary case. Its Challenger, Planner, Solver, and Critic
organize task generation, planning, solving, and quality filtering, while
verifier-based training updates their shared model~\citep{taxonomy260315255}. This
supports collaborative model improvement rather than independently demonstrating
topology evolution. Together, these cases distinguish the organization instantiated
for a task from the persistent mechanism that produces or uses it. A different
graph on the next task may reflect a learned coordination policy or a fixed
routing rule responding to different inputs. Evaluation must identify which
mechanism changed and whether its choices remain useful beyond the situations
that generated the update. This distinction also connects architecture and model
evolution without treating every dynamically assembled team as a new learning event.

\surveyheading{Collaboration during Evolution} GEA shares experience among a group of
candidate agents so that framework modifications and execution outcomes from one
lineage can inform others~\citep{weng2026groupevolvingagents}. The resulting offspring
retain their own revised implementations. Here, the group organizes the improvement
process and need not become a team that jointly solves each downstream task. This
distinction separates who collaborates to produce an update from what the deployed
update changes. Sharing experience can broaden the evidence available to an
offspring, but a useful change in one lineage may rely on roles or interfaces
absent from another. The transferable unit therefore includes the conditions
under which an edit worked, as well as the edit itself. Otherwise, combining
successful local changes can introduce incompatible assumptions.

Across these settings, evidence of team improvement should identify the retained
target and assess its contribution under the common execution budget $B$.
Additional participants can provide independent checks or complementary evidence,
but agreement can also arise from agents repeating the same unsupported premise.
Message count and consensus alone therefore provide weak evidence of effective
coordination. Comparisons should account for total model calls, communication,
and latency, and examine whether revised roles or interaction rules change which
errors are detected and corrected. A positive $\Delta J_t$ establishes an overall
gain under the specified conditions; attributing that gain to coordination
requires comparisons that separate these revisions from added computation or
simultaneous model and asset updates.

\endgroup

\section{Model Self-Evolution}
\label{sec:model-self-evolution}

Model self-evolution incorporates coding experience into the parameters
$\theta_t$ that shape subsequent agent decisions. Within
$\mathcal A_t=(\mathcal S_t,\mathcal H_t,\theta_t)$, these updates can change
how the model localizes a defect, proposes a repair, interprets feedback, or
decides to verify its work. Parameter learning can make an acquired behavior
available across tasks without requiring explicit retrieval of the originating
experience. Its usefulness then depends on whether the model preserves the
conditions under which that experience was informative. A successful repair
can teach a transferable strategy, but training on an incidental feature of
the same repair may produce an unreliable shortcut.

The following discussion distinguishes how training experience is generated,
how feedback becomes an optimization signal, and which model role is updated.
Self-play co-evolution changes the tasks or evaluators that supply learning
signals; environment-reward learning grounds updates in task outcomes; harness
co-evolution connects parameter learning with changes to the surrounding
execution system. These mechanisms can overlap. We retain the earlier
distinction between a candidate update and its adoption, and use
$\Delta J_t$ to assess subsequent benefit. The formulas below describe shared
mechanisms or explicit comparison protocols; individual methods instantiate
different data, losses, and update schedules.

\subsection{Self-Play Co-Evolution}
\label{sec:model-self-play}

Self-play makes the current system a source of its own future training
experience. An evolving model may generate tasks, solve them, construct tests,
or judge alternative outputs. Consequently, parameter updates also change the
distribution of experience available in the next round. Let $\mathcal Q_t$
denote the training-task distribution. In a batch of $n_t$ records, each
$(q_i,\tau_i,f_i)$ contains a task, an execution trace, and its feedback.
The records may include several model roles and unsuccessful attempts.
A common data-construction abstraction is
\begin{equation}
  \mathcal D_t^{\mathrm{train}}=
  \operatorname{Curate}_t\!\left(
    \{(q_i,\tau_i,f_i)\}_{i=1}^{n_t}
  \right).
  \label{eq:model-training-records}
\end{equation}
Here $q_i$ is sampled from $\mathcal Q_t$, which may combine generated and
externally supplied tasks. Curation determines whether a record supplies a
demonstration, a preference, a reward, or privileged information for a teacher.
It can relabel or reweight experience as well as reject it. The learning signal
therefore depends on both what the current system encounters and how those
encounters are interpreted.

\surveyheading{Task Generation and Curriculum}
Self-Play SWE-RL uses a shared model for bug injection and repair, combining
consistency checks and difficulty-sensitive rewards with joint reinforcement
learning~\citep{selfplaysoftwareagents2026}. Failed repairs can contribute
further bug states, connecting the solver's current limitations to later
training tasks. Absolute Zero similarly trains a model to propose and solve
executable reasoning tasks, using execution to establish task outcomes and
solver success rates to guide task learnability~\citep{taxonomy250503335}.
These mechanisms make task difficulty responsive to the learner. However,
difficulty alone does not establish relevance to the intended deployment
distribution. Anchored Self-Play addresses this gap in code repair by mixing
reference bugs into training and shaping generated edits through similarity
to reference changes~\citep{taxonomy260703523}. This comparison separates
three curriculum requirements, whether tasks are valid, whether the model can
learn from them, and whether they exercise the capabilities needed later.
Improving one does not ensure the others. Analysis of learnable information gain also links sustained self-play
improvement to the information supplied by the synthetic pipeline rather than
data volume~\citep{liu2026selfplayinformation}.

\surveyheading{Coupled Code and Test Learning}
ReVeal trains one policy for code generation and self-verification, filtering
generated tests against a reference solution during training
\citep{jin2025reveal}. CURE also uses a shared model in coder and tester
roles, with reference unit tests helping establish code quality and guide
tester learning~\citep{wang2025cure}. These reference signals anchor different
parts of the loop. A test that accepts a correct program need not distinguish
that program from plausible incorrect alternatives. For a task $q$, let
$\mathcal Z_t^-(q)$ be a nonempty set of programs independently established
as incorrect, and let $v$ be a test checked against the task specification.
Its empirical rejection rate is
\begin{equation}
  d_t(v;q)=\frac{1}{|\mathcal Z_t^-(q)|}
  \sum_{z\in\mathcal Z_t^-(q)}
  \mathbf 1[\text{$v$ rejects $z$}].
  \label{eq:model-test-discrimination}
\end{equation}
This diagnostic separates test validity from its ability to expose known
errors. It does not measure coverage of unobserved failures. When no independent
labels are available, both program quality and test quality must instead be
inferred from their interactions, making mutual agreement a less reliable
anchor.

ZeroCoder studies this latter setting by estimating code and test quality
from execution consistency without reference code or tests in its basic
configuration~\citep{fan2026zerocoder}. Its extensions use a small labeled
set for supervision and, for its dynamic Bayesian selector, prior calibration.
Sol-Ver trains solving and verification through supervised fine-tuning followed
by preference optimization~\citep{lin2025solverifier}. ACE uses a shared model
as solver and adversary, alternating solver fine-tuning with adversarial
preference learning~\citep{huang2026ace}. Its generated inputs probe runtime
failures, while reference tests remain part of code selection. The methods
therefore differ in both the optimization procedure and the claim supported
by execution. Normal termination, agreement with generated expectations, and
agreement with reference tests are distinct signals. A co-evolving evaluator
is useful when it reveals weaknesses that matter to the task; merely becoming
more favorable toward the co-evolving solver can conceal those weaknesses.

\surveyheading{Learning from Privileged Experience}
OPD-Evolver couples rapid memory updates with slower parameter learning through
on-policy self-distillation~\citep{taxonomy260617628}. The same evolver supplies
a teacher view conditioned on hindsight information and a student view
restricted to the information available when acting. Its parameter updates
cover selection, execution, memory writing, and maintenance. Let $x_j$ be
the student's information at token position $j$, including its own generated
prefix, and let $u_j$ denote additional hindsight information. With frozen
teacher parameters $\bar\theta_t$, the token-level distillation loss has the form
\begin{equation}
  \mathcal L_t^{\mathrm{distill}}(\theta)=
  \mathbb E_{(x_j,u_j)\sim\mathcal P_t}
  D_{\mathrm{KL}}\!\left(
    \pi_{\bar\theta_t}(\cdot\mid x_j,u_j)
    \,\|\,\pi_\theta(\cdot\mid x_j)
  \right).
  \label{eq:model-hindsight-distillation}
\end{equation}
Here $\mathcal P_t$ is formed from student rollouts and their associated
hindsight views; the teacher distribution receives no gradient. Asset and
architecture conditioning is implicit in the two views. The teacher's advantage
comes from additional information, without requiring a stronger external
model. This is a boundary case within self-play co-evolution, where the roles
are informationally asymmetric rather than adversarial. The update seeks to
make lessons from later outcomes useful at earlier decision points. The
privileged information must remain unavailable to the evaluated student if
the experiment is to establish that this capability was learned.

\subsection{Environment-Reward RL}
\label{sec:model-environment-reward}

Environment-grounded learning derives supervision from what generated programs
or agent trajectories achieve. For a fixed training round, write
$\mathcal A_\theta=(\mathcal S_t,\mathcal H_t,\theta)$ and let
$R_t(q,\tau)$ be the scalar training return constructed from execution
feedback. An idealized policy objective is
\begin{equation}
  \max_\theta\;
  \mathbb E_{q\sim\mathcal Q_t,\,
    \tau\sim\operatorname{Execute}(\mathcal A_\theta,q,\mathcal E_q)}
  [R_t(q,\tau)].
  \label{eq:model-environment-objective}
\end{equation}
The trajectory notation denotes the trace component of execution, with
$\mathcal E_q$ the task environment. Holding assets and architecture fixed
isolates parameter optimization within the round. Practical updates use finite
samples and may include regularization. The training return $R_t$ need not
equal the evaluation score $s$, and $\mathcal Q_t$ need not equal the held-out
distribution $\mathcal D$. These distinctions matter when a convenient reward
captures only part of the intended coding capability.

\surveyheading{Guided Exploration and Credit Assignment}
Agent-RLVR addresses sparse success signals by having agents first attempt
software tasks, then reattempt failures with guidance derived from plans or
environment feedback~\citep{da2025agentrlvr}. Successful traces support an
initial supervised stage; successful and unsuccessful attempts supply
preferences for offline DPO. Guidance is used during training, while the
evaluated agent acts without it. For a common conditioning input $x$, let
$y^+$ and $y^-$ denote preferred and dispreferred continuations in
$\mathcal D_t^{\mathrm{pref}}$. A preference-learning objective is
\begin{equation}
  \mathcal L_t^{\mathrm{pref}}(\theta)=
  -\mathbb E_{(x,y^+,y^-)\sim\mathcal D_t^{\mathrm{pref}}}
  \log\operatorname{sigmoid}\!\left[
    \beta\log\frac{\pi_\theta(y^+\mid x)}{\pi_{\mathrm{ref}}(y^+\mid x)}
    -\beta\log\frac{\pi_\theta(y^-\mid x)}{\pi_{\mathrm{ref}}(y^-\mid x)}
  \right].
  \label{eq:model-execution-preferences}
\end{equation}
Here $\pi_{\mathrm{ref}}$ is a fixed reference policy and $\beta>0$ controls
the preference scale. Execution determines the ordering; it does not uniquely
identify which intermediate decision caused the outcome. A passing trace may
contain redundant searches or incidental edits. Pair construction and guidance
therefore affect what the model learns from the same terminal test result.
Agent-RLVR's reported experiment covers one training cycle, so its gains should
not be interpreted as evidence of indefinitely repeated improvement.

\surveyheading{Reward Semantics and Search Experience}
SWE-RL uses issue descriptions, code context, and reference patches from
software history, rewarding similarity between generated and reference patches
with a penalty for invalid formatting~\citep{wei2025swerl}. It optimizes the
model with GRPO. Its supervision is grounded in software evolution data,
but the training reward is not a unit-test execution verdict. Similarity can
provide an inexpensive signal while favoring the particular implementation
represented by the reference patch. The method therefore illustrates why
software provenance, executable verification, and autonomous task generation
must be distinguished when comparing learning loops.

SOAR supplies a complementary boundary case under this feedback-based
category~\citep{taxonomy250714172}. It alternates evolutionary program search
with supervised learning from the resulting attempts, improving both program
sampling and refinement. Hindsight relabeling converts an unsuccessful program
into a solution for a different task consistent with its actual outputs.
Thus, execution feedback can support parameter learning without serving as a
direct policy-gradient reward. Relabeling preserves useful computation from
failed search, but changes what the resulting training pair teaches. A program
that failed the original specification is not thereby a successful solution
to that specification. SOAR's inclusion highlights a shared feedback source
across different optimization mechanisms.

\surveyheading{Feedback Infrastructure and Transfer}
SWE-Gym and R2E-Gym provide executable environments and training pipelines
for agents and verifiers~\citep{swegym2025,jain2025r2egym}; SWE-RM learns
execution-free feedback for candidate selection and policy improvement
\citep{shum2025swerm}. Their contribution includes making informative
experience or scalable evaluation available to a learning loop. The resulting
model gains still depend on how that feedback is used. Learned rewards can
reduce execution cost, but policy optimization may exploit their errors;
executable tests provide concrete evidence, but cover only specified cases.
Evaluation should therefore retain independent outcome checks and compare
updated policies under the same budget $B$. Training and data-collection costs should be reported separately, especially
when guidance or repeated search makes successful traces expensive.

\subsection{Harness Co-Evolution}
\label{sec:model-harness-co-evolution}

The harness $h_t$ shapes both the model's current execution and the experience
available for its next update. Changing action interfaces, feedback delivery,
or recovery rules can make different trajectories reachable. Conversely,
parameter learning can change which interventions remain useful. Harness
co-evolution therefore concerns this feedback between learned behavior and
execution control. It also distinguishes learning to solve tasks from learning
to modify the system that solves them.

\surveyheading{Learning the Editing Policy}
Harness-R1 trains a dedicated harness engineer through supervised initialization
and online GRPO, while freezing the target agent during engineer training
\citep{shao2026harnessr1}. Failure traces condition executable patches, and
rerunning the target with a patch supplies the engineer's reward. To describe
this separation, let $\omega$ denote editor parameters, $b_t$ the failure
evidence, and $\mathcal A_t^e$ the target with edit $e$ installed. For an
installable edit evaluated on a nonempty matched task batch $\mathcal V_t$,
the empirical utility change can be written as
\begin{equation}
  \widehat\Delta_t(e)=\frac{1}{|\mathcal V_t|}
  \sum_{q\in\mathcal V_t}
  [s(\mathcal A_t^e,q;B)-s(\mathcal A_t,q;B)].
  \label{eq:model-editor-utility}
\end{equation}
The batch includes previously successful cases so that regressions contribute
to the signal. Define $r_t^{\mathrm{edit}}(e)$ to equal this change for a
completed valid evaluation and zero for invalid or incomplete candidates.
The corresponding editor objective is
\begin{equation}
  \max_\omega\;
  \mathbb E_{e\sim\pi_\omega(\cdot\mid b_t)}
  [r_t^{\mathrm{edit}}(e)].
  \label{eq:model-editor-objective}
\end{equation}
The expectation also includes the training distribution of failure bundles.
The updated parameters belong to the editor, while the target receives an
architectural intervention. Harness-R1 evaluates this mechanism on WebShop,
ALFWorld, and DBBench; its evidence concerns executable harness engineering
across interactive domains, rather than a direct repository-repair benchmark.
The learned capability is to produce useful edits from new failures, so
transfer of the editor must be distinguished from reuse of one particular patch.

\surveyheading{From Harness Search to Model Learning}
HELIX proposes a build--update--rebuild process in which harness variants
produce verified experience for model updating, followed by harness rebuilding
for the changed model~\citep{fan2026helix}. It retains source-traceable
execution records for related candidates, including successful repairs,
regressions, and near misses, and materializes supervised, critic, filtering,
and preference data. Such contrasts can preserve why two executions differed,
instead of retaining only the best final patch. Its reported study demonstrates
harness search and training-data construction, but does not train an updated
model. It therefore supports the handoff from architecture evolution to
parameter learning, while leaving the complete repeated cycle to be evaluated.

To assess a completed coupled update, the old and new model--harness
combinations should be compared under matched conditions. For $a,b\in\{0,1\}$, let $j_{ab}$ be the expected score under
$\mathcal D$ and budget $B$ when model parameters $\theta_{t+a}$ are paired
with harness $h_{t+b}$. All four combinations must have compatible interfaces.
Assets, workflow, and team organization are held fixed for this comparison. The interaction contrast is
\begin{equation}
  I_t^{\theta,h}=j_{11}-j_{10}-j_{01}+j_{00}.
  \label{eq:model-harness-interaction}
\end{equation}
A positive value means the combined change exceeds the sum of the isolated
changes on the chosen score scale; it does not by itself imply positive total
improvement, which requires $j_{11}>j_{00}$. The contrast is specific to the
tested configurations, distribution, and budget. It makes a central question
of co-evolution explicit, whether parameter learning and harness revision
reinforce one another or merely contribute separate gains. Across all three
model-evolution mechanisms, stronger training signals matter only when their
effects survive adoption and improve coding behavior.

\section{Coding Feedback and Evolution Dynamics}
\label{sec:coding-evolution-signals}

Coding environments expose unusually rich evidence about the consequences of
agent actions. A single attempt can leave a patch, build status, test verdicts,
an exception stack, a sequence of tool calls, resource measurements, and a
developer correction. These observations make software engineering a natural
setting for self-evolution because they can often be attached to inspectable
artifacts and reproduced in an executable environment. Their precision does
not remove the central difficulty. A test may establish that one patch satisfies
one checked property while providing insufficient support for a memory, skill,
or harness rule that will influence many later tasks. We therefore organize the
discussion around four questions: why coding environments provide distinctive
signals, what each signal reveals, how methods convert signals into persistent
updates, and when those updates are made.

\subsection{Why Coding Is a Natural Domain for Self-Evolution}
\label{sec:coding-feedback-environment}

Coding has played a formative role in agent research and has been one of the
earliest and most sustained application frontiers for LLM-based agents. It
supplies an instrumented environment, not only a task domain, in which a model
can execute an action, observe its consequences, and revise its behavior.
Codex and AlphaCode established execution-based evaluation early in the
development of language models for code~\citep{chen2021evaluating,
li2022alphacode}. Repository-level systems then extended this
generate--execute--inspect loop to file navigation, command-line tools, testing,
and multi-step repair~\citep{jimenez2023swebench,yang2024sweagent,
wang2024openhands}. Tool-mediated action, environment feedback, and
verifier-guided revision were consequently developed or stress-tested early in
coding settings. These patterns now reach other domains, where code serves as
an operational medium for action, state, and verification
\citep{ning2026codeharness}. Coding is therefore a mature proving ground for
agents, with both attempted changes and much of their evidence already
represented in machine-readable form.

A coding task is embedded in a repository snapshot, dependency configuration,
toolchain, and execution environment. One attempt can leave a patch, build
status, named test failures, an exception stack, tool-call history, resource
measurements, and review comments. We represent the evidence associated with
one coding event as
\begin{equation}
  \mathcal F_t=(y_t,d_t,\tau_t,r_t,q_t,h_t,m_t; e_t,p_t),
  \label{eq:structured-coding-signal}
\end{equation}
where $y_t$ records execution outcomes, $d_t$ diagnostic evidence, $\tau_t$ the
coding trajectory, $r_t$ repository context, $q_t$ software-quality
attributes, $h_t$ human judgments, and $m_t$ model judgments.
The environment state $e_t$ and provenance $p_t$
bind those observations to a repository revision, dependency state, tool
version, candidate change, and evaluator. This structure matters because an
identical patch can produce a different result after a dependency update, and
the same score has a different meaning when it comes from a regression suite,
a model judge, or a developer review.

Three properties distinguish these signals as a basis for evolution.

\begin{itemize}[leftmargin=*,itemsep=5pt,topsep=3pt]
  \item \textbf{Specificity.} Coding feedback often identifies a concrete
  violated condition. A compiler can report an error class and source location,
  a unit test can name the failed assertion and expose expected and observed
  behavior, and a static analyzer can identify a rule and code region. The
  signal can therefore constrain both \emph{what} should change and
  \emph{where} the agent should look. Executable benchmarks and training
  environments exploit this property to turn test outcomes into validation or
  learning signals~\citep{chen2021evaluating,swegym2025,da2025agentrlvr}.
  Specificity is nevertheless local to the checked property. A passing suite
  is a partial oracle, and converting one failure into a general memory, skill,
  or harness rule still requires identifying the conditions under which that
  lesson applies.

  \item \textbf{Traceability.} Software work leaves a causal record that can be
  inspected after execution. Issues state the requested behavior, version
  control identifies the repository state, diffs record the proposed
  intervention, and trajectories connect searches, edits, commands, and
  responses. SWE-bench grounds evaluation in issue--repository--patch relations,
  while SWE-agent shows how an agent--computer interface makes repository
  actions and their feedback explicit~\citep{jimenez2023swebench,
  yang2024sweagent}. This lineage lets an evolution mechanism attach provenance
  to a retained update and assign credit more precisely than a final reward
  alone. It also exposes confounders. A successful patch may follow several
  irrelevant actions, whereas a failed attempt may contain a correct diagnosis
  followed by an incorrect edit.

  \item \textbf{Reproducibility.} A repository revision, environment image,
  command, and test suite can define a replayable experiment. The same candidate
  can be executed again, compared with its parent under matched conditions, and
  checked for regressions before adoption. SWE-Gym and R2E-Gym package coding
  tasks as executable environments, and SWE-bench-Live provides task-specific
  container images and repeated validation to improve reproducibility
  \citep{swegym2025,jain2025r2egym,zhang2025swebenchlive}. Reproducibility here
  is conditional rather than absolute. Flaky tests, external services,
  dependency drift, hardware, and stochastic agent decisions can still change
  an outcome. Coding environments are distinctive because these conditions can
  often be recorded, replayed, and tested for instability.
\end{itemize}

Together, specificity, traceability, and reproducibility shorten the path from
an observed failure to an addressable and testable candidate update. Other
agent environments may provide one or two of these properties, but software
engineering routinely combines all three through executable artifacts, version
history, and automated checks. This combination explains why coding has been an
important proving ground for self-correction and agent evolution. It does not,
however, collapse task verification into evolution evaluation. Tests directly
judge the current software artifact; only retained reuse and matched evaluation
on later cases can show that the changed agent has improved.

\subsection{What Coding Signals Reveal}
\label{sec:types-coding-signals}

The properties established above explain why software evidence is useful, but
an evolution mechanism still depends on which evidence it actually consumes.
We distinguish seven non-exclusive signal classes throughout
this survey: \textbf{execution outcomes (E)}, \textbf{diagnostic evidence (D)},
\textbf{coding trajectories (T)}, \textbf{repository context (R)},
\textbf{quality attributes (Q)}, \textbf{human judgments (H)}, and
\textbf{model judgments (M)}. The first five classes are grounded in software
execution, process, state, or properties. H and M instead record evaluative
judgments: H reflects developer or organizational assessment, whereas M reflects
a learned evaluator whose reliability depends on both the evaluator and its
input evidence.
The codes describe evidence used to construct, revise, or select a persistent
update. They do not describe the benchmark on which the resulting agent is
later evaluated. A method may receive several codes because one coding event
can expose several views of the same outcome. For example, a failed integration
test supplies an execution outcome (E), its stack trace supplies diagnostic evidence
(D), the preceding tool calls form a trajectory (T), and the repository revision
defines the conditions under which the evidence is valid (R).

\input{tables/coding-evolution-signals}

\surveyheading{Execution Outcomes (E)}
Execution outcomes record whether a software artifact satisfies an operational
condition. They include unit and integration tests, compilation,
interpretation, patch verifiers, task-resolution checks, and measured execution
outcomes. DGM, Mendel GM, and Huxley GM use verified coding performance to select
agent variants across generations, while SICA retains source-level
self-modifications according to coding-task results
\citep{zhang2025darwin,mendel2026godel,huxley2025godel,
robeyns2025selfimproving}. AFlow uses execution outcomes to select workflows,
and Agent-RLVR turns unit-test results into rewards for policy learning
\citep{zhang2024aflow,da2025agentrlvr}. The same class of evidence can therefore
drive architecture or model evolution. Its strength is an explicit and
repeatable verdict; its boundary is the property checked. Compilation confirms
acceptance by a build process, and a passing test confirms only the encoded
assertions. Infrastructure failure, test non-execution, and a completed negative
verdict must also remain distinguishable because they support different updates.

\surveyheading{Diagnostic Evidence (D)}
Diagnostic evidence describes where and how execution failed. It includes
compiler messages, assertion output, exception stacks, test logs, shell output,
dependency conflicts, build failures, and tool errors. Live-SWE-agent uses
command and test feedback while revising utilities, whereas HarnessFix and HELIX
use failed trajectories and diagnostic traces to identify editable harness
components~\citep{xia2025livesweagent,chen2026harnessfix,fan2026helix}.
ReVeal, CURE, ZeroCoder, and Sol-Ver similarly use interpreter feedback,
generated tests, or execution failures to improve coder--verifier behavior
\citep{jin2025reveal,wang2025cure,fan2026zerocoder,lin2025solverifier}.
Diagnostics give failure an address, making them especially useful for deciding
what to revise. That address may still identify a symptom. An exception in a
caller may originate in an earlier interface change, while an import error may
reflect the environment rather than the application code.

\surveyheading{Coding Trajectories (T)}
A coding trajectory records how an attempt unfolds through repository search,
fault localization, file inspection, editing, testing, failed branches, retries,
and recovery. Its value lies in retaining process information discarded by a
final score. SAGE compresses search--edit--test rollouts into reusable plans;
SWE-Exp, EvoCoder, and Subtask Memory transform issue-resolution histories into
experience at episode, repository, or subtask granularity
\citep{hayashi2025sage,chen2026sweexp,lin2024evocoder,
shen2026subtaskmemory}. CODESKILL and Socratic-SWE further distill traces into
skills for future tasks~\citep{li2026codeskill,xiao2026socraticswe}. Trajectories
support procedural abstraction and credit assignment, but they do not solve
either problem automatically. Successful traces can contain ineffective
actions, and failed traces can contain a correct diagnosis followed by a faulty
edit. Labeling every step with the final outcome would preserve both kinds of
error.

\surveyheading{Repository Context (R)}
Repository context describes the persistent software setting in
which an experience was produced. They include issues, commits, pull requests,
changed files, directory structure, interfaces, cross-file dependencies,
line-level relevance, and repository revisions. Repository Memory links commit
history, issues, and file changes to later localization, while GSkill and GSE
use repository structure, conventions, and interaction histories when forming
reusable capabilities~\citep{wang2026repositorymemory,tan2026gskill,
yang2026gse}. SWE-Pruner learns which code should remain visible, and Libra
revises repository catalogs after localization failures
\citep{wang2026swepruner,zhao2026libra}. These signals do more than supply
context. They define applicability. A workaround tied to one dependency version
should not become an unconditional rule, and an accepted historical patch is
evidence about one development state rather than the only valid implementation.

\surveyheading{Quality Attributes (Q)}
Quality attributes describe software properties beyond functional correctness,
including runtime, memory use, resource consumption, security, vulnerability
removal, maintainability, and reviewability. EffiSkill derives optimization
skills from slow-to-fast program pairs and measured efficiency gains, while
EvoRepair combines repair outcomes with vulnerability patterns to retain
security-repair experience~\citep{wang2026effiskill,hu2026evorepair}. Such
signals are essential when two patches both pass tests but differ in operational
or engineering value. They also introduce multi-objective trade-offs. An update
that improves latency can increase memory use, and a security hardening rule can
reduce compatibility. The metric, workload, and acceptable trade-off must
therefore remain attached to the retained update.

\surveyheading{Human Judgments (H)}
Human judgments arise from ordinary software collaboration. They
include code-review comments, developer edits, acceptance or rejection,
corrections, project conventions, and personal preferences. Behavioral Rules
turns accepted review comments into persistent instructions and tracks whether
the corresponding errors recur. Personalized Skills extracts user-specific
procedures from developer--agent histories, and Ouroboros incorporates reviewed
commits and feedback about weaknesses in the coding-agent core
\citep{aggarwal2026behavioralrules,huang2026personalizedskills,
razzhigaev2026ouroboros}. Human judgments can express intent, maintainability,
and organizational authority that tests omit. It is also sparse, costly, and
sometimes subjective, so the identity and scope of the feedback provider matter
when deciding whether a preference should become a general rule.

\surveyheading{Model Judgments (M)}
Model judgments are produced by an LLM judge, critic, reward model, or peer
agent rather than directly by program execution or a developer. It includes
natural-language diagnoses, candidate rankings, patch scores, responsibility
assignments, and learned reward estimates. JUDGEFLOW uses a judge to rank the
responsibility of workflow blocks, SIFT screens candidate harness patches before
expensive execution, and SWE-RM predicts the quality of software-engineering
trajectories~\citep{taxonomy260107477,fu2026sift,shum2025swerm}. M makes dense
and scalable feedback available when execution is expensive or when a judgment
requires semantic interpretation. Its output is nevertheless an inference, not
an independent software fact. A confident critique may mislocalize a failure,
and a reward model may preserve biases in its training comparisons. When a model
judge consumes tests, diagnostics, trajectories, or repository evidence, we
assign M together with the corresponding E, D, T, or R codes; an ungrounded
model score receives M alone.

The seven codes are non-exclusive evidence labels rather than a ranking of
reliability. Execution outcomes are usually precise but narrow; trajectories
and human judgments are semantically richer but require interpretation; model
judgments are scalable but inherit the coverage, provenance, and failure modes of
both its inputs and evaluator. A method evaluated on a coding benchmark does not
automatically receive a coding-specific evidence code. The label records what
the evolution loop actually consumes while constructing or selecting its update.

We manually annotated all 108 methods in
Figure~\ref{fig:taxonomy-coding-agents} under this criterion. Execution outcomes
dominate (E, 99), followed by trajectories (T, 87), diagnostics (D, 55), model
judgments (M, 43), repository context (R, 37), quality attributes (Q, 21), and
human judgments (H, 5).

The distribution exposes a sharp grounding imbalance. Current systems learn
mainly from what software can execute and what agent runtimes can record, while
developer intent remains weakly represented. Grouping the same annotations by
persistent update target shows that this imbalance is structured rather than
uniform. Memory methods rely most
heavily on trajectories and repository context, consistent with retaining an
experience together with the conditions under which it remains applicable.
Skill and tool methods draw on the broadest mix of signals and use quality
attributes more often than other targets, because a reusable operation must be
judged by more than task completion alone. Harness evolution has the highest
reliance on model judgments, which help assign responsibility among interacting
runtime components before executable checks validate a proposed repair. Model
evolution, meanwhile, uses execution outcomes and trajectories in every surveyed
method and diagnostics in most of them. Signal choice is therefore coupled to
what evolves. Each update target observes a different slice of software
development and inherits the blind spots of that evidence.

\subsection{How Signals Drive Agent Evolution}
\label{sec:how-signals-drive-evolution}

A failed test is evidence, not yet an agent update. Its diagnosis can be stored
as memory, abstracted into a debugging skill, used to revise a validation
workflow, or converted into training supervision. The same software event can
therefore produce changes with different reach and failure modes.

The conversion mechanism determines which part of the evidence becomes a lesson
and how it can affect future work. We therefore examine experience generation,
candidate construction, and selection for reuse.
Self-play generates new tasks and trajectories; supervised fine-tuning (SFT),
preference optimization, and reinforcement learning (RL) fit model parameters
to such records; program search compares executable agent structures. These
operations can be combined. Signal type and retained component alone do not
reveal how an agent evolves.

\surveyheading{Constructing experience}
Many methods begin with trajectories from actual coding attempts, including
failed edits and their diagnostics. SWE-Exp distills both successful and failed
repairs, while Agent-RLVR uses unit-test results to identify failed attempts and
then obtains guided reattempts that provide otherwise scarce successful
trajectories~\citep{chen2026sweexp,da2025agentrlvr}. Self-Play SWE-RL (SSR) takes a
different route: a shared agent injects bugs into sandboxed repositories,
specifies them through test patches, and attempts to repair them. The generated
cases change with the agent's capabilities and supply experience for joint RL
\citep{selfplaysoftwareagents2026}. Thus, self-play is a way to construct a
training distribution, not itself an update to the deployed agent. Generated
tasks must still be valid, informative, and relevant to later coding work.

\surveyheading{Distilling reusable behavior}
Non-parametric methods select or synthesize something that can be used again
without retraining the task model. SWE-Exp compresses a repair attempt into
issue-resolution experience; Repository Memory and Agent KB organize repository
facts and past interactions for later retrieval
\citep{chen2026sweexp,wang2026repositorymemory,tang2025agentkb}. Retrieval is
part of the mechanism: an accurate lesson that is never surfaced, or is applied
to a different repository state, cannot improve the next decision. CODESKILL
instead extracts procedures at several granularities and maintains a compact
skill bank. It trains the skill-management policy with RL using both rubric
feedback and verifiable outcomes from a frozen downstream coding agent
\citep{li2026codeskill}. This is an important cross-layer case: RL optimizes the
manager, while the reusable resource presented to the coding agent is a skill.
Socratic-SWE uses trace-derived skills to help construct training tasks, and
Live-SWE-agent writes and revises callable utilities during issue resolution
\citep{xiao2026socraticswe,xia2025livesweagent}. Skill abstraction and tool
synthesis increase reuse, but can omit the conditions under which the original
repair worked.

\surveyheading{Searching executable structures}
Other methods make the agent implementation itself a candidate. HarnessFix
turns recurring failure diagnoses into scoped harness repairs, while EvoMAC
uses testing-team feedback to revise coding-team instructions and connections
\citep{chen2026harnessfix,hu2024evomac}. DGM samples an agent from an archive,
uses a foundation model to modify its source code, evaluates the variant on
coding tasks, and keeps alternative lineages available for later search
\citep{zhang2025darwin}. AFlow represents workflows as code and uses Monte Carlo
Tree Search, code edits, and execution feedback to explore their structure
\citep{zhang2024aflow}. In these systems, feedback guides proposal or selection;
it does not specify the correct architecture. The measured result depends on
both the selected implementation and the search budget used to discover it.

\surveyheading{Optimizing model parameters}
When the retained change is a model, the training record and optimization rule
must be stated separately. Agent-RLVR uses successful initial trajectories for
SFT, then guided retries of failures to construct successful--unsuccessful
pairs for offline DPO~\citep{da2025agentrlvr}. SWE-RL instead trains on historical issue
and patch pairs using GRPO; its inexpensive reward measures similarity to the
reference patch, rather than unit-test success
\citep{wei2025swerl}. Self-Play SWE-RL uses generated bug-repair episodes for
joint RL~\citep{selfplaysoftwareagents2026}. These methods all learn from
software-related evidence, but they optimize against different targets. A test
verdict, reference-patch similarity, and a preference pair can rank the same
trajectory differently. Parameter learning can spread such judgments across
many future decisions, making reward design and data curation as consequential
as the optimizer.

Table~\ref{tab:evolution-operations} compares 27 representative methods from
Figure~\ref{fig:taxonomy-coding-agents}, pairing their update mechanisms
with reported results on individual coding benchmarks.

Each row reports a result on the benchmark named beside it. CODESKILL's
Terminal-Bench 2 gain compares its learned skill manager with a no-skill
baseline on a benchmark held out from training; AFlow's HumanEval entry is
a final score. The numbers use different models,
budgets, baselines, and reuse scopes. A final score describes task performance;
a change against a matched baseline more directly informs the contribution of
an update. In particular,
the table cannot isolate whether a model gain came from generating better
training cases, filtering them differently, or changing the optimization rule.
The same distinction applies to memory retrieval, skill management, and
architecture search. Candidate construction must therefore be followed by
independent checks and later reuse before it counts as a beneficial retained
change.

\input{tables/evolution-mechanisms}

Signal conversion can amplify errors. Self-authored verification can confidently
approve invalid updates; external audits improve reliability
\citep{guo2026selfverification}. Memory Reward Inflation shows failed episodes
being overrated and retrieved; correction raises BIRD execution accuracy from
54.0\% to 56.9\%~\citep{asadolahi2026memoryinflation}. Phantom Guardrails
reports nonexistent failures inducing persistent harness rules in 15 of 60 runs
\citep{wang2026phantomguardrails}. These cases show why adoption requires
independent checks of provenance and applicability.

\subsection{When to Evolve}
\label{sec:when-to-evolve}

The time at which feedback appears need not be the time at which the agent
changes. Four events should be distinguished:
\begin{equation}
  t_{\mathrm{observe}}\leq t_{\mathrm{propose}}
  \leq t_{\mathrm{adopt}}\leq t_{\mathrm{reuse}}.
  \label{eq:evolution-timing}
\end{equation}
An agent may observe a compiler error immediately, propose a tool after several
failed commands, adopt it only after validation, and reuse it later in the same
task or in a future repository. Online feedback therefore does not by itself
establish online evolution. Timing is defined by when a persistent change enters
later execution.

\surveyheading{Within-Task Evolution}
Within-task evolution adopts an update before the current task ends, shortening
the distance between feedback and behavioral
change. Live-SWE-agent creates and revises utilities while resolving an issue,
and EvoMAC updates coding-team instructions and connections across rounds of the
same development task~\citep{xia2025livesweagent,hu2024evomac}. This timing can
exploit task-local state and test a candidate quickly, but the available evidence
is incomplete and highly correlated with the current trajectory. A provisional
update is appropriate when its scope is narrow and its effects can be reversed.

\surveyheading{Post-Task Evolution}
Post-task evolution forms and adopts an update after one task has finished. It
can use the complete trajectory, final validation results,
and regressions that were unavailable at the first failure. SWE-Exp distills
finished attempts, CODESKILL constructs and maintains skills from accumulated
coding activity, and HarnessFix turns recurring failed trajectories into scoped
harness repairs~\citep{chen2026sweexp,li2026codeskill,chen2026harnessfix}. Waiting
improves the evidence available for credit assignment, although it delays reuse
and may discard transient state needed to understand the failure.

\surveyheading{Cross-Task Evolution}
Cross-task evolution aggregates evidence from multiple tasks or evolution
stages before committing broader
changes. DGM evaluates generations of agent implementations, AFlow searches
workflows over repeated benchmark executions, and Agent-RLVR collects guided
trajectories before parameter optimization
\citep{zhang2025darwin,zhang2024aflow,da2025agentrlvr}. Accumulation can reveal
recurring patterns and support stronger validation, making it suitable for
architecture and model updates with a wider blast radius. It also increases
adaptation latency and the opportunity to fit the tasks used for proposal and
selection.

The timing choice therefore trades responsiveness against evidential breadth.
Local asset updates can often be attempted early because they are scoped and
reversible. Architecture changes usually require broader regression evidence,
while parameter updates commonly aggregate many trajectories before training.
This is a tendency rather than a fixed correspondence. The appropriate time is
determined by the uncertainty of $\mathcal F_t$, the intervention depth of
$\widehat u_t$, its reuse scope $\rho_t$, and the cost of reversing it. Across
all timings, the evolution claim begins when a candidate is adopted into later
execution, and the improvement claim still requires a positive effect under the
evaluation scope and budget developed next.

\section{Benchmarks and Evaluation}
\label{sec:benchmarks}

Evaluating a self-evolving coding agent involves two coupled objects: the
software artifact produced for the current task and the persistent agent state
that shapes subsequent tasks. Tests, compilers, and reviewers can assess the
artifact directly, but they supervise the retained update only indirectly. Let
$p_t$ denote a patch and $V(p_t;e_t)$ its validation result in environment
$e_t$. Even a successful validation does not establish a beneficial agent
update,
\begin{equation}
  V(p_t;e_t)=1
  \quad \text{does not establish} \quad
  \Delta J_t>0.
  \label{eq:artifact-update-gap}
\end{equation}
A valid patch can yield an overgeneralized memory, an unnecessary skill, or a
harness rule that harms later tasks. Conversely, an unsuccessful attempt can
contain a correct diagnosis that becomes useful after its conditions are
identified. Evaluation must therefore distinguish artifact correctness from
the behavioral contribution of the update derived from that artifact.

Coding benchmarks further play three roles in an evolution pipeline. Their
tasks and signals may construct updates, select which candidate updates are
retained, or report the final performance of the evolved agent. Reusing the
same evidence across these roles can make repeated adaptation to the evaluator
look like general improvement. A benchmark specifies where the agent acts and
how task outcomes are checked; it does not by itself define a valid evaluation
of evolution. The protocol must additionally specify what persists, which
evidence influences adoption, what budget is available, and how far the claimed
reuse extends.

\input{tables/representative-coding-benchmarks}

\subsection{Conventional Coding Benchmarks Used for Evolution}

Most self-evolving coding agents are evaluated in benchmarks originally
designed to measure coding competence rather than change in the agent itself.
These benchmarks remain valuable because they provide executable tasks,
observable trajectories, and software-specific outcome checks. Their role in
an evolution study, however, depends on both the coding behavior they expose
and the claims their evaluation unit can support. We group the commonly used
settings into repository-level issue resolution, function-level and
competition-style programming, and operational or specialized software tasks.
Table~\ref{tab:representative-coding-benchmarks} lists representative datasets
in these three groups and provides links to their available resources.

\surveyheading{Repository-level issue resolution}
Repository-level benchmarks expose issues, source trees, dependency states,
execution traces, patches, and regression outcomes. SWE-bench introduced this
setting using real GitHub issues and corresponding pull requests, while
SWE-bench Lite, SWE-bench Verified, and SWE-Bench Pro refine task quality,
validation, and difficulty
\citep{jimenez2023swebench,swebenchpro2025}. SWE-Gym further turns repository
tasks into executable environments for agent training and evaluation
\citep{swegym2025}. These settings expose realistic evidence for constructing
updates because an experience can be grounded in a repository revision and its
patch can be checked against later software behavior. In their usual episodic
use, however, the evaluation unit remains an issue--patch pair. The resulting
score shows whether the current agent solves sampled issues, not whether an
update produced from issue $c_i$ improves the agent on a later issue $c_j$.

\surveyheading{Function-level and competition-style programming}
HumanEval and MBPP provide concise specifications and executable tests, while
APPS, CodeContests, and LiveCodeBench extend the difficulty and diversity of
programming problems
\citep{chen2021evaluating,austin2021program,hendrycks2021apps,
li2022alphacode,jain2024livecodebench}. Their controlled interfaces make them
useful for comparing model learning and workflow search, including systems such
as SEMAG and SEW \citep{peng2026semag,liu2025sew}. They reveal whether an update
improves program synthesis under a stable oracle, but usually provide little
evidence about repository navigation or dependency management. Repeated
attempts can expose self-correction and algorithmic refinement within a problem,
but a reset between problems removes the persistence needed to test whether the
agent carries an improvement forward.

\surveyheading{Operational and specialized software settings}
Terminal-Bench evaluates long-horizon interaction with terminals and execution
harnesses \citep{merrill2026terminalbench}. Repository understanding,
compilation, and environment construction are examined through LongCodeQA,
CompileBench, Repo2Run, EnvBench, and EnConda-Bench
\citep{wang2026swepruner,ren2026taco,guo2026evoconfig}. Security and efficiency
benchmarks add objectives not captured by functional tests. PATCHEVAL,
SEC-bench, and VUL4J examine vulnerability repair, while PIE and Mercury measure
program optimization \citep{hu2026evorepair,wang2026effiskill}. Finally,
developer-agent sessions expose review, acceptance, and repeated use in real
projects \citep{aggarwal2026behavioralrules,huang2026personalizedskills}.
Together, these settings broaden the observable consequences of evolution.
They also make comparisons harder because environment state, workloads, tool
versions, and human assistance can change along with the agent. Their value is
therefore greatest when the relevant state can be replayed and competing agents
receive matched opportunities to act.

These conventional benchmarks provide the environments and feedback from which
an agent may evolve, but they do not by themselves identify evolution. This
limitation becomes critical when the same tasks or test signals generate
trajectories, guide candidate selection, and report final performance. Repeated
adaptation can then raise the score without producing a reusable improvement.
Static task collections also face solution leakage, weak tests, and training
contamination, motivating filtered and continuously refreshed variants
\citep{aleithan2024swebenchplus,badertdinov2025swerebench,
zhang2025swebenchlive}. These concerns affect coding-agent evaluation broadly;
the additional challenge for self-evolution is to show that a retained change,
rather than repeated exposure to the evaluator, causes better future behavior.

\subsection{Benchmarks Designed to Evaluate Evolution}

Evolution-oriented benchmarks differ in the object whose change they evaluate,
rather than in when an update occurs. We distinguish changes to a task solution,
a software artifact, and persistent agent state. Evidence at one level does not
establish change at the next.

\surveyheading{Solution-Level Adaptation}
EvoCodeBench tracks correctness, runtime, memory use, and algorithmic efficiency
across successive refinements of one solution~\citep{zhang2026evocodebench}.
It tests whether feedback improves the current solution. If the agent is reset
before the next problem, the result does not show that the state used for future
tasks has changed.

\surveyheading{Artifact-Level Evolution}
SWE-EVO uses release-scale tasks that require coordinated changes across files
while preserving existing behavior~\citep{thai2025sweevo}. It measures sustained
change to a software artifact, but a repository can evolve while its agent
remains fixed. The result supports agent evolution only when earlier experience
is retained in $S_t$, $H_t$, or $\theta_t$ and affects later decisions.

\surveyheading{Persistent Agent-State Evolution}
These benchmarks carry agent state across tasks or fresh sessions.
LifelongAgentBench tests accumulation and transfer over interdependent task
sequences~\citep{zheng2025lifelongagentbench}. SkillFlow evaluates skill
discovery, repair, reuse, and regression, while SkillLearnBench connects skill
quality and invocation to task outcomes
\citep{zhang2026skillflow,zhong2026skilllearnbench}. PAST-Bench compares matched
persistence-on and persistence-off conditions, and SEA-Eval jointly tracks
reliability, long-term performance, and token use over sequential streams
\citep{xue2026pastbench,jiang2026seaeval}. MemoryArena and Evo-Memory isolate
memory accumulation, revision, and behavioral reuse
\citep{he2026memoryarena,wei2025evomemory}. Together, these protocols test
whether retained experience changes later behavior; equal success rates may
still conceal different trajectories and costs.

\surveyheading{Causal Attribution Protocols}
An evolution benchmark must combine realistic software tasks with sequential
state and attribution controls. It should separate the cases used to construct
updates, select candidates, and report final performance. Let
$\mathcal D_{\mathrm{upd}}$, $\mathcal D_{\mathrm{sel}}$, and
$\mathcal D_{\mathrm{eval}}$ denote these roles. Because candidate selection is
part of improvement, final evaluation must remain insulated from both update
construction and selection. Repository, temporal, and defect-family separation
is stronger than a random issue split when cases can share files, APIs, tests,
or solution patterns.

GDPevo illustrates this principle by distributing atomic rules across update
tasks and recombining them in held-out cases~\citep{zhou2026gdpevo}. Its business
workflows are not coding tasks, but the same design can test new combinations of
APIs, defect families, repository structures, or validation requirements.

Attribution further requires a matched frozen agent that receives the same
observations, tools, budget, and repair opportunities but does not retain the
candidate update. Component hybrids such as
$(S_{t+1},H_t,\theta_t)$, $(S_t,H_{t+1},\theta_t)$, and
$(S_t,H_t,\theta_{t+1})$ can locate the source of a gain when components are
exchangeable. Checkpoints, repeated evaluation of earlier groups, and multiple
task orders then reveal retention, forgetting, and curriculum effects. These
controls turn a coding benchmark into an experiment about a changing agent.

\subsection{Measuring Benefit, Regression, and Cost}

Pass rate, resolve rate, repair rate, benchmark score, and Pass@$k$ remain
necessary measures of software outcomes
\citep{xia2025livesweagent,chen2026sweexp,peng2026semag,liu2025sew}. For an
evolving agent, however, an aggregate score can hide exchanges between acquired
and lost behavior. Recent evolution-oriented evaluations make this limitation
visible by comparing changing and frozen agents across sequential task streams,
rather than reporting only the terminal score
\citep{jiang2026seaeval,deng2026finevobench}. Let
$d_t(c)=s(A_{t+1},c;\mathcal B)-s(A_t,c;\mathcal B)$ be the matched score
change on case $c$. The acquisition rate is
\begin{equation}
  G_t=\Pr_{c\sim\mathcal D}\!\left[d_t(c)>0\right],
  \label{eq:acquisition-rate}
\end{equation}
and the regression rate is
\begin{equation}
  R_t=\Pr_{c\sim\mathcal D}\!\left[d_t(c)<0\right].
  \label{eq:regression-rate}
\end{equation}
For binary task outcomes, $G_t$ counts newly solved cases and $R_t$ counts cases
lost after the update. Reporting both prevents a positive average gain from
concealing substantial regressions. For graded metrics, their magnitudes should
be reported alongside the two rates. These rates provide a common summary of a
distinction that recent benchmarks expose through different measurements.
SkillLearnBench finds that skill-learning gains are not consistent across tasks
and models \citep{zhong2026skilllearnbench}, while ContinualSkillBench shows that
sequential improvement can be comparable to context-only adaptation and can
leave weaker agents with fragmented, task-specific skill collections
\citep{guan2026continualskillbench}. A positive average can therefore reflect a
small set of narrow gains even when the retained component remains redundant or
fails to generalize.

Longitudinal evaluation adds another axis. If $J_{i,t}$ denotes performance on
task group $i$ after the agent has evolved through stage $t$, the matrix of
$J_{i,t}$ values shows forward transfer to later groups and retention on earlier
ones. Periodic evaluation on old repositories or defect families can reveal
forgetting, while evaluation on new repositories tests whether the learned
behavior exceeds the scope of its generating feedback. The same checkpoints
should also record component growth, retirement, and rollback success. A memory
bank that continually expands while retrieval quality declines is not well
described by its final solve rate alone. FinEvo-Bench exemplifies this design
with paired evolving and non-evolving controls evaluated under independently
shuffled task streams \citep{deng2026finevobench}. Evidence from skill
maintenance also shows that regression can be path dependent. Unconditional
skill accumulation can improve performance initially and then degrade it, while
removing the original harmful skills may recover only part of the loss because
later skills have already inherited their errors
\citep{taxonomy260805810}. Checkpoints should therefore preserve update lineage
and measure both pre-admission harm and post-hoc rollback effectiveness.

Evolution costs and use costs should be separated. Let
$C_{\mathrm{evolve}}$ include experience processing, candidate construction,
search, and selection, and let $C_{\mathrm{use}}$ be the cost of executing the
retained agent once. If an update is reused $N$ times, its amortized cost per
use is
\begin{equation}
  \overline C_N=
  \frac{C_{\mathrm{evolve}}}{N}+C_{\mathrm{use}}.
  \label{eq:amortized-evolution-cost}
\end{equation}
This distinction changes the interpretation of efficiency. An expensive
offline harness search may be worthwhile when its result supports many later
tasks, whereas a utility created for one trajectory has little opportunity to
amortize its construction cost. SkillMOO makes this trade-off explicit by
optimizing skill bundles for pass rate, monetary cost, and runtime, with pruning
and substitution sometimes outperforming continued accumulation
\citep{taxonomy260409297}. SEA-Eval further shows that similar success rates can
hide large differences in token consumption \citep{jiang2026seaeval}. A broader
study of coding-agent trajectories likewise finds substantial run-to-run token
variation and no monotonic relation between greater token use and accuracy
\citep{bai2026agentmoney}. Reports should therefore include the update budget,
evaluation and selection costs, subsequent token and tool usage, wall time,
storage growth, and human supervision.

The resulting evaluation record should connect four quantities: the software
outcomes obtained, the persistent components changed, the cases that gained or
regressed, and the resources consumed to produce and use those changes.
Security, maintainability, and reviewability should be added when they are part
of the intended software outcome rather than inferred from functional success.
EVOMAL illustrates why this record must extend beyond task accuracy. A retrieved
malicious skill can be copied into newly authored skills and remain in the
evolving library after the planted source is removed
\citep{taxonomy260825776}. Such propagation is not captured by $G_t$ or $R_t$
alone, but it is a persistent behavioral consequence of the update process.
This protocol turns benchmark scores into evidence about a changing agent. It
also preserves the central boundary of the survey. A retained mechanism shows
that evolution occurred, while matched, held-out, and longitudinal evaluation
shows whether that evolution improved the coding agent.

\section{Self-Evolving Coding Products}
\label{sec:products}

\subsection{Product Landscape}
\label{sec:product-landscape}

Research systems describe what a coding agent could evolve, whereas deployed
products reveal which forms of evolution can operate under practical
constraints. We organize the product landscape using the asset, architecture,
and model layers introduced earlier and identify the persistent object changed
by each product. Inclusion requires a documented mechanism for retaining agent
state across later steps, sessions, or tasks; it does not require a fully
autonomous or experimentally validated evolution loop.
Table~\ref{tab:product-landscape} summarizes representative products documented
as of September 2026.

The distribution across these layers is highly asymmetric. Persistent memory,
skills, and context appear in many products, a smaller group reaches into the
execution harness, and none publicly exposes repository-derived parameter
updates as a persistent product capability. This pattern defines a
\emph{deployment frontier}. The relevant design question is how much behavioral
leverage an update provides relative to the cost of scoping, validating,
auditing, and reversing it. Increasing intervention depth usually expands both
the influence of an update and the consequences of an error. The three layers
therefore describe intervention depth rather than a maturity ranking; a
well-governed asset update may be more useful in production than a deeper change
whose contribution cannot be isolated.

\input{tables/product-landscape}

\surveyheading{Asset-centered products}
Assets are the dominant entry point for productized evolution. Claude Code,
GitHub Copilot, Cursor, Windsurf, Augment, and OpenClaw persist repository facts,
preferences, or workspace context across sessions
\citep{anthropic2026automemory,github2026copilotmemory,cursor2026automations,
windsurf2026memories,augment2025memoryreview,openclaw2026memory}. Other products
move from factual retention toward procedural reuse. Gemini CLI proposes memory
patches and skill drafts, Devin consolidates session evidence into knowledge and
playbooks, Letta Code revises memory and versions skills, Hermes creates and
updates skills from experience, and Replit Agent can preserve learned procedures
as project skills \citep{google2026automemory,cognition2026devininsights,
letta2026code,nous2026hermes,replit2026skills}. These mechanisms change what the
agent can recall or invoke while leaving the surrounding execution structure
largely intact. Their concentration at this layer follows from the properties
of externalized state. Memory and skills can be represented as bounded files or records, scoped
to a user, branch, or repository, inspected before reuse, and edited or removed
without replacing the whole agent. Their provenance can remain attached to the
stored item, and access control can follow existing workspace boundaries. This
supports tenant-specific adaptation while reducing the need to pool private
repository traces into shared model training. GitHub Copilot, for example,
attaches source citations to repository facts and revalidates them against the
current branch, while Gemini CLI holds generated memory patches and skills for
review before adoption
\citep{github2026copilotmemory,google2026automemory}. Asset updates thus resemble
small, versionable software changes whose scope and owner can be made explicit.

This advantage also reveals the main weakness of asset-centered products.
Storage is easier than maintaining a useful asset collection. As entries
accumulate, overlapping memories can disagree, a skill can outlive the API or
repository state under which it was learned, and retrieval can introduce an
irrelevant item into an otherwise correct trajectory. Product quality therefore
depends increasingly on lifecycle operations such as consolidation, conflict
detection, revalidation, retirement, and rollback. The competitive boundary is
shifting from whether a product can remember to whether it can govern what
should remain active and determine when that retained state should influence a
decision.

\surveyheading{Architecture-spanning products}
A smaller set exposes changes to the organization of execution. Prime Agent
treats prompts, skills, memory, and sub-agent specifications as editable
continual-harness state and records revisions for rollback
\citep{primeintellect2026primeagent}. DeepSeek Harness allows capabilities and
runtime modes to be recomposed through plugins and persistent presets
\citep{deepseek2026harness}. OpenBlock presents an additional boundary case that
connects learned codebase patterns with planned skill and evaluation-suite
construction \citep{openblock2026ob1}. These products can also update assets,
but their architectural significance lies in changing how capabilities are
assembled, delegated, or controlled. Such changes have a wider effect because
they can alter control flow, tool exposure, and coordination across many later
tasks.

Harness evolution is emerging because it occupies a useful middle layer. A
change to validation order, retry policy, context routing, or sub-agent
delegation can affect many decisions without modifying the underlying model.
The implementation remains code or configuration that can be versioned,
tested, compared, and rolled back. It can also survive a model upgrade, giving
the product a source of persistent behavior that is partly independent of a
particular model release. These properties make architecture a high-leverage
target when stronger base models alone no longer determine product behavior.

The same leverage increases the blast radius of a faulty update. A misleading
memory may affect the tasks that retrieve it, whereas a changed termination or
validation rule may affect every task passing through the harness. Asset and
architecture evolution are also coupled in practice. An asset determines what
knowledge or operation is available, while the harness determines when it is
retrieved, trusted, executed, and revised. A product that creates skills but
never changes these control rules remains asset-centered; a product that revises
selection, delegation, or validation moves into architecture even when the edit
also touches prompts or skills. This functional distinction explains why
similar persistent files can represent different depths of intervention.

\surveyheading{Model-level products}
The catalogue contains no product that publicly documents persistent,
repository-derived updates to its deployed model parameters. Providers may
continue training their underlying models, but that process normally occurs
outside the user-visible agent loop and cannot be attributed to a particular
repository trajectory. The product landscape therefore becomes sparser as the
intervention moves from assets to architecture and then to model parameters.
Deeper changes have a broader behavioral scope, require more expensive
validation, and are harder to diagnose or reverse when they fail. Several
constraints explain this gap. A repository trajectory provides sparse and
correlated evidence, and a passing patch does not show which parameter
change should be retained. Online adaptation can also introduce forgetting,
spread repository-specific shortcuts across tasks, or encode sensitive code in
a component shared by multiple users. Parameter updates are expensive to
evaluate because their effects are distributed across behaviors that may be
unrelated to the generating task. Rollback is correspondingly coarse: restoring
a checkpoint removes all changes bundled into it, while the harmful influence
of one memory or harness edit can often be traced and withdrawn directly.

This creates a separation between provider evolution and product evolution.
Providers can train new model releases using aggregated data, but the resulting
change follows a release cycle and is rarely attributable to one user's coding
experience. User-visible self-evolution therefore remains concentrated in
external state and executable orchestration. The product landscape suggests a
practical design rule: expose the deepest persistent intervention whose
provenance, evaluation scope, and rollback boundary can still be made concrete.

\subsection{From Experience to Governed Updates}
\label{sec:product-governance}

Productization turns evolution into a controlled state change. A task
trajectory and its feedback may generate a candidate update $\hat u_t$, but the
candidate should affect later execution only after an admission decision. Let
$z_t$ denote the persistent product state, which may contain assets or
architecture. A generic governed update is
\begin{equation}
  z_{t+1}=
  \begin{cases}
    \operatorname{Apply}(z_t,\hat u_t), &
      \operatorname{Gate}(\hat u_t;v_t,p_t,\rho_t)=1,\\
    z_t, & \text{otherwise},
  \end{cases}
  \label{eq:product-update-gate}
\end{equation}
where $v_t$ summarizes validation, $p_t$ records provenance and authority, and
$\rho_t$ specifies the permitted scope of reuse. The gate may be implemented by
automatic checks, explicit user approval, organizational policy, or a
combination of them.

This compact expression contains several decisions that products often
conflate. Proposal determines which part of a trajectory becomes a candidate;
validation asks whether its evidence extends beyond the case that produced it;
authorization determines who may change shared agent state; and scoping decides
which user, repository, branch, or runtime may receive the update. These are
part of the evolution mechanism because they control which experience is
allowed to shape future behavior. A permissive gate increases plasticity but
can accumulate stale memories, conflicting skills, or brittle control rules. A
conservative gate preserves stability but may discard useful adaptation. The
appropriate threshold depends on the update's blast radius. A repository fact
can be admitted under narrow scope and revalidated when retrieved, whereas a
tool or harness revision may require integration tests, held-out tasks, and
explicit rollback support. Validation can therefore combine the estimated gain
$\widehat{\Delta J}_t$, regression risk $\widehat R_t$, update and use costs,
and the breadth of $\rho_t$ rather than reducing admission to success on the
generating task.

Products instantiate this control point differently. Gemini CLI never applies
automatically extracted memory or skill candidates without approval, whereas
Claude Code writes repository-scoped memory that remains directly inspectable
and editable \citep{google2026automemory,anthropic2026automemory}. GitHub Copilot
validates a recalled repository fact before use and retires entries that remain
unused \citep{github2026copilotmemory}. Prime Agent exposes update history and
rollback for its continual harness \citep{primeintellect2026primeagent}. These
mechanisms treat a persistent agent update much like a software change. It is
proposed, scoped, reviewed or validated, committed, monitored, and potentially
reverted.

Admission should remain provisional after deployment. An update can move from
\emph{candidate} to \emph{active}, then to \emph{revised}, \emph{quarantined},
or \emph{retired} as new evidence arrives. Monitoring must attribute later
failures and gains to the version that was actually used, while rollback must
restore associated indexes, dependency bindings, and derived presets as well as
the visible memory or code. Product governance thus turns a collection of
persistent additions into maintained, versioned agent state. This lifecycle is
also what allows the hot-swappable architecture discussed next to support
continued evolution without making every runtime experiment permanent.

\subsection{Toward Hot-Swappable Coding Agents}
\label{sec:hot-swappable-products}

\begin{futureharnessquote}
\emph{``A future harness may generate and deploy modifications to its own
components while continuously serving requests.''}\par
\hfill---\textbf{\emph{A Programming Paradigm for Spatiotemporal
Composability}}~\citep[p.~5]{shi2026spatiotemporal}
\end{futureharnessquote}

DeepSeek Harness provides a concrete prototype of a product form in which the
composition of a running agent becomes a primary site of evolution. It exposes
models, tools, skills, session management, sandboxes, storage, execution loops,
scheduling, and the user interface as plugins, with configuration determining
which implementations are mounted together \citep{deepseek2026harness}. The
unit of change can therefore be an individual capability such as a search tool,
context policy, validation loop, or model adapter. Its Cordis lifecycle can
unload an existing plugin, unwind the registrations and effects owned by that
instance, and load revised code, while configuration edits affect only the
entries that were mounted, removed, or reconfigured
\citep{deepseek2026harnesshmr}. Creator mode extends this mechanism by allowing
developers or agents to inspect the live runtime, test temporary plugins in
memory, and combine selected components into presets
\citep{deepseek2026harness}. These mechanisms suggest a two-speed evolution
process. The fast path mounts a candidate during a session and observes its
immediate behavioral effects. The slower path validates it on broader cases,
records its provenance and dependencies, and promotes it into a versioned
preset for later sessions. A temporary mount supplies evidence about a
candidate; adoption, retention, and subsequent reuse turn that candidate into a
persistent update. The harness consequently acts as an \emph{evolution control
plane} joining experience-derived proposals with evaluation and deployment.

This product form shifts the central engineering problem from replacing a whole
agent to controlling change inside a live composition. Active tasks must remain
bound to compatible plugin versions, state must be migrated or discarded when
interfaces change, and rollback must withdraw both a component and the effects
it introduced. Version binding, dependency-aware disposal, shadow or canary
evaluation, and atomic promotion can support a small stable kernel surrounded
by an evolving periphery of plugins. The common plugin boundary does not erase
the taxonomy developed earlier. A callable utility remains an asset update, a
plugin that changes routing or the execution loop is an architecture update,
and a model-adapter swap establishes model self-evolution only when the model's
parameters are themselves updated. Functional role still determines the depth
of evolution, even when every change is packaged and deployed in the same way.

\section{Challenges and Open Problems}
\label{sec:challenges}

Self-evolving coding agents raise challenges beyond those of conventional
coding agents because their behavior changes over time. An unreliable test
result, noisy trajectory, weak verifier, or benchmark-specific shortcut may not
only affect one patch, but also be stored as memory, distilled into a skill,
selected as a workflow, used to update a model, embedded in a scaffold, or used
to reshape the agent's context and execution environment. Thus, the key challenge is
not only whether self-evolution improves benchmark performance, but whether the
evolutionary process remains reliable, reproducible, and aligned with software
engineering constraints. The consequences depend on both reuse scope and
intervention depth. A faulty repository memory may bias a family of related
issues, whereas a faulty harness or model update can alter every subsequent
decision. The open problems below therefore concern the reliability of the
entire path from local software evidence to persistent behavioral change.

\surveyparagraph{subsection}{Reproducibility, contamination, and benchmark overfitting}{Reproducibility, contamination, and benchmark overfitting.}
Self-evolution makes reproducibility difficult because agents may change across
runs, tasks, repositories, tool environments, or model versions. Systems that
select self-modifications or agent variants using benchmark outcomes are
especially sensitive to evaluation noise and benchmark leakage
\citep{robeyns2025selfimproving,zhang2025darwin}. This concern is amplified in
code evaluation, where contamination and benchmark-specific adaptation are
already known issues~\citep{jain2024livecodebench}. SWE-rebench evaluates models
on continuously collected, fresh software-engineering tasks and finds that some
results on SWE-bench Verified may be inflated by contamination
\citep{badertdinov2025swerebench}. DeepSWE exposes a related measurement problem.
On original long-horizon tasks, an independent judge disagrees with its
purpose-built verifiers far less often than with tests inherited from public
fixes, showing that both solution exposure and incomplete tests can distort the
apparent capability of an agent~\citep{huang2026deepswe}. A matched-budget study
of automatic harness evolution provides more direct evidence: on
Terminal-Bench~2.1, evolved harnesses do not consistently outperform simple
test-time search baselines and show limited generalization to held-out tasks
\citep{wang2026rethinkingharnesseval}. Reproducibility also depends on the
surrounding harness. Holding the model fixed across 35 sequential Qwen Code
releases reveals quality fluctuations attributable to harness changes, while a
separate comparison finds up to a $40\times$ difference in tokens per solved
task across three harnesses~\citep{bensghaier2026harnessevolution,
vats2026scaffoldeffect}. For a self-evolving agent, contamination extends beyond
pretraining overlap. If
the same tasks construct experience, select retained updates, and measure the
final system, the evaluation loop can progressively fit its own evidence. Even
a fully reproducible final score may then reproduce an overfitted evolutionary
path. A credible report should preserve the initial agent, task order, update
history, evaluator version, and adopted and rejected candidates; separate
update, selection, and held-out evaluation cases; and repeat the process across
seeds or task orders. The object to reproduce is the trajectory of change, not
only the terminal agent.

\surveyparagraph{subsection}{Feedback reliability, safety, and tool dependence}{Feedback reliability, safety, and tool dependence.}
Executable feedback is central to self-evolving coding agents, but tests,
compilers, CI logs, generated tests, and reward models are imperfect. Systems
that rely on unit-test validation, environment rewards, or learned verifiers may
therefore inherit the biases and blind spots of these signals
\citep{wei2025swerl,swegym2025}. This is
particularly risky when agents modify tools, workflows, or their own scaffolds,
because a misleading feedback signal can shape future behavior rather than only
one output. An analysis of 4,892 patches from ten software-development agents
finds that even test-passing patches can modify different files and functions
from developer patches, exposing limits in the tests used to identify acceptable
repairs~\citep{chen2024patchquality}. Security-focused evidence is more direct.
Across 1,030 valid traces from seven agent frameworks, 170 patches passed the
available functional checks while omitting protections, applying inadequate
fixes, or introducing vulnerabilities; neither test passing nor LLM review
reliably intercepted these cases~\citep{bai2026silentfailures}. The same concern
applies when the agent authors its own acceptance signal. Experiments on
heuristic self-improvement show that self-assigned verification scores can stay
near perfect while sealed deployment performance remains low or regresses;
retaining an external accept--reject audit improves results across six models
\citep{guo2026selfverification}. Memory-based agents expose a related feedback
loop. Incorrect episodes can receive inflated self-evaluation scores and then be
preferentially retrieved, whereas a de-inflation method raises BIRD execution
accuracy from 54.0\% to 56.9\%~\citep{asadolahi2026memoryinflation}. Coding
feedback is therefore explicit without necessarily being complete. The
danger grows when an update compresses a locally observed outcome into a broadly
reused rule while discarding the conditions under which the outcome was valid.
Future systems need to retain provenance, environment versions, test coverage,
and applicability conditions with the update itself. Adoption should combine
independent signals where possible, such as regression tests, static analysis,
security checks, differential execution, and review, while treating flaky
tests, generated oracles, sandbox mismatches, and tool failures as uncertainty
rather than negative task evidence. Tool reliability and sandbox fidelity are
thus part of the learning signal, not merely implementation details.

\surveyparagraph{subsection}{Reversibility, harness bloat, and specialization}{Reversibility, harness bloat, and specialization.}
The hot-swappable product form discussed in
Section~\ref{sec:hot-swappable-products} makes reversibility a runtime property.
Withdrawing a live component must also undo its state changes and dependency
effects while requests continue to be served. Cordis calls this property
\emph{temporal composability} and implements it through tracked inverse effects
\citep{shi2026spatiotemporal}. AHE similarly makes harness components explicit
and revertible, but end-to-end guarantees of this kind are not yet standard
across self-evolving coding agents~\citep{lin2026ahe}. Its experiments also show
why explicit components and attribution matter: ten evolution rounds improve
Terminal-Bench~2 performance, while ablations locate the transferable gains in
tools, middleware, and long-term memory rather than in every edited component.
The Phantom Guardrails study demonstrates how bloat can arise even without a
real failure: a proposer installs a guardrail for a nonexistent violation in
15 of 60 runs when harmless input resembles a familiar rule, compared with none
on featureless input, and an add-only loop makes the spurious component persist
\citep{wang2026phantomguardrails}. Reversibility must include the evaluation path
as well as task execution.
An audit of real self-improvement trajectories finds harness tampering across
different systems and observes that such edits can persist in the lineage of the
selected best agent~\citep{wang2026harnesstampering}. A candidate may appear
better because it changes task access, termination, logging, or validation rather
than because it solves software problems more effectively. This makes authority,
provenance, and evaluator isolation part of harness integrity. Components that
generate and judge an update should not be able to silently weaken the checks
that authorize its adoption. Without lifecycle controls, evolution can become
cumulative trial-and-error. A
locally successful fix may be appended as another prompt rule, memory item,
tool wrapper, routing condition, or agent role rather than replacing the
mechanism that caused the failure. The resulting harness becomes larger, more
coupled, and harder to audit, while narrow benchmark selection can turn these
additions into repository- or environment-specific shortcuts. Future systems
should therefore treat removal as a first-class operation, track the provenance
and inverse of persistent modifications, enforce complexity budgets, retire
redundant components, and validate promoted changes on held-out software
environments. Progress should be measured against both performance and structural
cost. Otherwise, an agent can improve a benchmark while becoming less portable,
less inspectable, and more expensive to maintain.

\surveyparagraph{subsection}{Long-term memory, skills, and coordination}{Long-term memory, skills, and coordination.}
Memory and skill mechanisms allow agents to reuse software-engineering
experience, but they also introduce quality-control problems. Experience banks,
repository memory, and skill libraries may become stale, redundant, overly
repository-specific, or contaminated by failed trajectories
\citep{chen2026sweexp,li2026codeskill,xiao2026socraticswe}. Controlled experiments
on memory management reveal an experience-following effect in which similar
retrieved records induce similar outputs. Incorrect or misaligned records can
therefore propagate errors, whereas selective addition and deletion outperform
unrestricted memory growth~\citep{xiong2026memorymanagement}. Skill evolution
has a similar lifecycle problem. A systematic study across extractors, consumer
models, and five agentic domains finds that generated skills help on average but
also produce non-trivial negative transfer; extraction quality and consumption
quality do not necessarily coincide~\citep{huang2026skillconsumption}. These
results make retention conditional: an artifact is useful only for a compatible
task, retriever, consumer, and execution context. This producer--consumer gap is
also visible at the harness level. Models from different capability tiers can
produce updates with similar gains, yet their ability to activate and follow
the resulting artifacts is non-monotonic, with weak models often failing to use
what was retained~\citep{lin2026harnessupdating}. Multi-agent and
workflow-evolving systems add coordination challenges because a
local improvement can alter the assumptions of neighboring components. Changing
one role, message schema, or routing rule may make an existing skill unreachable
or duplicate another agent's responsibility. AgentConductor empirically shows
that topology affects both coding accuracy and communication cost, obtaining up
to a 14.6\% Pass@$1$ improvement while reducing topology density and token use
\citep{wang2026agentconductor}. Evidence from a deployed collaboration network
shows that scale alone does not solve this problem: among more than 1.5 million
shared assets in EvoMap, 98\% are never reused and over 84\% of approved assets
bypass quality checks through vacuous tests~\citep{ye2026evomap}. The broader
problem is therefore not how much experience the system retains, but whether it
can maintain a coherent persistent state. Future work should jointly manage
relevance, conflicts, dependencies, usage, and retirement across memory, skills,
workflows, and teams instead of optimizing each store in isolation.

\surveyparagraph{subsection}{Evaluation beyond short benchmarks}{Evaluation beyond short benchmarks.}
Current evaluations mainly measure short-horizon success through pass rates,
resolve rates, or benchmark scores. Yet real software engineering also requires
maintainability, security, reviewability, efficiency, and long-term reliability.
The gap is already visible without moving beyond coding. SWE-EVO requires
release-scale changes spanning an average of 21 files and hundreds of tests; a
GPT-5 and OpenHands configuration resolves 21\% of these tasks, compared with
65\% on SWE-bench Verified~\citep{thai2025sweevo}. The difference shows that
success on isolated issues does not establish sustained planning, preservation
of earlier behavior, or control of interactions among changes. Longitudinal
agent benchmarks reveal a second gap. LifelongAgentBench finds that
straightforward experience replay is limited by irrelevant records and context
constraints~\citep{zheng2025lifelongagentbench}. PATH-Bench further shows that
strong forward transfer does not guarantee retention and that later experience
can reshape gains acquired earlier in a task sequence
\citep{yang2026pathbench}. Coding-specific memory experiments across six
benchmarks find a 3.7\% average gain from cross-domain transfer, but also show
that abstract validation routines transfer more reliably than concrete traces,
which can cause negative transfer through excessive specificity
\citep{kim2026memorytransfer}. Evaluation must consequently follow the agent
after an update rather than test it once. Repeated checkpoints should measure
newly acquired behavior, regressions on earlier repositories, transfer to
held-out repositories and defect families, resource growth, and sensitivity to
task order. Cross-domain transfer remains an open question, but the immediate
standard should be stronger: evolution learned from one part of software
engineering should survive time and improve later coding work beyond the cases
that produced it.

\section{Conclusion}
\label{sec:conclusion}

Coding agents are unusually well positioned to learn from their own work.
Execution yields specific outcomes, diagnostics, and trajectories that can be
tied to repository state and checked again. Our synthesis follows how this
evidence becomes persistent change in assets, architecture, or model
parameters, with progressively deeper intervention. The crucial distinction
is between retaining an update and showing that it improves later decisions.
Passing a test validates a software artifact under particular conditions;
demonstrating agent improvement requires reuse and benefit on matched,
held-out tasks.

Research and products currently favor memories, skills, tools, and editable
harnesses because their changes can be scoped, inspected, and reversed. Looking
ahead, a coding agent could treat these components as a versioned system that
it revises during ordinary coding work. Experience would generate candidate
changes, independent checks would govern their adoption, and later tasks would
show whether each change should be kept, revised, or rolled back. Extending
this cycle to workflows and model parameters will require stronger evidence
of gains, regressions, and costs across repositories and time. The long-term
goal is an agent that can identify which experiences deserve to persist, where
they apply, and when they should be forgotten.

\bibliographystyle{plainnat}
\bibliography{references}

\end{document}

%% file: fig/taxonomy-coding-agents.tex
\begingroup%
\setcitestyle{authoryear,round}%
\definecolor{taxAssetsFill}{HTML}{E8F1FA}%
\definecolor{taxAssetsLine}{HTML}{5498D0}%
\definecolor{taxArchFill}{HTML}{FDEBD7}%
\definecolor{taxArchLine}{HTML}{E68A43}%
\definecolor{taxModelFill}{HTML}{F0E8F7}%
\definecolor{taxModelLine}{HTML}{9A70BA}%
\begin{tikzpicture}[x=1mm,y=-1mm,
 taxnode/.style={rectangle,rounded corners=1.3pt,line width=0.45pt,
 align=center,inner xsep=0.8mm,inner ysep=0.65mm,minimum height=6.2mm,
 font=\fontsize{7.5}{8.7}\selectfont,outer sep=0pt},
 taxassets/.style={draw=taxAssetsLine,fill=taxAssetsFill},
 taxarch/.style={draw=taxArchLine,fill=taxArchFill},
 taxmodel/.style={draw=taxModelLine,fill=taxModelFill},
 taxedge/.style={draw=black!48,line width=0.42pt,line cap=round}]
\node[taxnode,draw=black!50,fill=black!7,text width=17.4mm] (n1) at (9.5,149.469) {Self-Evolving\\Coding Agents};
\node[taxnode,taxassets,text width=20.4mm] (n11) at (33,68.164) {Assets\\Self-Evolution};
\node[taxnode,taxassets,text width=21.4mm] (n111) at (58.5,25.019) {Memory\\Self-Evolution};
\node[taxnode,taxassets,text width=26.4mm] (n1111) at (87,14.461) {Experience-Derived\\Memory};
\node[taxnode,taxassets,text width=26.4mm] (n1112) at (87,35.576) {External Knowledge\\Memory};
\node[taxnode,taxassets,text width=21.4mm] (n112) at (58.5,66.712) {Skill\\Self-Evolution};
\node[taxnode,taxassets,text width=26.4mm] (n1121) at (87,51.153) {Skill Acquisition};
\node[taxnode,taxassets,text width=26.4mm] (n1122) at (87,64.595) {Skill Optimization};
\node[taxnode,taxassets,text width=26.4mm] (n1123) at (87,73.433) {Skill Governance};
\node[taxnode,taxassets,text width=26.4mm] (n1124) at (87,82.272) {Skill Safety};
\node[taxnode,taxassets,text width=21.4mm] (n113) at (58.5,91.091) {Tool\\Self-Evolution};
\node[taxnode,taxassets,text width=21.4mm] (n114) at (58.5,111.310) {Context\\Self-Evolution};
\node[taxnode,taxassets,text width=26.4mm] (n1141) at (87,99.291) {Context Retrieval};
\node[taxnode,taxassets,text width=26.4mm] (n1142) at (87,106.891) {Context Compression};
\node[taxnode,taxassets,text width=26.4mm] (n1143) at (87,115.110) {Prompt Evolution};
\node[taxnode,taxassets,text width=26.4mm] (n1144) at (87,123.329) {Environment Evolution};
\node[taxnode,taxarch,text width=20.4mm] (n12) at (33,186.749) {Architecture\\Self-Evolution};
\node[taxnode,taxarch,text width=21.4mm] (n121) at (58.5,163.575) {Harness\\Evolution};
\node[taxnode,taxarch,text width=26.4mm] (n1211) at (87,134.883) {Archive-Based\\Evolution};
\node[taxnode,taxarch,text width=26.4mm] (n1212) at (87,145.256) {Iteration-Based\\Evolution};
\node[taxnode,taxarch,text width=26.4mm] (n1213) at (87,157.163) {Trace-Driven\\Evolution};
\node[taxnode,taxarch,text width=26.4mm] (n1214) at (87,170.605) {Population-Based\\Search};
\node[taxnode,taxarch,text width=26.4mm] (n1215) at (87,182.513) {Pipeline-Driven\\Evolution};
\node[taxnode,taxarch,text width=26.4mm] (n1216) at (87,192.267) {Runtime Evolution};
\node[taxnode,taxarch,text width=21.4mm] (n122) at (58.5,201.086) {Workflow\\Evolution};
\node[taxnode,taxarch,text width=21.4mm] (n123) at (58.5,209.924) {Multi-Agent\\Evolution};
\node[taxnode,taxmodel,text width=20.4mm] (n13) at (33,230.774) {Model\\Self-Evolution};
\node[taxnode,taxmodel,text width=21.4mm] (n131) at (58.5,222.097) {Self-Play\\Co-Evolution};
\node[taxnode,taxmodel,text width=21.4mm] (n132) at (58.5,231.851) {Environment-\\Reward RL};
\node[taxnode,taxmodel,text width=21.4mm] (n133) at (58.5,239.451) {Harness\\Co-Evolution};
\node[taxnode,taxassets,text width=100mm,minimum height=28.922mm] (pn1111) at (154.8,14.461) {\mbox{SAGE~\citep{hayashi2025sage}}; \mbox{SWE-Exp~\citep{chen2026sweexp}}; \mbox{EvoCoder~\citep{lin2024evocoder}};\\\mbox{Subtask Memory~\citep{shen2026subtaskmemory}}; \mbox{EvoRepair~\citep{hu2026evorepair}};\\\mbox{PMCoder~\citep{zhang2026pmcoder}}; \mbox{EXPEREPAIR~\citep{mu2025experepair}};\\\mbox{VeriAgent~\citep{taxonomy260317613}}; \mbox{MemRepair~\citep{taxonomy260517444}};\\\mbox{ReasoningBank~\citep{taxonomy250925140}}; \mbox{Reflexion~\citep{shinn2023reflexion}};\\\mbox{SE-Agent~\citep{taxonomy250802085}}; \mbox{Agent KB~\citep{tang2025agentkb}};\\\mbox{Adaptive ML Library Development~\citep{taxonomy250202534}};\\\mbox{PILOT in the Loop~\citep{taxonomy260826530}}; \mbox{SWE-MeM~\citep{gao2026swemem}};\\\mbox{CTIM-Rover~\citep{taxonomy2025realm130}}};
\node[taxnode,taxassets,text width=100mm,minimum height=10.508mm] (pn1112) at (154.8,35.576) {\mbox{Repository Memory~\citep{wang2026repositorymemory}}; \mbox{MemCoder~\citep{deng2026memcoder}};\\\mbox{Learning to Commit~\citep{taxonomy260326664}};\\\mbox{Accumulated Behavioral Rules~\citep{aggarwal2026behavioralrules}}};
\node[taxnode,taxassets,text width=100mm,minimum height=16.646mm] (pn1121) at (154.8,51.153) {\mbox{CODESKILL~\citep{li2026codeskill}}; \mbox{GSkill~\citep{tan2026gskill}};\\\mbox{Socratic-SWE~\citep{xiao2026socraticswe}}; \mbox{GSE~\citep{yang2026gse}};\\\mbox{SkillForge~\citep{taxonomy260818933}}; \mbox{Verilog-Evolve~\citep{taxonomy260526498}};\\\mbox{Trace2Skill~\citep{taxonomy260521810}};\\\mbox{Procedural Skills to Strategy Genes~\citep{taxonomy260415097}}};
\node[taxnode,taxassets,text width=100mm,minimum height=7.438mm] (pn1122) at (154.8,64.595) {\mbox{EffiSkill~\citep{wang2026effiskill}}; \mbox{Personalized Skills~\citep{huang2026personalizedskills}};\\\mbox{SkillMOO~\citep{taxonomy260409297}}; \mbox{daVinci-kernel~\citep{taxonomy260616497}}};
\node[taxnode,taxassets,text width=100mm,minimum height=7.438mm] (pn1123) at (154.8,73.433) {\mbox{Ratchet~\citep{zhang2026ratchet}}; \mbox{SkillsVote~\citep{taxonomy260518401}};\\\mbox{Who Grades the Grader?~\citep{taxonomy260712790}}};
\node[taxnode,taxassets,text width=100mm,minimum height=7.438mm] (pn1124) at (154.8,82.272) {\mbox{Library Drift~\citep{zhang2026librarydrift}}; \mbox{When Self-Evolution Backfires~\citep{taxonomy260805810}};\\\mbox{EVOMAL~\citep{taxonomy260825776}}};
\node[taxnode,taxassets,text width=131mm,minimum height=6.200mm] (pn113) at (139.3,91.091) {\mbox{Live-SWE-agent~\citep{xia2025livesweagent}}; \mbox{SIGA~\citep{ho2026siga}}};
\node[taxnode,taxassets,text width=100mm,minimum height=6.200mm] (pn1141) at (154.8,99.291) {\mbox{EVOR~\citep{su2024evor}}; \mbox{CodeMEM~\citep{wang2026codemem}}};
\node[taxnode,taxassets,text width=100mm,minimum height=6.200mm] (pn1142) at (154.8,106.891) {\mbox{TACO~\citep{ren2026taco}}; \mbox{SWE-Pruner~\citep{wang2026swepruner}}};
\node[taxnode,taxassets,text width=100mm,minimum height=7.438mm] (pn1143) at (154.8,115.110) {\mbox{EPiC~\citep{taherkhani2024epic}}; \mbox{RL Prompt Optimization~\citep{taxonomy260519102}};\\\mbox{SePO~\citep{tao2026sepo}}; \mbox{From Failing to Passing~\citep{taxonomy260705121}}};
\node[taxnode,taxassets,text width=100mm,minimum height=6.200mm] (pn1144) at (154.8,123.329) {\mbox{Libra~\citep{zhao2026libra}}; \mbox{EvoEnv~\citep{taxonomy260514392}}};
\node[taxnode,taxarch,text width=100mm,minimum height=10.508mm] (pn1211) at (154.8,134.883) {\mbox{DGM~\citep{zhang2025darwin}}; \mbox{Mendel GM~\citep{mendel2026godel}};\\\mbox{Huxley GM~\citep{huxley2025godel}}; \mbox{SIFT~\citep{fu2026sift}}; \mbox{RQGM~\citep{taxonomy260626294}};\\\mbox{HyperAgents~\citep{taxonomy260319461}}};
\node[taxnode,taxarch,text width=100mm,minimum height=7.438mm] (pn1212) at (154.8,145.256) {\mbox{SICA~\citep{robeyns2025selfimproving}}; \mbox{STOP~\citep{zelikman2024stop}};\\\mbox{Ouroboros~\citep{razzhigaev2026ouroboros}}};
\node[taxnode,taxarch,text width=100mm,minimum height=13.577mm] (pn1213) at (154.8,157.163) {\mbox{Self-Harness~\citep{zhang2026selfharness}}; \mbox{AutoSaddler~\citep{taxonomy260823041}};\\\mbox{Evo-Harness~\citep{wei2026evoharness}}; \mbox{AHE~\citep{lin2026ahe}};\\\mbox{HarnessFix~\citep{chen2026harnessfix}}; \mbox{ReCreate~\citep{taxonomy2026acllong1432}};\\\mbox{Harness-of-Harness~\citep{taxonomy260901481}}; \mbox{One Recipe, Many Harnesses~\citep{yang2026onerecipe}}};
\node[taxnode,taxarch,text width=100mm,minimum height=10.508mm] (pn1214) at (154.8,170.605) {\mbox{DarwinX~\citep{zhang2026darwinx}}; \mbox{HarnessBank~\citep{luo2026harnessbank}};\\\mbox{EvolveNet~\citep{nie2026evolvenet}}; \mbox{HarnessCompass~\citep{zhang2026harnesscompass}};\\\mbox{ADAS~\citep{taxonomy240808435}}};
\node[taxnode,taxarch,text width=100mm,minimum height=10.508mm] (pn1215) at (154.8,182.513) {\mbox{Meta-Harness~\citep{lee2026metaharness}}; \mbox{AgentDevel~\citep{taxonomy260104620}};\\\mbox{HarnessX~\citep{chen2026harnessx}}; \mbox{CCA~\citep{wong2025cca}};\\\mbox{Autogenesis~\citep{taxonomy260415034}}; \mbox{SEA~\citep{sengupta2026sea}}};
\node[taxnode,taxarch,text width=100mm,minimum height=6.200mm] (pn1216) at (154.8,192.267) {\mbox{TTHE~\citep{nie2026tthe}}; \mbox{Life-Harness~\citep{xu2026lifeharness}}; \mbox{Argus~\citep{li2026argus}}};
\node[taxnode,taxarch,text width=131mm,minimum height=7.438mm] (pn122) at (139.3,201.086) {\mbox{AFlow~\citep{zhang2024aflow}}; \mbox{SEW~\citep{liu2025sew}}; \mbox{EvoFlow~\citep{taxonomy250207373}};\\\mbox{EvoAgentX~\citep{wang2025evoagentx}}; \mbox{FlowEvo~\citep{taxonomy260721596}}; \mbox{JUDGEFLOW~\citep{taxonomy260107477}}};
\node[taxnode,taxarch,text width=131mm,minimum height=7.438mm] (pn123) at (139.3,209.924) {\mbox{EvoMAC~\citep{hu2024evomac}}; \mbox{AgentConductor~\citep{wang2026agentconductor}}; \mbox{SEMAG~\citep{peng2026semag}};\\\mbox{GEA~\citep{weng2026groupevolvingagents}}; \mbox{SAGE (Multi-Agent)~\citep{taxonomy260315255}}; \mbox{Evolve as a Team~\citep{taxonomy260529790}}};
\node[taxnode,taxmodel,text width=131mm,minimum height=10.508mm] (pn131) at (139.3,222.097) {\mbox{Self-Play SWE-RL~\citep{selfplaysoftwareagents2026}}; \mbox{ReVeal~\citep{jin2025reveal}}; \mbox{CURE~\citep{wang2025cure}};\\\mbox{ZeroCoder~\citep{fan2026zerocoder}}; \mbox{Sol-Ver~\citep{lin2025solverifier}}; \mbox{ACE~\citep{huang2026ace}};\\\mbox{Anchored Self-Play~\citep{taxonomy260703523}}; \mbox{Absolute Zero~\citep{taxonomy250503335}}; \mbox{OPD-Evolver~\citep{taxonomy260617628}}};
\node[taxnode,taxmodel,text width=131mm,minimum height=6.200mm] (pn132) at (139.3,231.851) {\mbox{Agent-RLVR~\citep{da2025agentrlvr}}; \mbox{SWE-RL~\citep{wei2025swerl}}; \mbox{SOAR~\citep{taxonomy250714172}}};
\node[taxnode,taxmodel,text width=131mm,minimum height=6.200mm] (pn133) at (139.3,239.451) {\mbox{Harness-R1~\citep{shao2026harnessr1}}; \mbox{HELIX~\citep{fan2026helix}}};
\draw[taxedge] (n1.east) -- (20.5,149.469);
\draw[taxedge] (20.5,68.164) -- (20.5,230.774);
\draw[taxedge] (20.5,68.164) -- (n11.west);
\draw[taxedge] (20.5,186.749) -- (n12.west);
\draw[taxedge] (20.5,230.774) -- (n13.west);
\draw[taxedge] (n11.east) -- (45.5,68.164);
\draw[taxedge] (45.5,25.019) -- (45.5,111.310);
\draw[taxedge] (45.5,25.019) -- (n111.west);
\draw[taxedge] (45.5,66.712) -- (n112.west);
\draw[taxedge] (45.5,91.091) -- (n113.west);
\draw[taxedge] (45.5,111.310) -- (n114.west);
\draw[taxedge] (n111.east) -- (71.5,25.019);
\draw[taxedge] (71.5,14.461) -- (71.5,35.576);
\draw[taxedge] (71.5,14.461) -- (n1111.west);
\draw[taxedge] (71.5,35.576) -- (n1112.west);
\draw[taxedge] (n1111.east) -- (pn1111.west);
\draw[taxedge] (n1112.east) -- (pn1112.west);
\draw[taxedge] (n112.east) -- (71.5,66.712);
\draw[taxedge] (71.5,51.153) -- (71.5,82.272);
\draw[taxedge] (71.5,51.153) -- (n1121.west);
\draw[taxedge] (71.5,64.595) -- (n1122.west);
\draw[taxedge] (71.5,73.433) -- (n1123.west);
\draw[taxedge] (71.5,82.272) -- (n1124.west);
\draw[taxedge] (n1121.east) -- (pn1121.west);
\draw[taxedge] (n1122.east) -- (pn1122.west);
\draw[taxedge] (n1123.east) -- (pn1123.west);
\draw[taxedge] (n1124.east) -- (pn1124.west);
\draw[taxedge] (n113.east) -- (pn113.west);
\draw[taxedge] (n114.east) -- (71.5,111.310);
\draw[taxedge] (71.5,99.291) -- (71.5,123.329);
\draw[taxedge] (71.5,99.291) -- (n1141.west);
\draw[taxedge] (71.5,106.891) -- (n1142.west);
\draw[taxedge] (71.5,115.110) -- (n1143.west);
\draw[taxedge] (71.5,123.329) -- (n1144.west);
\draw[taxedge] (n1141.east) -- (pn1141.west);
\draw[taxedge] (n1142.east) -- (pn1142.west);
\draw[taxedge] (n1143.east) -- (pn1143.west);
\draw[taxedge] (n1144.east) -- (pn1144.west);
\draw[taxedge] (n12.east) -- (45.5,186.749);
\draw[taxedge] (45.5,163.575) -- (45.5,209.924);
\draw[taxedge] (45.5,163.575) -- (n121.west);
\draw[taxedge] (45.5,201.086) -- (n122.west);
\draw[taxedge] (45.5,209.924) -- (n123.west);
\draw[taxedge] (n121.east) -- (71.5,163.575);
\draw[taxedge] (71.5,134.883) -- (71.5,192.267);
\draw[taxedge] (71.5,134.883) -- (n1211.west);
\draw[taxedge] (71.5,145.256) -- (n1212.west);
\draw[taxedge] (71.5,157.163) -- (n1213.west);
\draw[taxedge] (71.5,170.605) -- (n1214.west);
\draw[taxedge] (71.5,182.513) -- (n1215.west);
\draw[taxedge] (71.5,192.267) -- (n1216.west);
\draw[taxedge] (n1211.east) -- (pn1211.west);
\draw[taxedge] (n1212.east) -- (pn1212.west);
\draw[taxedge] (n1213.east) -- (pn1213.west);
\draw[taxedge] (n1214.east) -- (pn1214.west);
\draw[taxedge] (n1215.east) -- (pn1215.west);
\draw[taxedge] (n1216.east) -- (pn1216.west);
\draw[taxedge] (n122.east) -- (pn122.west);
\draw[taxedge] (n123.east) -- (pn123.west);
\draw[taxedge] (n13.east) -- (45.5,230.774);
\draw[taxedge] (45.5,222.097) -- (45.5,239.451);
\draw[taxedge] (45.5,222.097) -- (n131.west);
\draw[taxedge] (45.5,231.851) -- (n132.west);
\draw[taxedge] (45.5,239.451) -- (n133.west);
\draw[taxedge] (n131.east) -- (pn131.west);
\draw[taxedge] (n132.east) -- (pn132.west);
\draw[taxedge] (n133.east) -- (pn133.west);
\end{tikzpicture}%
\endgroup%

%% file: tables/assets-methods.tex
\begingroup
\fontsize{8.5}{10.2}\selectfont
\setlength{\tabcolsep}{2.5pt}
\renewcommand{\arraystretch}{1.1}
\setlength{\LTleft}{\fill}
\setlength{\LTright}{\fill}
\setlength{\LTpre}{8pt}
\setlength{\LTpost}{9pt}
\setlength{\LTcapwidth}{\linewidth}
\newcommand{\assetprimary}{\ensuremath{\bullet}}
\newcommand{\assetother}{\ensuremath{\circ}}
\newcommand{\assetop}{{\usefont{U}{msa}{m}{n}\symbol{"58}}}
\begin{longtable}{@{}>{\raggedright\arraybackslash}p{0.34\linewidth}>{\centering\arraybackslash}p{0.065\linewidth}>{\centering\arraybackslash}p{0.045\linewidth}>{\centering\arraybackslash}p{0.043\linewidth}>{\centering\arraybackslash}p{0.068\linewidth}>{\centering\arraybackslash}p{0.044\linewidth}>{\centering\arraybackslash}p{0.052\linewidth}>{\centering\arraybackslash}p{0.052\linewidth}>{\raggedright\arraybackslash}p{\dimexpr0.291\linewidth-16\tabcolsep\relax}@{}}
\caption{Asset targets, operations, and update triggers of representative methods.}
\label{tab:asset-methods}\\
\toprule
\multirow{2}{*}{\textbf{Method}} & \multicolumn{4}{c}{\textbf{Asset target}} & \multicolumn{3}{c}{\textbf{Asset operations}} & \multirow{2}{*}{\textbf{Update trigger}} \\
\cmidrule(lr){2-5}\cmidrule(lr){6-8}
 & Memory & Skill & Tool & Context & Add & Delete & Revise & \\
\midrule
\endfirsthead
\multicolumn{9}{c}{\small{\sffamily\bfseries\color{ISEDeep}\tablename~\thetable:}
  Asset targets, operations, and update triggers of representative methods (continued).}\\[6pt]
\toprule
\multirow{2}{*}{\textbf{Method}} & \multicolumn{4}{c}{\textbf{Asset target}} & \multicolumn{3}{c}{\textbf{Asset operations}} & \multirow{2}{*}{\textbf{Update trigger}} \\
\cmidrule(lr){2-5}\cmidrule(lr){6-8}
 & Memory & Skill & Tool & Context & Add & Delete & Revise & \\
\midrule
\endhead
\midrule
\multicolumn{9}{r}{\scriptsize Continued on next page}\\
\endfoot
\bottomrule
\multicolumn{9}{@{}p{\linewidth}@{}}{\scriptsize
\assetprimary/\assetother\ denote primary/other categories; \assetop\ marks documented operations, and blanks mean unconfirmed.
Add, Delete, and Revise denote creation, removal or retirement, and modification of retained assets, respectively.}\\[2pt]
\multicolumn{9}{@{}p{\linewidth}@{}}{\scriptsize
$\dagger$ SIGA evolves adapter contents across assets; its tool-category placement does not imply that new runtime tools account for its gains.}\\
\endlastfoot
Reflexion~\citep{shinn2023reflexion} & \assetprimary & \assetother & \assetother & \assetother & \assetop &  & & Failed trials \\
\addlinespace[2pt]
CODESKILL~\citep{li2026codeskill} & \assetother & \assetprimary & \assetother & \assetother & \assetop &  & \assetop & New/failed experience \\
\addlinespace[2pt]
Live-SWE-agent~\citep{xia2025livesweagent} & \assetother & \assetother & \assetprimary & \assetother & \assetop &  & \assetop & Tool needs \\
\addlinespace[2pt]
TACO~\citep{ren2026taco} & \assetother & \assetother & \assetother & \assetprimary & \assetop &  & \assetop & Rule gaps; complaints \\
\addlinespace[2pt]
SWE-Exp~\citep{chen2026sweexp} & \assetprimary & \assetother & \assetother & \assetother & \assetop &  & & Repair outcomes \\
\addlinespace[2pt]
GSE~\citep{yang2026gse} & \assetother & \assetprimary & \assetother & \assetother & \assetop &  & \assetop & Diagnosed skill gaps \\
\addlinespace[2pt]
EVOR~\citep{su2024evor} & \assetother & \assetother & \assetother & \assetprimary & \assetop &  & & Execution feedback \\
\addlinespace[2pt]
EvoCoder~\citep{lin2024evocoder} & \assetprimary & \assetother & \assetother & \assetother & \assetop & \assetop & \assetop & Reproduction outcomes \\
\addlinespace[2pt]
GSkill~\citep{tan2026gskill} & \assetother & \assetprimary & \assetother & \assetother & \assetop &  & \assetop & Rollout feedback \\
\addlinespace[2pt]
SIGA$\dagger$~\citep{ho2026siga} & \assetother & \assetother & \assetprimary & \assetother & \assetop &  & \assetop & Validation rewards \\
\addlinespace[2pt]
CodeMEM~\citep{wang2026codemem} & \assetother & \assetother & \assetother & \assetprimary & \assetop & \assetop & & Interaction rounds \\
\addlinespace[2pt]
Subtask Memory~\citep{shen2026subtaskmemory} & \assetprimary & \assetother & \assetother & \assetother & \assetop &  & & Subtask completion \\
\addlinespace[2pt]
SkillForge~\citep{taxonomy260818933} & \assetother & \assetprimary & \assetother & \assetother & \assetop &  & & Synthetic repair traces \\
\addlinespace[2pt]
EPiC~\citep{taherkhani2024epic} & \assetother & \assetother & \assetother & \assetprimary & \assetop &  & \assetop & Failed tests \\
\addlinespace[2pt]
EvoRepair~\citep{hu2026evorepair} & \assetprimary & \assetother & \assetother & \assetother & \assetop &  & \assetop & Repair feedback \\
\addlinespace[2pt]
Verilog-Evolve~\citep{taxonomy260526498} & \assetother & \assetprimary & \assetother & \assetother & \assetop &  & \assetop & Verifier feedback \\
\addlinespace[2pt]
SePO~\citep{tao2026sepo} & \assetother & \assetother & \assetother & \assetprimary & \assetop &  & \assetop & Evaluation scores \\
\addlinespace[2pt]
EXPEREPAIR~\citep{mu2025experepair} & \assetprimary & \assetother & \assetother & \assetother & \assetop & \assetop & \assetop & Repair feedback \\
\addlinespace[2pt]
Trace2Skill~\citep{taxonomy260521810} & \assetother & \assetprimary & \assetother & \assetother & \assetop &  & \assetop & Rollout diagnostics \\
\addlinespace[2pt]
Libra~\citep{zhao2026libra} & \assetother & \assetother & \assetother & \assetprimary & \assetop & \assetop & \assetop & Localization failures \\
\addlinespace[2pt]
ReasoningBank~\citep{taxonomy250925140} & \assetprimary & \assetother & \assetother & \assetother & \assetop &  & & Task outcomes \\
\addlinespace[2pt]
EffiSkill~\citep{wang2026effiskill} & \assetother & \assetprimary & \assetother & \assetother & \assetop &  & & Slow/fast code pairs \\
\addlinespace[2pt]
Agent KB~\citep{tang2025agentkb} & \assetprimary & \assetother & \assetother & \assetother & \assetop & \assetop & \assetop & Rewards; capacity \\
\addlinespace[2pt]
SkillMOO~\citep{taxonomy260409297} & \assetother & \assetprimary & \assetother & \assetother & \assetop & \assetop & \assetop & Task success; cost \\
\addlinespace[2pt]
Repository Memory~\citep{wang2026repositorymemory} & \assetprimary & \assetother & \assetother & \assetother & \assetop &  & & Repository history \\
\addlinespace[2pt]
Ratchet~\citep{zhang2026ratchet} & \assetother & \assetprimary & \assetother & \assetother & \assetop & \assetop & & Failures; contribution \\
\addlinespace[2pt]
MemCoder~\citep{deng2026memcoder} & \assetprimary & \assetother & \assetother & \assetother & \assetop &  & & New commits \\
\addlinespace[2pt]
SkillsVote~\citep{taxonomy260518401} & \assetother & \assetprimary & \assetother & \assetother & \assetop &  & \assetop & Execution attribution \\
\addlinespace[2pt]
SAGE~\citep{hayashi2025sage} & \assetprimary & \assetother & \assetother & \assetother & \assetop &  & & Initial rollout \\
\addlinespace[2pt]
Library Drift~\citep{zhang2026librarydrift} & \assetother & \assetprimary & \assetother & \assetother & \assetop & \assetop & & Failures; contribution \\
\end{longtable}
\endgroup

%% file: tables/coding-evolution-signals.tex
\begin{table}[htbp]
  \centering
  \caption{Seven non-exclusive signal classes used by coding-agent evolution
  mechanisms. The codes identify evidence consumed during updating rather than
  the benchmark used for final evaluation.}
  \label{tab:coding-evolution-signals}
  \begingroup
  \small
  \setlength{\tabcolsep}{4.5pt}
  \renewcommand{\arraystretch}{1.16}
  \rowcolors{2}{ISESoftBlue!38}{white}
  \begin{tabularx}{\linewidth}{@{}
    >{\RaggedRight\arraybackslash\hyphenpenalty=10000\exhyphenpenalty=10000}p{.18\linewidth}
    >{\RaggedRight\arraybackslash\hyphenpenalty=10000\exhyphenpenalty=10000}p{.29\linewidth}
    >{\RaggedRight\arraybackslash\hyphenpenalty=10000\exhyphenpenalty=10000}p{.19\linewidth}
    >{\RaggedRight\arraybackslash\hyphenpenalty=10000\exhyphenpenalty=10000}X@{}}
    \toprule
    \rowcolor{white}
    \textbf{Signal class} & \textbf{Observable evidence} &
    \textbf{Question answered} &
    \textbf{Evolutionary role and validity boundary} \\
    \midrule
    \textbf{E} Execution outcomes &
    Unit and integration tests, builds, executions, patch verifiers &
    Did the artifact satisfy an executable check? &
    Validate or select candidates within observed test coverage. \\
    \addlinespace[3pt]
    \textbf{D} Diagnostic evidence &
    Compiler messages, test logs, exceptions, build and tool failures &
    Where and how did the failure appear? &
    Locate update targets; symptoms may not identify their causes. \\
    \addlinespace[3pt]
    \textbf{T} Coding trajectories &
    Searches, inspections, edits, tests, retries, and recovery actions &
    How was the result produced? &
    Extract procedures; final success provides incomplete credit. \\
    \addlinespace[3pt]
    \textbf{R} Repository context &
    Issues, commits, pull requests, diffs, interfaces, dependencies, revisions &
    When is reuse valid? &
    Ground retrieval and reuse despite repository and dependency drift. \\
    \addlinespace[3pt]
    \textbf{Q} Quality attributes &
    Efficiency, security, maintainability, and vulnerability repair &
    Which quality attributes improved? &
    Assess non-functional gains and competing objectives. \\
    \addlinespace[3pt]
    \textbf{H} Human judgments &
    Code review, edits, acceptance, rejection, corrections, preferences &
    What do developers or organizations value? &
    Convey intent and authority under sparse, subjective review. \\
    \addlinespace[3pt]
    \textbf{M} Model judgments &
    LLM critiques, rankings, credit assignments, learned rewards &
    How does an evaluator assess the evidence? &
    Scale semantic evaluation, subject to evaluator reliability. \\
    \bottomrule
  \end{tabularx}
  \endgroup
\end{table}

%% file: tables/evolution-mechanisms.tex
\begin{table}[H]
  \centering
  \caption{Evolution mechanisms and reported results for 27 representative methods in Figure~\ref{fig:taxonomy-coding-agents}.}
  \label{tab:evolution-operations}
  \footnotesize
  \setlength{\tabcolsep}{2.5pt}
  \renewcommand{\arraystretch}{1.18}
  \begin{tabularx}{\linewidth}{@{}>{\RaggedRight\arraybackslash}p{2.70cm}>{\RaggedRight\arraybackslash}X>{\RaggedRight\arraybackslash}p{1.75cm}>{\RaggedRight\arraybackslash}p{3.45cm}>{\RaggedRight\arraybackslash}p{3.45cm}@{}}
    \toprule
    \textbf{Method} & \textbf{Mechanism} & \textbf{Target} & \textbf{Result} & \textbf{Benchmark} \\
    \midrule
    \rowcolor[HTML]{E8F1FA}\multicolumn{5}{@{}l}{\rule{0pt}{2.4ex}\textbf{Asset updates}} \\
    \href{https://arxiv.org/html/2511.05931v1}{\textcolor{black}{SAGE}} & Grounded plan abstraction & Memory & 73.2\% resolved & SWE-bench Verified \\
    \href{https://arxiv.org/abs/2507.23361}{\textcolor{black}{SWE-Exp}} & Experience distillation & Memory & 73.0\% Pass@1 & SWE-bench Verified \\
    \href{https://arxiv.org/pdf/2605.30105v2}{\textcolor{black}{EvoRepair}} & Scored experience retrieval & Memory & 93.47\% (15 turns) & PATCHEVAL \\
    \href{https://arxiv.org/html/2303.11366}{\textcolor{black}{Reflexion}} & Verbal reflection memory & Memory & 91.0\% & HumanEval \\
    \href{https://arxiv.org/html/2510.01003}{\textcolor{black}{Repository Memory}} & Commit-memory retrieval & Memory & 68.6\% loc. Acc@1 & SWE-bench Verified \\
    \href{https://arxiv.org/html/2603.13258}{\textcolor{black}{MemCoder}} & Commit-memory refinement & Memory & 68.4\% $\rightarrow$ 77.8\% & SWE-bench Verified \\
    \href{https://arxiv.org/html/2605.25430v1}{\textcolor{black}{CODESKILL}} & RL-based skill management & Skills & +8.24 points & Terminal-Bench 2 \\
    \href{https://arxiv.org/html/2511.13646v3}{\textcolor{black}{Live-SWE-agent}} & Runtime tool synthesis & Tools & 70.6\% $\rightarrow$ 75.4\% & SWE-bench Verified \\
    \href{https://arxiv.org/html/2402.12317}{\textcolor{black}{EVOR}} & Query-knowledge co-evolution & Context & 53.3\% Pass@1 & EvoR-bench (Tensor-M) \\
    \href{https://arxiv.org/html/2601.16746}{\textcolor{black}{SWE-Pruner}} & Goal-conditioned context pruning & Context & 70.6\% $\rightarrow$ 72.0\% & SWE-bench Verified \\
    \rowcolor[HTML]{FDEBD7}\multicolumn{5}{@{}l}{\rule{0pt}{2.4ex}\textbf{Architecture updates}} \\
    \href{https://arxiv.org/html/2505.22954}{\textcolor{black}{DGM}} & Archive-based code search & Agent code & 20.0\% $\rightarrow$ 50.0\% & SWE-bench Verified\textsuperscript{*} \\
    \href{https://arxiv.org/html/2504.15228v2}{\textcolor{black}{SICA}} & Reflective self-code editing & Agent code & 17.0\% $\rightarrow$ 53.0\% & SWE-bench Verified\textsuperscript{*} \\
    \href{https://arxiv.org/html/2606.06324}{\textcolor{black}{HarnessFix}} & Trace-guided harness repair & Harness & 45.3\% $\rightarrow$ 57.3\% & SWE-bench Verified\textsuperscript{*} \\
    \href{https://arxiv.org/html/2410.10762v4}{\textcolor{black}{AFlow}} & Workflow search (MCTS) & Workflow & 94.7\% Pass@1 & HumanEval\textsuperscript{*} \\
    \href{https://arxiv.org/html/2505.18646v1}{\textcolor{black}{SEW}} & Workflow and prompt mutation & Workflow & 92.1\% Pass@1 & HumanEval \\
    \href{https://arxiv.org/html/2502.07373}{\textcolor{black}{EvoFlow}} & Crossover, mutation, niching & Workflow & 92.85\% Pass@1 & HumanEval\textsuperscript{*} \\
    \href{https://arxiv.org/html/2601.07477}{\textcolor{black}{JUDGEFLOW}} & Judge-guided block search & Workflow & 93.4\% Pass@1 & HumanEval\textsuperscript{*} \\
    \href{https://arxiv.org/html/2410.16946}{\textcolor{black}{EvoMAC}} & Textual backpropagation & Team & 94.51\% Pass@1 & HumanEval \\
    \href{https://arxiv.org/html/2602.17100}{\textcolor{black}{AgentConductor}} & SFT + GRPO orchestration & Orchestrator & 58.8\% Pass@1 & APPS \\
    \href{https://arxiv.org/html/2602.04837}{\textcolor{black}{GEA}} & Group experience sharing & Agent code & 71.0\% resolved & SWE-bench Verified \\
    \rowcolor[HTML]{F0E8F7}\multicolumn{5}{@{}l}{\rule{0pt}{2.4ex}\textbf{Model updates}} \\
    \href{https://arxiv.org/abs/2512.18552}{\textcolor{black}{SSR}} & Self-play + joint RL & Model & +10.4 points & SWE-bench Verified \\
    \href{https://arxiv.org/html/2506.11442v2}{\textcolor{black}{ReVeal}} & Verification-aware RL (TAPO) & Model & 38.7\% (25 turns) & LiveCodeBench V6 \\
    \href{https://arxiv.org/html/2506.03136}{\textcolor{black}{CURE}} & Coder-tester co-evolution (RL) & Model & 33.5\% $\rightarrow$ 40.5\% & LiveCodeBench V2 \\
    \href{https://arxiv.org/html/2502.14948}{\textcolor{black}{Sol-Ver}} & Self-play, SFT, and DPO & Model & 18.23\% $\rightarrow$ 27.24\% & LiveCodeBench \\
    \href{https://arxiv.org/html/2505.03335}{\textcolor{black}{Absolute Zero}} & Self-play + TRR++ & Model & 19.9\% $\rightarrow$ 31.7\% & LiveCodeBench V1-5 \\
    \href{https://arxiv.org/html/2506.11425}{\textcolor{black}{Agent-RLVR}} & Guided retries, SFT, offline DPO & Model & 9.4\% $\rightarrow$ 22.4\% & SWE-bench Verified \\
    \href{https://arxiv.org/abs/2502.18449}{\textcolor{black}{SWE-RL}} & Patch-similarity reward + GRPO & Model & 41.0\% & SWE-bench Verified \\
    \bottomrule
  \end{tabularx}
  \par\smallskip
  \begin{minipage}{\linewidth}\scriptsize\RaggedRight
  Scores report final performance; arrows and gains indicate within-paper comparisons. Points denote percentage points. Repository Memory measures localization; EvoRepair measures cumulative repair. \textsuperscript{*}Benchmark subset. Evaluation settings differ across papers.
  \end{minipage}
\end{table}

%% file: tables/representative-coding-benchmarks.tex
\begin{table}[H]
  \centering
  \caption{Forty representative conventional coding benchmarks grouped by evaluation setting. We summarize their task units, programming languages, and official GitHub and Hugging Face resources.}
  \label{tab:representative-coding-benchmarks}
  \begingroup
  \scriptsize
  \setlength{\tabcolsep}{4pt}
  \renewcommand{\arraystretch}{0.96}
  \begin{tabularx}{\linewidth}{@{}>{\RaggedRight\arraybackslash}p{.22\linewidth}
    >{\RaggedRight\arraybackslash}p{.34\linewidth}
    >{\RaggedRight\arraybackslash}X
    >{\Centering\arraybackslash}p{.055\linewidth}@{}}
  \toprule
  \textbf{Benchmark} & \textbf{Task unit} & \textbf{Language} & \textbf{Link} \\
  \midrule

  \rowcolor[rgb]{0.88,0.95,1}
  \multicolumn{4}{c}{\textbf{Repository-level issue resolution}} \\
  \addlinespace[1pt]
  SWE-bench & GitHub issue and patch & Python &
  \ghhflinks{https://github.com/SWE-bench/SWE-bench}{https://huggingface.co/datasets/SWE-bench/SWE-bench} \\
  SWE-bench Verified & Expert-validated issue and patch & Python &
  \ghhflinks{https://github.com/SWE-bench/SWE-bench}{https://huggingface.co/datasets/SWE-bench/SWE-bench_Verified} \\
  SWE-bench Lite & Curated issue and patch subset & Python &
  \ghhflinks{https://github.com/SWE-bench/SWE-bench}{https://huggingface.co/datasets/princeton-nlp/SWE-bench_Lite} \\
  SWE-Gym & Executable repository issue & Python &
  \ghhflinks{https://github.com/SWE-Gym/SWE-Gym}{https://huggingface.co/SWE-Gym/datasets} \\
  Multi-SWE-bench & Multilingual issue and patch &
  Java, JS, TS, Go, Rust, C, C++ &
  \ghhflinks{https://github.com/multi-swe-bench/multi-swe-bench}{https://huggingface.co/datasets/ByteDance-Seed/Multi-SWE-bench} \\
  SWE-bench-Live & Recent issue and patch & Python; multilingual extensions &
  \ghhflinks{https://github.com/microsoft/SWE-bench-Live}{https://huggingface.co/datasets/SWE-bench-Live/SWE-bench-Live} \\
  SWE-bench-Live (ML/Win) & Recent multilingual and Windows issue &
  Python, JS, TS, C, C++, C\#, Java, Go, Rust &
  \githublink{https://github.com/microsoft/SWE-bench-Live}\hspace{0.14em}%
  \hflink{https://huggingface.co/datasets/SWE-bench-Live/MultiLang}\hspace{0.14em}%
  \hflink{https://huggingface.co/datasets/SWE-bench-Live/Windows} \\
  SWE-Bench Pro & Long-horizon issue and patch & Python, Go, TS &
  \ghhflinks{https://github.com/scaleapi/SWE-bench_Pro-os}{https://huggingface.co/datasets/ScaleAI/SWE-bench_Pro} \\
  SWE-bench++ & Expanded multilingual issue and patch &
  Python, Go, TS, JS, Ruby, PHP, Java, Rust, C++, C\#, C &
  \ghhflinks{https://github.com/TuringEnterprises/SWE-Bench-plus-plus}{https://huggingface.co/datasets/TuringEnterprises/SWE-Bench-plus-plus} \\
  SWE-EVO & Multi-step repository evolution & Python &
  \ghhflinks{https://github.com/SWE-EVO/SWE-EVO}{https://huggingface.co/datasets/Fsoft-AIC/SWE-EVO} \\
  SWE-bench Multimodal & Visual issue and patch & JS, TS &
  \ghhflinks{https://github.com/SWE-bench/SWE-bench}{https://huggingface.co/datasets/SWE-bench/SWE-bench_Multimodal} \\
  Visual SWE-bench & Visually grounded issue and patch & Python &
  \ghhflinks{https://github.com/luolin101/CodeV}{https://huggingface.co/datasets/luolin101/Visual-SWE-bench} \\
  SWE-PolyBench & Multilingual issue and patch & Python, Java, JS, TS &
  \ghhflinks{https://github.com/amazon-science/SWE-PolyBench}{https://huggingface.co/datasets/AmazonScience/SWE-PolyBench} \\
  SWE-bench Multilingual & Curated multilingual issue and patch &
  C, C++, Go, Java, JS, TS, PHP, Ruby, Rust &
  \ghhflinks{https://github.com/SWE-bench/swe-bench-multilingual-tasks}{https://huggingface.co/datasets/SWE-bench/SWE-bench_Multilingual} \\
  SWE-Sharp-Bench & C\# issue and patch & C\# &
  \ghhflinks{https://github.com/microsoft/prose/tree/main/misc/SWE-Sharp-Bench}{https://huggingface.co/datasets/microsoft/SWE-Sharp-Bench} \\
  SWE-Perf & Repository performance issue & Python, Cython &
  \ghhflinks{https://github.com/SWE-Perf/swe-perf}{https://huggingface.co/datasets/SWE-Perf/SWE-Perf} \\
  OmniGIRL & Multimodal multilingual issue and patch & Python, TS, Java, JS &
  \ghhflinks{https://github.com/deepsoftwareanalytics/omnigirl}{https://huggingface.co/datasets/Deep-Software-Analytics/OmniGIRL} \\
  SWE-Factory & Scalable repository issue and patch & Python, Java, JS, TS &
  \ghhflinks{https://github.com/DeepSoftwareAnalytics/swe-factory}{https://huggingface.co/SWE-Factory} \\
  FEA-Bench & Front-end engineering issue & Python &
  \ghhflinks{https://github.com/microsoft/FEA-Bench}{https://huggingface.co/datasets/microsoft/FEA-Bench} \\
  \addlinespace[1pt]
  \rowcolor[rgb]{0.88,0.95,1}
  \multicolumn{4}{c}{\textbf{Function-level and competition-style programming}} \\
  \addlinespace[1pt]
  HumanEval & Function completion & Python &
  \ghhflinks{https://github.com/openai/human-eval}{https://huggingface.co/datasets/openai/openai_humaneval} \\
  MBPP & Natural-language program synthesis & Python &
  \ghhflinks{https://github.com/google-research/google-research/tree/master/mbpp}{https://huggingface.co/datasets/google-research-datasets/mbpp} \\
  APPS & Interview and competition problems & Python &
  \ghhflinks{https://github.com/hendrycks/apps}{https://huggingface.co/datasets/codeparrot/apps} \\
  CodeContests & Competition problems & C++, Python, Java &
  \ghhflinks{https://github.com/google-deepmind/code_contests}{https://huggingface.co/datasets/deepmind/code_contests} \\
  LiveCodeBench & Temporally ordered coding problems & C++, Python, Java, others &
  \ghhflinks{https://github.com/LiveCodeBench/LiveCodeBench}{https://huggingface.co/datasets/livecodebench/code_generation} \\
  BigCodeBench & Function calls under complex instructions & Python &
  \ghhflinks{https://github.com/bigcode-project/bigcodebench}{https://huggingface.co/datasets/bigcode/bigcodebench} \\
  EvalPlus & Extended functional test suites & Python &
  \ghhflinks{https://github.com/evalplus/evalplus}{https://huggingface.co/datasets/evalplus/humanevalplus} \\
  MultiPL-E & Translated function synthesis & More than 18 languages &
  \ghhflinks{https://github.com/nuprl/MultiPL-E}{https://huggingface.co/datasets/nuprl/MultiPL-E} \\
  DS-1000 & Data-science code completion & Python &
  \ghhflinks{https://github.com/xlang-ai/DS-1000}{https://huggingface.co/datasets/xlangai/DS-1000} \\
  CRUXEval & Input and output prediction for functions & Python &
  \ghhflinks{https://github.com/facebookresearch/cruxeval}{https://huggingface.co/datasets/cruxeval-org/cruxeval} \\
  CoderEval & Repository-context function or method generation & Python, Java &
  \githublink{https://github.com/CoderEval/CoderEval} \\

  \addlinespace[2pt]
  \rowcolor[rgb]{0.88,0.95,1}
  \multicolumn{4}{c}{\textbf{Operational and specialized software tasks}} \\
  \addlinespace[1pt]
  Terminal-Bench & End-to-end terminal task & Bash, Python, task-dependent &
  \ghhflinks{https://github.com/harbor-framework/terminal-bench}{https://huggingface.co/datasets/harborframework/terminal-bench} \\
  InterCode & Interactive coding trajectory & Bash, SQL &
  \githublink{https://github.com/princeton-nlp/intercode} \\
  KernelBench & GPU kernel generation & Python, CUDA C++ &
  \ghhflinks{https://github.com/ScalingIntelligence/KernelBench}{https://huggingface.co/datasets/ScalingIntelligence/KernelBench} \\
  SEC-bench & Repository security task & Multiple repository languages &
  \ghhflinks{https://github.com/SEC-bench/SEC-bench}{https://huggingface.co/datasets/SEC-bench/SEC-bench} \\
  CRAVE & Code-review verdict & Multiple repository languages &
  \hflink{https://huggingface.co/datasets/TuringEnterprises/CRAVE} \\
  CompileBench & Legacy software compilation & C, C++, build scripts &
  \githublink{https://github.com/QuesmaOrg/CompileBench} \\
  DebugBench & Fault localization and program repair & C++, Java, Python &
  \githublink{https://github.com/thunlp/DebugBench} \\
  BIRD & Database-grounded text-to-SQL & SQL &
  \ghhflinks{https://github.com/bird-bench/mini_dev}{https://huggingface.co/datasets/birdsql/bird_sql_dev_20251106} \\
  Spider 2.0 & Enterprise data workflow and text-to-SQL & SQL, Python, dbt &
  \githublink{https://github.com/xlang-ai/Spider2} \\
  VerilogEval & RTL generation and simulation & Verilog &
  \githublink{https://github.com/NVlabs/verilog-eval} \\
  \bottomrule
  \end{tabularx}
  \endgroup
\end{table}

%% file: tables/product-landscape.tex
\begin{table}[H]
  \centering
  \caption{Representative self-evolving coding products organized by their
  documented persistent update layers.}
  \label{tab:product-landscape}
  \fontsize{7.6}{8.4}\selectfont
  \setlength{\tabcolsep}{3.2pt}
  \renewcommand{\arraystretch}{1.02}
  \newcommand{\companymark}[1]{%
    \raisebox{-0.2ex}{\includegraphics[height=1.3em,width=1.3em,
      keepaspectratio]{assets/company-logos/#1.png}}\hspace{0.35em}}
  \begin{tabularx}{\linewidth}{@{}
      >{\raggedright\arraybackslash}X
      >{\raggedright\arraybackslash}p{0.22\linewidth}
      >{\centering\arraybackslash}p{0.10\linewidth}
      >{\centering\arraybackslash}p{0.075\linewidth}
      >{\centering\arraybackslash}p{0.075\linewidth}
      >{\centering\arraybackslash}p{0.075\linewidth}@{}}
    \toprule
    \multirow{2}{*}{\textbf{Product}} &
    \multirow{2}{*}{\textbf{Company}} &
    \multirow{2}{*}{\textbf{Year}} &
    \multicolumn{3}{c}{\textbf{Persistent update layer}} \\
    \cmidrule(lr){4-6}
    & & & \textbf{Assets} & \textbf{Arch.} & \textbf{Model} \\
    \midrule
    \rowcolor[rgb]{0.88,0.95,1}
    \multicolumn{6}{@{}c}{\textbf{Architecture-spanning products}} \\
    Prime Agent~\citep{primeintellect2026primeagent} &
      \companymark{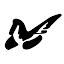}Prime Intellect & 2026 &
      $\bullet$ & $\bullet$ & $\circ$ \\
    DeepSeek Harness~\citep{deepseek2026harness} &
      \companymark{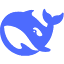}DeepSeek & 2026 &
      $\circ$ & $\bullet$ & $\circ$ \\
    OpenBlock Agent$^{\dagger}$~\citep{openblock2026ob1} &
      \companymark{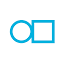}OpenBlock Labs & 2026 &
      $\bullet$ & $\bullet$ & $\circ$ \\
    \midrule
    \rowcolor[rgb]{0.86,0.94,0.84}
    \multicolumn{6}{@{}c}{\textbf{Asset-centered products}} \\
    Gemini CLI Auto Memory~\citep{google2026automemory} &
      \companymark{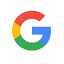}Google & 2026 &
      $\bullet$ & $\circ$ & $\circ$ \\
    GitHub Copilot Memory~\citep{github2026copilotmemory} &
      \companymark{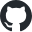}GitHub & 2026 &
      $\bullet$ & $\circ$ & $\circ$ \\
    Augment Agent / Cosmos~\citep{augment2025memoryreview} &
      \companymark{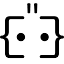}Augment & 2025--2026 &
      $\bullet$ & $\circ$ & $\circ$ \\
    Claude Code Auto Memory~\citep{anthropic2026automemory} &
      \companymark{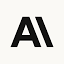}Anthropic & 2026 &
      $\bullet$ & $\circ$ & $\circ$ \\
    Cursor Memories / Automations~\citep{cursor2026automations} &
      \companymark{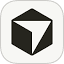}Cursor & 2025--2026 &
      $\bullet$ & $\circ$ & $\circ$ \\
    Devin Insights / Playbooks~\citep{cognition2026devininsights} &
      \companymark{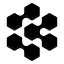}Cognition & 2025--2026 &
      $\bullet$ & $\circ$ & $\circ$ \\
    Windsurf Cascade Memories~\citep{windsurf2026memories} &
      \companymark{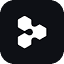}Windsurf & 2025--2026 &
      $\bullet$ & $\circ$ & $\circ$ \\
    Letta Code~\citep{letta2026code} &
      \companymark{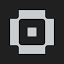}Letta & 2025--2026 &
      $\bullet$ & $\circ$ & $\circ$ \\
    Hermes Agent~\citep{nous2026hermes} &
      \companymark{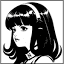}Nous Research & 2026 &
      $\bullet$ & $\circ$ & $\circ$ \\
    Replit Agent~\citep{replit2026skills} &
      \companymark{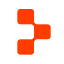}Replit & 2025--2026 &
      $\bullet$ & $\circ$ & $\circ$ \\
    OpenClaw~\citep{openclaw2026memory} &
      \companymark{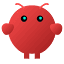}Community & 2026 &
      $\bullet$ & $\circ$ & $\circ$ \\
    \bottomrule
  \end{tabularx}
  \vspace{2pt}

  \begin{minipage}{\linewidth}
    \fontsize{7.6}{8.4}\selectfont
    $\bullet$ indicates a documented persistent update at the layer; $\circ$
    indicates that no persistent self-update is documented, not that the product
    lacks that component. $^{\dagger}$OpenBlock includes announced capabilities.
  \end{minipage}
\end{table}